\documentclass[epj-spec]{svjour}
\usepackage{graphics}
\usepackage{graphicx}
\usepackage{epsfig}
\usepackage{amssymb}

\newcommand{\be}{\begin{equation}}
\newcommand{\ee}{\end{equation}}
\newcommand{\ba}{\begin{eqnarray}}
\newcommand{\ea}{\end{eqnarray}}
\newcommand{\nn}{\nonumber}
\newcommand{\ex}{{\rm e}}
\newcommand{\Tr}  {\mathop{\rm Tr}}

\def\lsi{\raise0.3ex\hbox{$<$\kern-0.75em\raise-1.1ex\hbox{$\sim$}}}
\def\gsi{\raise0.3ex\hbox{$>$\kern-0.75em\raise-1.1ex\hbox{$\sim$}}}
\newcommand{\lsim}{\mathop{\lsi}}
\newcommand{\gsim}{\mathop{\gsi}}
\newcommand{\eq}{Eq.~}

\def\bfx{{\bf x}}
\def\bfy{{\bf y}}

\def\slash#1{#1\!\!\!\!/\!\,\,} 
\def\Dslash{\slash D}
\begin{document}

\title{Lattice QCD at finite temperature and density}
\author{Owe Philipsen\thanks{\email{o.philipsen@uni-muenster.de}}}
\institute{Institut f\"ur Theoretische Physik, Universit\"at M\"unster, 
D-48149 M\"unster, Germany}
\abstract{
QCD at finite temperature and density is becoming increasingly important for various experimental
programmes, ranging from heavy ion physics to astro-particle physics. 
The non-perturbative nature of non-abelian quantum field theories at finite temperature
leaves lattice QCD as the only tool by which we may hope to come to reliable predictions
from first principles. This requires careful extrapolations to the thermodynamic, chiral and continuum limits in order to eliminate systematic effects introduced by the discretization procedure.
After an introduction to lattice QCD at finite temperature and density, its possibilities and current systematic 
limitations, a review of present numerical results is given. In particular, plasma properties such as
the equation of state, screening masses, static quark free energies and spectral functions are discussed, as well as the critical temperature and the QCD phase structure at zero and finite density.  
} 
\maketitle
\section{Introduction}
\label{sec:intro}
Quantumchromodynamics (QCD) is the theory of the strong interactions, describing nuclear matter 
in terms of its constituent quarks and gluons and their interactions.  
One of its key features is asymptotic freedom, {\it i.e.}~the fact that the coupling 
strength is decreasing as a function of momentum transfer of an interaction. 
Thus, while the theory is amenable to perturbation theory at large momenta, it
is non-perturbative for energy scales $\lsim 1$ GeV and lattice QCD is the only known method for
first principles calculations in this regime. 
 
At a critical temperature $T_c\approx 200$ MeV, QCD predicts a transition between the familiar confined
hadron physics and a deconfined phase of quark gluon plasma (QGP).
At the same temperature, the weakly broken chiral symmetry, responsible for the three light pions, gets
restored. QCD at high temperatures is of outstanding importance in today's theoretical and experimental nuclear and particle physics programmes. A thermal environment with sufficiently high temperatures 
for a QCD plasma has certainly existed during the early stages of the universe, 
which passed through the quark hadron transition on its way to its present state. 
Current and future heavy ion collision experiments are
recreating this primordial plasma 
at RHIC (BNL), LHC (CERN) and FAIR (GSI).
These studies will have a bearing far beyond QCD in the context of 
early universe  and astro-particle physics.
Many prominent features of the observable universe, such as the baryon asymmetry or the seeding
for structure formation, have been determined primordially 
in hot plasmas described by non-abelian gauge theories. The QCD plasma serves as a prototype 
for those, since it is the only one we can hope to produce in 
laboratory experiments.
Moreover, 
certain QCD plasma properties entering calculations of dark matter abundances 
need to be known at percent level accuracy in order to match the ever 
more precise astro-physical data expected in the near future \cite{cosmo}.  
On the other hand, for high densities and low temperatures, exotic non-hadronic 
phases
such as a color superconductor have been predicted \cite{wilc}, and 
such physics might be realized in the cores of compact stars.

For these applications we need to know how the
properties of QCD change under extreme conditions of temperature and density.
This entails a determination of the QCD phase diagram and the associated critical 
properties, a quantitative understanding of the equation of state, the way
in which hadron properties get modified as well as the nature of the lightest excitations
in non-hadronic phases. 

Early hopes that the high temperature plasma phase might be accessible to perturbation theory
have proved wrong, as we shall see. Indeed, today one rather speaks of the strongly coupled
quark gluon plasma (sQGP). This leaves lattice QCD as the main computational tool even in that
regime. However, lattice simulations are still struggling with limitations and systematic errors.
For example, only in the last few years has it become possible to perform simulations for QCD
at small baryon density, while the case of cold and high density matter remains unaccessible to date.  
These lectures are intended to provide an introduction to the problems tackled by lattice simulations,
their potential and limitations, as well as an overview over current results.

\section{Finite temperature QCD in the continuum and on the lattice}
\label{sec:theo}

\subsection{QCD at finite temperature and density}

The thermodynamics of QCD is most conveniently described by the 
grand canonical partition function \cite{tbooks}
\be  
Z(V,{\mu_f},T;g,{m_f},)=\Tr \left(\ex^{-(H-\mu_f {Q_f})/T}\right)=
\int DA \,D\bar{\psi}\,D\psi \; \ex^{-S_g[A_\mu]} \,\ex^{-S_f[\bar{\psi},\psi,A_\mu]},
\label{part}
\ee
with the euclidean gauge and fermion actions
\ba
S_g[A_\mu]&=& \int\limits_0^{1/T} dx_0 \int\limits_V d^3x \;
\frac{1}{2} {\rm Tr}\; F_{\mu\nu} F_{\mu\nu}, \nn\\
S_f[\bar{\psi},\psi,A_\mu]&=& \int\limits_0^{1/T} dx_0 \int\limits_V d^3x \;
 \sum_{f=1}^{N_f} \bar{\psi}_f \left( \gamma_\mu
D_\mu+ m_f- \mu_f \gamma_0 \right) \psi_f .
\label{lagrangian}
\ea
The partition function depends on 
the external macroscopic parameters $T,V,\mu_f$, as 
well as on the microscopic parameters like masses and the coupling constant. 
The index $f$ labels the different quark flavours, and the conserved quark 
numbers corresponding to the chemical potentials $\mu_f$ are 
$Q_f=\bar{\psi}\gamma_0\psi$.
In the following we will consider mostly two and three flavours of quarks,
and always take $m_u=m_d$. The case $m_s=m_{u,d}$ is then denoted by $N_f=3$, while $N_f=2+1$
implies $m_s\neq m_{u,d}$.
For our purposes we will couple all flavours to the same chemical potential
$\mu$ unless otherwise stated.  
The chemical potential for baryon number is then $\mu_B=3\mu$.
Once the partition function is known,
thermodynamic properties such as free energy, pressure, average particle numbers
or the thermal expectation value of an operator $O$ readily follow,
\be
F=-T \ln Z,\quad
P=\frac{\partial (T\ln Z)}{\partial V},\quad
\langle Q_f\rangle=\frac{\partial (T\ln Z)}{\partial \mu_f},\quad
\langle O \rangle = Z^{-1}\Tr(O \ex^{-(H-\mu_f Q_f)/T}). 
\ee 
In the thermodynamic limit $V\rightarrow \infty$, 
we are interested in the corresponding densities, {\it i.e.}~$f=F/V, p=P/V=-f,\ldots$. 

While the formulation of QCD thermodynamics is straightforward, 
controlled evaluations are prohibitively difficult.
Perturbation theory, the tool so successful for electroweak physics at zero temperature,  fails us 
completely in this context. The reasons are two-fold. 
Firstly, the running coupling is still not weak enough at temperature scales of interest, $g(100-300{\rm MeV})\sim 1$. Worse still is the fact that finite $T$
perturbation theory for non-abelian gauge theories displays an insurmountable infrared problem
\cite{linde}.  At finite T, a gauge boson self coupling gets dressed with 
a thermal distribution function which for small momenta behaves as
\be
\frac{g^2}{e^{E/T}-1}
\stackrel{E,p\ll T}{\sim}\;\frac{g^2T}{m}.
\ee
Thus, the effective expansion parameter diverges for perturbatively zero mass 
particles such as gluons, no matter how weak the coupling, {\it i.e.}~this problem
also exists in the symmetric electroweak phase. 
Higher orders do generate an effective gluonic mass
scale $m\sim g^2 T$ to cure this divergence, and resummation methods have
been devised to calculate it \cite{resum}. 
But now the coupling cancels out of the expansion parameter and all loop orders 
contribute in the same way, leaving convergence of such series unclear. 
Hence, a fully non-perturbative treatment of the problem is warranted.

\subsection{The Lattice Formulation}
\label{latte}

The thermal quantum field theory is discretized by introducing a 
euclidean space-time 
lattice $L^3\times N_t$ with lattice spacing $a$, such that volume and 
temperature are
\be \label{latvt}
V=(aL)^3,\qquad T=\frac{1}{aN_t}\,.
\ee
The fermion fields live on the lattice sites $x$, 
whereas the gauge fields are represented by
link variables $U_\mu(x)\in SU(3)$ connecting the sites. 
The simplest actions used in many thermodynamics calculations 
are the Wilson gauge action and the staggered, or Kogut-Susskind, fermion action
\be
S_g[U]=\sum_{x}\sum_{1\leq\mu<\nu\leq 4}\beta\left(1-\frac{1}{3}{\rm Re}
\Tr U_p\right),\qquad 
S_{f}^{KS}
= \sum_{x,y} \bar{\chi}(x) M_{xy}\, {\chi}(y),
\ee
respectively. Here
$U_p=U_\mu(x)U_\nu(x+a\hat{\mu})U^\dag_\mu(x+a\hat{\nu})U^\dag_\nu(x)$ is
the elementary plaquette and the bare lattice and continuum gauge couplings 
are related by $\beta=6/g^2$. The staggered fermion matrix is given by
\begin{eqnarray}
M_{x y}(m_f, \tilde{\mu}) &=&
{1\over 2} \sum_{\mu =1}^3 (-1)^{x_0+\dots+x_{\mu-1}}
( \delta_{x+\hat{\mu},y} U_\mu(x) - \delta_{x,y+\hat{\mu}}
  U_\mu(y)^{\dagger}) \nonumber \\
& &
+{1\over 2}(\delta_{x+\hat{0},y} U_0(x)\; \ex^{a\mu} -
\delta_{x,y+\hat{0}} U_0(y)^{\dagger}\ex^{-a\mu}) \;
+ \; \delta_{x y} am_f \quad .
\end{eqnarray}
Note that the chemical potential $a\mu$ enters through the 
temporal links only \cite{latmu}.
Now the Gaussian integral over the quark fields can be performed, leading to the
partition function
\be
Z(L,a\mu,N_t;\beta,N_f,am_f)=
        \int DU\, \prod_f \left(\det M(m_f,\mu)\right)^{1/4}\ex^{-S_g[U]}.
        \label{latact}        
\ee
The lattice action is not unique, any action reproducing the continuum action 
in the limit $a\rightarrow 0$ is in principle admissible. 
We refer to textbooks \cite{lbooks} for discussions of various actions
and their advantages/disadvantages, as well as 
the daunting task of a chiral fermion formulation. 
While the above staggered fermions
have a remnant U(1) chiral symmetry, they also exhibit spurious flavours 
(called tastes) which then have to be removed by taking the fourth root 
of the fermion determinant in \eq(\ref{latact}).
It is not yet settled whether this produces potentially hazardous 
non-localities or singularities, a subject of much current debate \cite{sharpe}.
Things empirically seem to work well as long as continuum limits are taken before
chiral limits.
On the other hand, Wilson fermions do not require roots of determinants,
but they break chiral symmetry completely at
finite lattice spacing and require additive quark mass renormalization. They 
also feature so-called exceptional configurations with negative eigenvalues
of the determinant, making simulations with light quarks very difficult.
There are also fermion actions avoiding those issues, such as domain wall 
fermions or overlap fermions, for reviews see \cite{dom,ovlp}. 
However, at the present stage these actions are an order of magnitude
more costly in simulation time and hence not yet widely used for
dynamical thermodynamics simulations.
It is for this reason that most results quoted in this review 
have been produced with staggered or Wilson fermions and their improved variants. 
The non-uniqueness of lattice actions can be exploited to construct
actions which remove the $O(a),O(a^2)\ldots$ corrections to continuum results,
thus improving the approach to the continuum limit. Introductions to
improvement can be found in \cite{impr}.

\subsection{Constraints on lattice simulations and systematic errors}
\label{sec:syst}

It is important to realize from the outset that current lattice simulations, and especially those
at finite temperature and density, are still hampered by systematic 
errors and uncertainties. Let us discuss the origins of those. The Compton wave length of a hadron is proportional to its inverse mass, $\sim m_H^{-1}$, and the largest of those constitutes the correlation length of the statistical system. 
To keep finite size as well as discretization errors under control, we need to require
\be
a\ll m_H^{-1}\ll aL.
\label{latconst}
\ee
Typical lattice sizes today are $(12-32)^3\times 4$, $16^3\times 8$ etc., depending on computing budgets and machines.
For temperatures around $T\sim 200$ MeV, \eq(\ref{latvt}) with $N_t=4-6$ implies
$a\sim 0.1-0.3$ fm and $aL\sim 1.5-3$ fm. For the low $T$ phase, \eq(\ref{latconst}) then tells us that the lightest hadron needs to be $m_{\pi}\gsim 250$ MeV. The push to do physical quark masses is only just beginning to be a possibility on the most powerful machines and with the cheapest actions
({\it i.e.}~staggered, with Wilson rapidly catching up). On the other hand, at high $T$
screening masses scale as $m_H\sim T$ requiring $N_t^{-1}\ll 1\ll LN_t^{-1}$. Hence, the spatial
lattice size should be significantly larger than the temporal one. 
This limits the feasible temperatures to several $T_c$. 

Once the simulations have been carried out, all quantities appear in lattice units,
{\it i.e.}~as dimensionless numbers, and need to be translated to physical units. 
This procedure of ``setting the scale'' introduces additional systematic 
errors, which are often much larger than the statistical errors of the simulations. 
Suppose we want to convert some measured critical
temperature to physical units. It thus needs to be related to a
quantity whose value in nature we know and which can also be obtained from a lattice simulation,  for example a hadron mass at zero temperature, 
\be
\frac{T}{m_H}= \frac{1}{N_t(am_H)}.
\ee
In practice this is difficult, because QCD with physical parameters is not 
quite doable numerically yet. Moreover, one is often interested in the 
limit of heavy or zero quark masses, or with two or three degenerate flavours. 
In such cases one uses quantities related to the static potential,
such as the string tension $\sigma$ or the Sommer 
scale $r_0$ \cite{sommer},
which strictly speaking only exist
in the pure gauge or heavy quark effective theories, 
\be
\frac{T}{\sqrt{\sigma}}=\frac{1}{(a\sqrt{\sigma})N_t},\quad \sqrt{\sigma}\approx 425 {\rm MeV};
\qquad Tr_0=\frac{r_0}{aN_t}\quad \mbox{with} \quad 
\left. r^2\frac{dV(r)}{dr}\right|_{r=r_0}=1.65.
\ee
The values are provided through analyses of 
heavy quark effective theories applied to
$b$-quark systems, {\it e.g.}~\cite{gray}. 
Fortunately the string tension turns out to be rather insensitive to the presence of light quarks, {\it i.e.}~it is
approximately the same independent of the quark content of QCD.

Another difficulty arises for extrapolations to the continuum limit.
Given some temperature $T$, the lattice spacing is set via $T=1/(a N_t)$, which then controls
the gauge coupling $\beta=6/g^2(a)$.  Thus, changing $\beta$ changes temperature 
for a lattice with fixed
$N_t$. As a consequence, the cut-off in a simulation at fixed $N_t$ varies
as a function of temperature.  {\it E.g.}, if a simulation is performed for a set of 
quark masses fixed in lattice units, $am_f=const$, then changing $a(T)$ implies 
that the quark mass in physical units changes as well! In principle this can be avoided by simulating along the ``lines of constant physics'', {\it i.e.}~tuning bare masses together with the lattice spacing
such that the masses in physical units stay
constant. In practice, this is a formidable task since the lines of constant physics are, of course, not
known and have to be mapped out non-perturbatively first. 
Again, simulations along lines of constant physics are only
just being started.

\subsection{Finite baryon density and the sign problem}

As soon as a chemical potential for quark number is switched on, $\mu\neq 0$, a Monte Carlo
evaluation of the partition function eq.~(\ref{latact}) is essentially impossible due to the so-called sign problem of QCD. The problem is encapsulated in the $\gamma_5$-hermiticity of the Dirac operator, 
\be
\Dslash(\mu)^\dagger = \gamma_5 \Dslash(-\mu^*) \gamma_5,
\label{gamma5}
\ee
which implies that $\det M$ is complex for the gauge group $SU(3)$ and $\mu\neq 0$.
Note that the partition function and all physical observables remain real, 
{\it i.e.}~the imaginary parts of the determinant cancel out once they are averaged over all gauge configurations.
However, in a Monte Carlo evaluation using importance sampling, 
the determinant is evaluated in the background of single gauge configurations 
and interpreted as part of the probability weight for that configuration. 
This is impossible if the determinant is complex. 
Methods to circumvent this problem
will be discussed in sec.~\ref{sec:mu}.

\subsection{The quenched limit, the chiral limit and physical QCD }
\label{sec:limits}

Much of our intuition regarding the quark hadron transition is based on the limiting 
cases of infinite or zero quark masses, {\it i.e.}~the quenched (or static quark) and chiral limits, respectively. In these cases there exist
true order parameters and the phase structure and transition can be discussed qualitatively
based on symmetry breaking and universality arguments.

For infinitely heavy quarks QCD reduces to Yang-Mills theory in the presence of static sources.
The static quark propagator is given by a Wilson line closing through the time boundary, {\it i.e.}~the Polyakov loop, 
\be
L({\bfx})=\prod_{x_0}^{N_t}U_{0}(x).
\ee 
The action displays a global symmetry under center transformations $U_0(x)\rightarrow z_nU_0(x)$, with $z_n=\exp{i2\pi n/3}\in Z(3)$. On the other hand, static sources transform non-trivially,
$L(\bfx)\rightarrow z_nL(\bfx)$. 
The free energy of a static quark is given by \cite{mcls}
\be
\langle \Tr L(\bfx) \rangle\sim \ex^{-\frac{F_q}{T}},
\ee
and constitutes an order parameter 
for confinement, {\it i.e.}~$\langle L\rangle =0$ for $T<T_c$, and $\langle L\rangle \neq 0$ for $T>T_c$. Hence there must be a true non-analytic phase transition corresponding to the breaking of the $Z(3)$ symmetry. 
One can construct an effective theory for the order parameter reflecting its $Z(3)$ symmetry \cite{sveya}, and universality arguments suggest a first order phase transition. This is indeed borne out
by simulations. 

In the limit of zero quark masses the classical QCD Lagrangian is invariant under global chiral symmetry
transformations, the total symmetry being $U_A(1)\times SU_L(N_f)\times SU_R(N_f)$. The
axial $U_A(1)$ is anomalous, quantum corrections break it down to $Z(N_f)$. The remainder
gets spontaneously broken to the diagonal subgroup, $SU_L(N_f)\times SU_R(N_f)\rightarrow 
SU_V(N_f)$, giving rise to $N_f^2-1$ massless Goldstone bosons, the pions. The order parameter
signalling chiral symmetry is the chiral condensate,
\be
\langle\bar{\psi}\psi\rangle =\frac{1}{L^3N_t}\frac{\partial}{\partial m_f} \ln Z.
\ee
It is nonzero for $T<T_c$, when chiral symmetry is spontaneously broken, and zero for $T>T_c$.
Effective models for the order parameter in this case are thus $O(2N_f)$ linear 
sigma models \cite{chiral}.
The finite temperature phase transition then corresponds to chiral symmetry restoration.

QCD with physical quark masses obviously does not correspond to either limit. The $Z(3)$ 
symmetry is explicitly broken once there are dynamical fermions, and the chiral symmetry
is broken by non-zero quark mass terms. 
Nevertheless, physical QCD displays confinement as well as three very light pions as ``remnants''
of those symmetries. In the presence of mass terms there is no true order parameter,
{\it i.e.}~the expectation values of the Polyakov loop as well as the chiral condensate are non-zero
everywhere. Hence the deconfined or chirally symmetric phase is analytically 
connected with 
the confined or chirally broken phase, and there is no need for a non-analytic phase transition. The following questions then arise, which should be answered by 
numerical simulations: for which parameter values of QCD is there
a true phase transition, and what is its order? Are confinement and chirality changing across the same single transition or are there different ones? If there is just one transition, which is the driving mechanism?
If there is only a smooth crossover, how do the properties of matter change?

\subsection{\label{sec:dr}Effective high T theory: dimensional reduction}

As we have seen, systematics presently 
constrain simulations of lattice QCD to temperatures below a few times 
$T_c$. However, for applications to early universe physics, 
as well as our own understanding and
comparisons with perturbation theory, 
we would also like to have non-perturbative information
at much higher temperatures. This can be achieved by means of effective field theories.
At large $T$, when the gauge coupling $g(T$) is sufficiently small, 
a hierarchy between different relevant scales of thermal QCD develops,
\be \label{eq:scale}
2\pi T \gg gT \gg g^2T.
\ee
The lowest non-vanishing bosonic Matsubara mode $\sim 2\pi T$ is characteristic for
non-interacting particles. The dynamics generates the Debye scale $m_E\sim gT$, over which
color electric fields $A_0$ are screened, and its non-perturbative analogue $m_M\sim g^2T$  
for color magnetic fields $A_i$ \cite{linde}.

For physics on scales larger than the inverse temperature, $|\bfx|\sim 1/gT\gg
1/T$, this allows an effective theory approach in which the
calculations are factorized:
integration over the hard modes may be performed perturbatively
by expanding in powers of the ratio of scales $gT/(2\pi T)\sim g/(2\pi)$.
This includes all non-zero Matsubara modes, in particular the fermions.
It results in a 3d effective theory for modes $\sim gT$ and softer, 
\be \label{actc}
        S_{eff} = \int d^{3}x \left\{ \frac{1}{2} \Tr(F_{ij}F_{ij})
        +\Tr(D_{i}A_0)^2 
+m_E^2 \Tr(A_0^2)
        +\lambda_3(\Tr(A_0^2)^{2} \right\} ,
\ee 
and is
known as dimensional reduction \cite{dr}. 
Since 4d euclidean time has been integrated over, 
$A_0$ now appears as an adjoint scalar, and the effective
parameters are functions of the original ones, 
$g_3^2=g^2T,m_E(N,N_f,g,m_q^f)\sim gT,\lambda_3(N,N_f,g,m_q^f)\sim g^4T$.

Discretization and simulation of this reduced problem is easy. 
Without fermions and in one dimension less,
much larger volumes and finer lattices can be employed.
Moreover, 3d gauge theories are superrenormalizable and the coupling scales linearly with
lattice spacing. Hence, very accurate continuum limits can be obtained and systematic
errors from simulations can be eliminated.
However, the reduction step
entails two approximations: the perturbative computation is limited to
a finite order in $g$ and neglects higher-dimensional operators, which are
suppressed by powers of the scale ratio.
The reduction step has been performed up to two-loop order \cite{ad} at which parameters
have relative accuracy ${\cal O}(g^4)$,
while for correlation functions the error is \cite{rules}
$\delta C/C\sim {\cal{O}}(g^3)$.
In the treatment of the electroweak phase transition,
this is less than 5\%\cite{ew},
for QCD it depends on the size of the coupling $g(T)$ and thus on $T$.  It has been
non-perturbatively checked by comparing simulations in 4d with those of the 3d effective theory,
that the latter 
accurately reproduces static correlation lengths
at temperatures as low as 
$T\gsim 2T_c$ \cite{hlp1,drtest},
thus allowing for a straightforward treatment of very large temperatures as
well as detailed dynamical investigations in the plasma phase.

\section{Numerical results at finite T  and $\mu=0$}

\subsection{The (pseudo-) critical temperature}
\label{sec:tc}

The first task when investigating finite temperature QCD is to find the phase boundary. Thus, for QCD
with a fixed quark content, we are interested in the critical (or pseudo-critical) temperature where the
transition from the confining regime to the plasma regime takes place.
The method to locate a transition usually is to look for
rapid changes of observables like the Polyakov loop $L$,
the chiral condensate or the plaquette,  and for peaks of their fluctuations, 
{\it i.e.}~the generalized susceptibilities
\be
\chi_O = V N_t \langle ( O - \langle O \rangle )^2 \rangle. 
\ee
The locations of these peaks define (pseudo-)critical
couplings $\beta_c$ which can be turned into temperatures through
the knowledge of a zero temperature quantity like a hadron
mass $am_H(\beta_c)$ at the same coupling,
$T_c / m_H = 1/(N_t am_H(\beta_c))$. In practice, often the two-loop beta function is used as a short cut,
although this becomes valid only when the lattice spacing is fine enough to be in the perturbative regime.

\begin{figure}[t]
\centerline{
\epsfig{bbllx=105,bblly=220,bburx=460,bbury=595,file=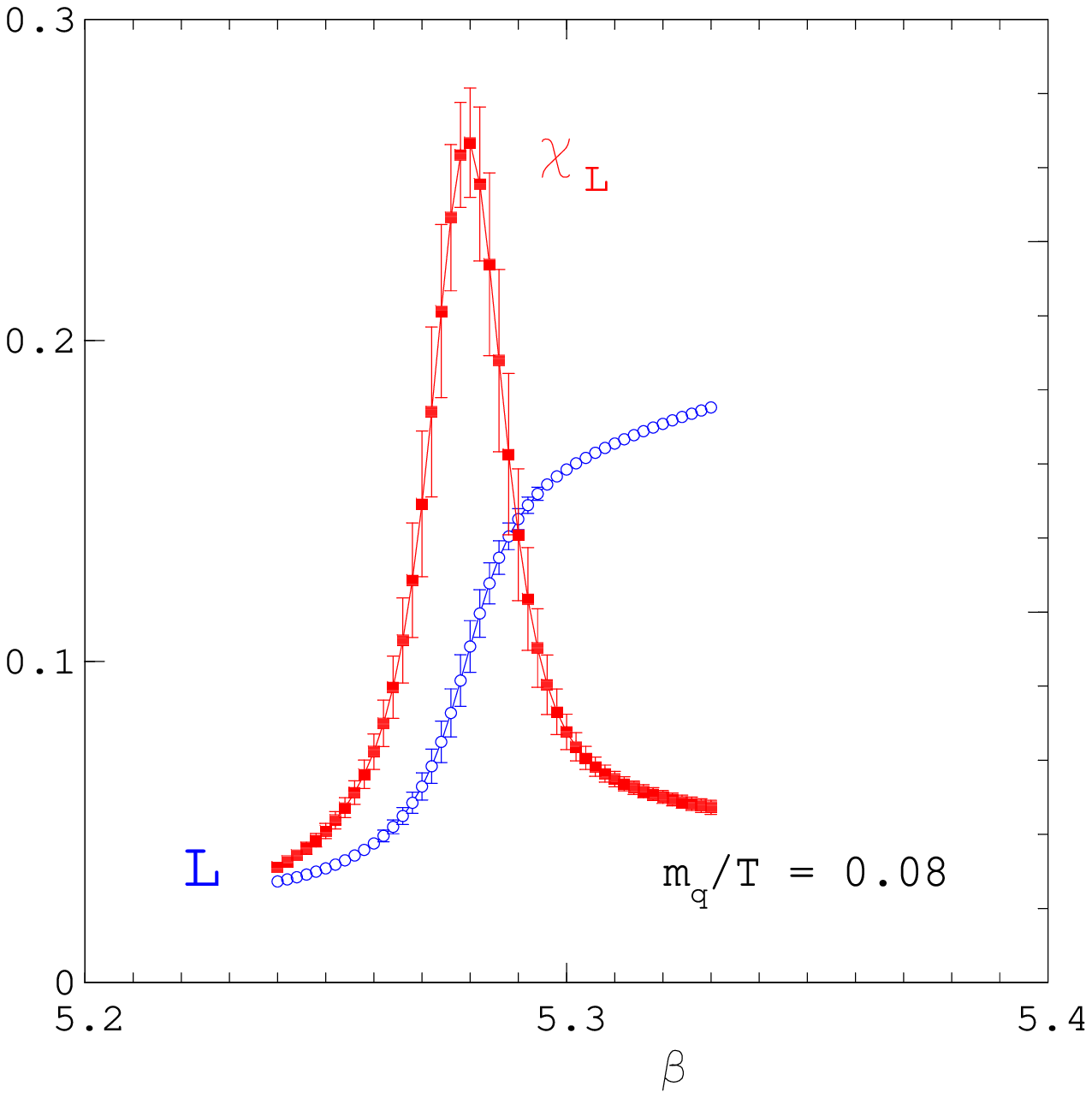,height=45mm} 
\epsfig{bbllx=105,bblly=220,bburx=460,bbury=595,file=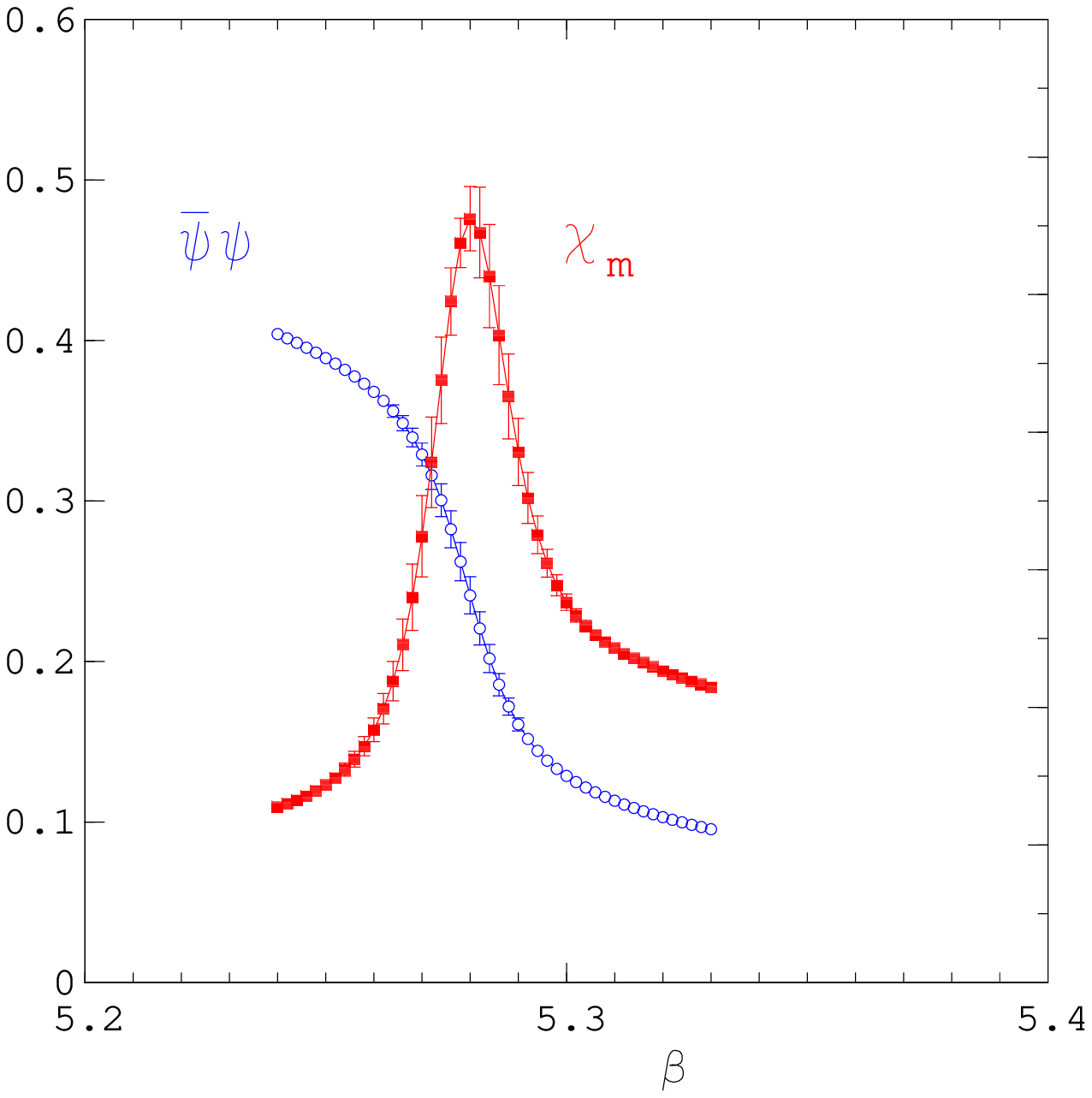,height=45mm} 
\epsfig{file=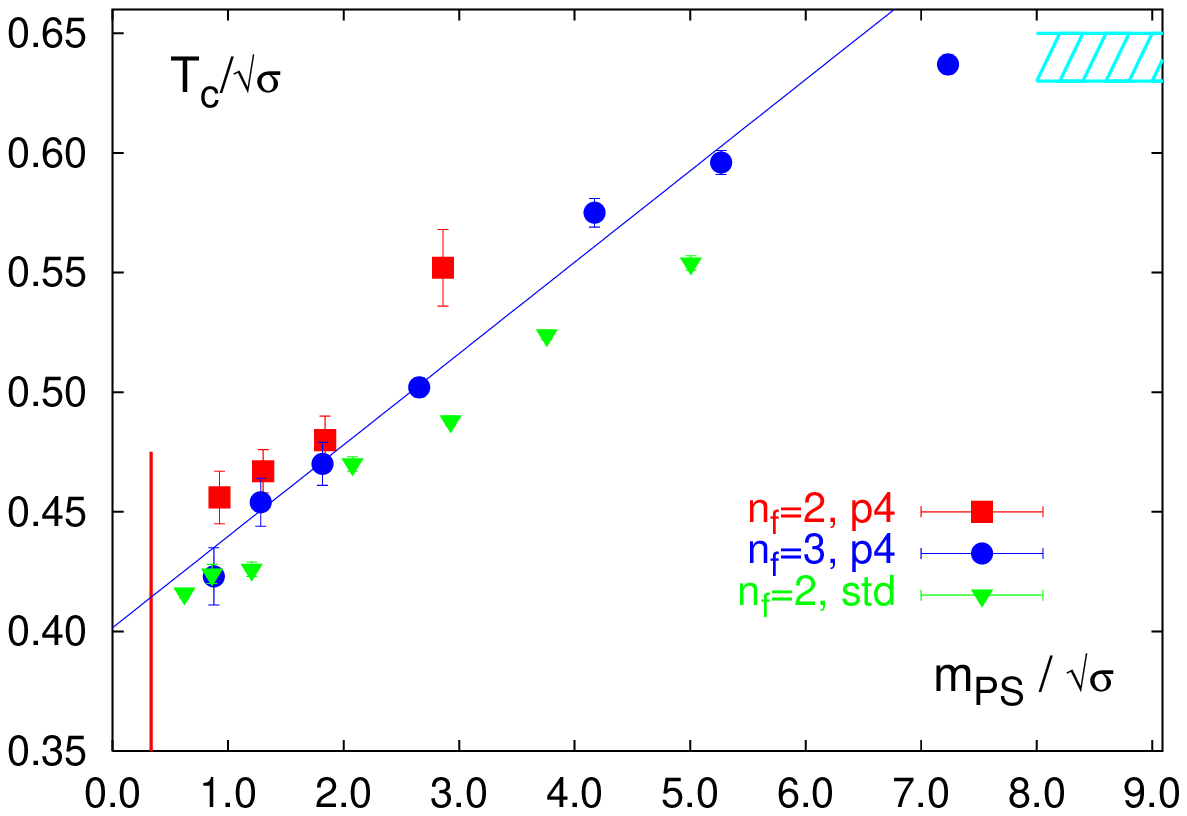,height=45mm}
}
\caption{The order parameters for deconfinement $\langle L\rangle$ (left) and chiral symmetry restoration $\langle \bar{\psi}\psi\rangle$ (middle), together with their susceptibilities as a function
of the gauge coupling for two-flavour QCD. From \cite{karsch}. Right: Quark mass dependence of
$T_c$ for $N_f=2,3$ with improved (p4) and standard staggered quarks. From \cite{Peikert}.}
\label{fig:tc}       
\end{figure}

On coarse lattices $N_t=4,6$, all numerical evidence is consistent with confining and chiral properties changing at approximately the same $\beta_c$. 
An example is shown in fig.~\ref{fig:tc} (left and middle) for two-flavour QCD.
On finite volumes, true non-analytic phase transitions do not exist and the identified phase boundary
represents only a crossover.
The nature and critical properties of the transition can be obtained
from scaling behaviour of various quantities with volume and quark mass,  
as will be discussed in sec.~\ref{sec:pd}.

$T_c$ for the pure $SU(3)$ gauge
theory has been known rather precisely for quite a while \cite{Boyd_eos,Iwa_tc}. 
It has since then been confirmed
by studies using various improved actions,
thus reducing
finite lattice spacing effects.
The number is most readily given in terms of the string tension, 
\begin{equation}
T_c / \sqrt{\sigma} = 0.632 \pm 0.002,
\end{equation}
which was obtained as a weighted average over all available lattice data \cite{anrv}.
For QCD with dynamical quarks, one finds $T_c$ to be increasing with quark mass,
as shown in fig.~\ref{fig:tc} (right). The data correspond to one lattice spacing 
only, {\it i.e.}~are not yet continuum results. 
Extrapolating the results to the chiral limit, 
one obtains
\begin{eqnarray}
\underline{\rm 2-flavor~ QCD:} &&
T_c  = \cases{
(171\pm 4)\; {\rm MeV}, & clover-improved Wilson \cite{japan_tc}     \nonumber \cr
(173\pm 8)\; {\rm MeV}, & improved staggered \cite{Peikert}     \nonumber \cr
             }  \nonumber \\
\underline{\rm 3-flavor~ QCD:} &&
T_c  = \; \; \;\, (154\pm 8)\; {\rm MeV}, \;\; \hspace*{0.1cm}
    \mbox{improved staggered \cite{Peikert}}
\nonumber
\end{eqnarray}
where $m_\rho$ has been used to set the scale. In view of the systematic errors introduced by
the discretization of the fermion sector, it is reassuring that consistent results are obtained from
staggered and Wilson fermions. 
Qualitatively we see that adding light degrees of freedom reduces
the critical temperature for the transition.

\begin{figure}[t]
\hspace*{-2cm}
\includegraphics[height=4cm]{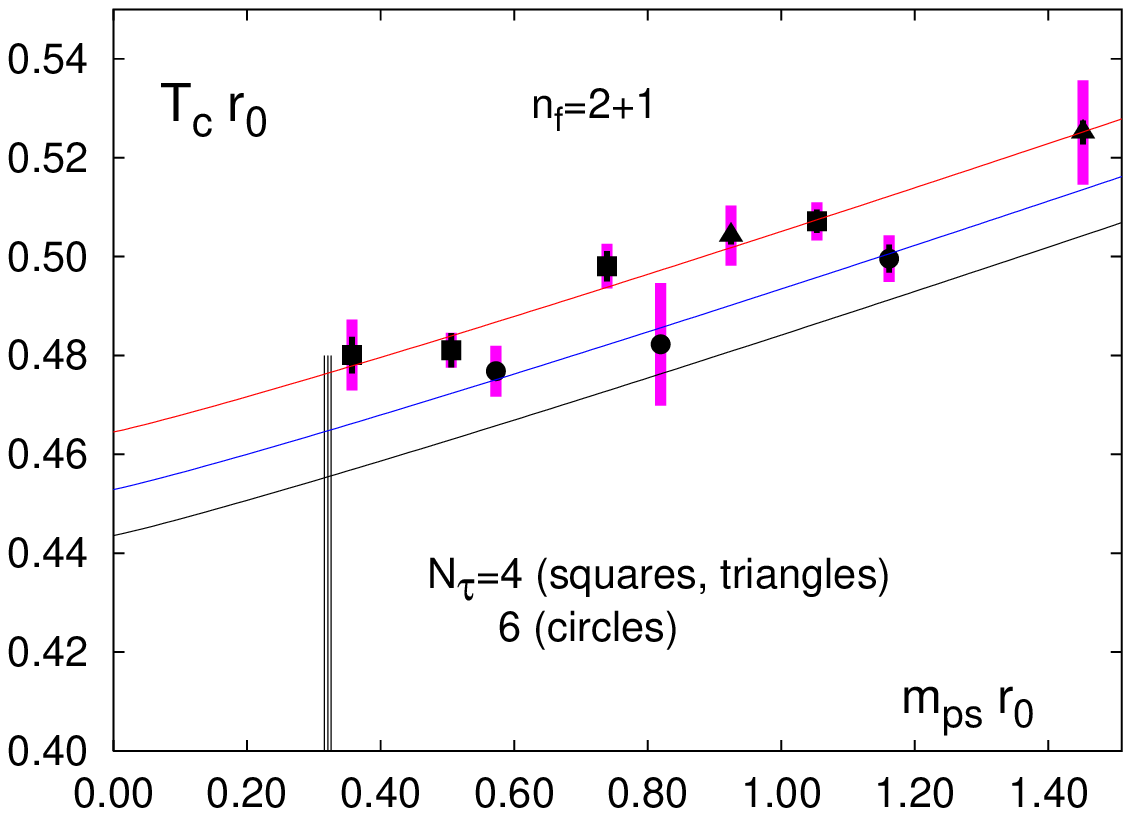}
\includegraphics*[height=4cm,bb=18 520 590 704]{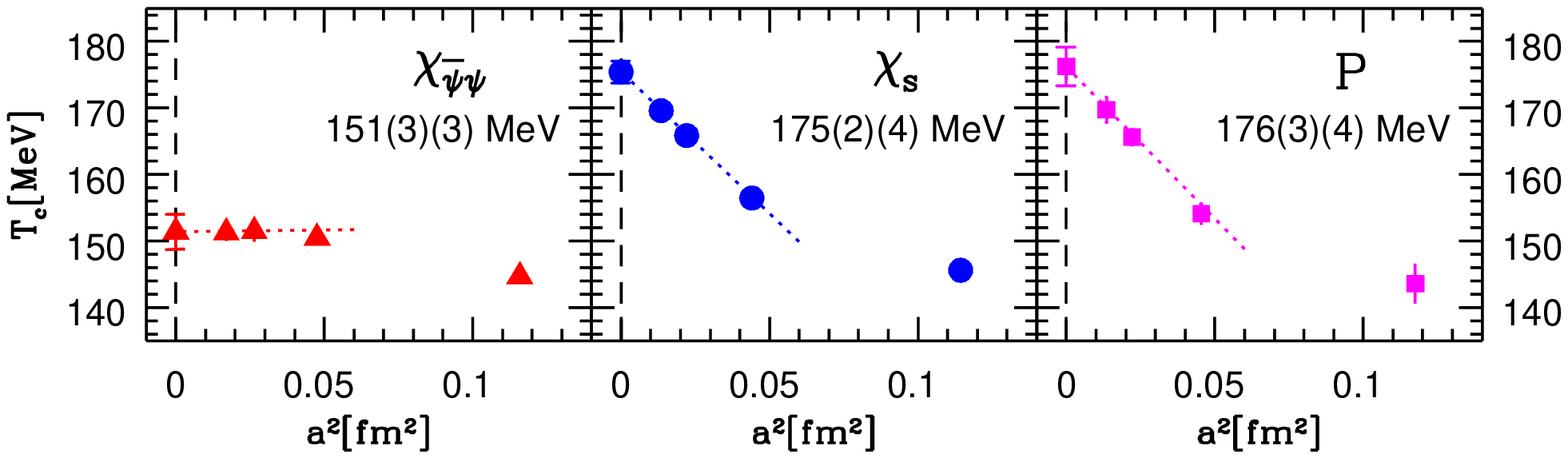}
\caption[]{Left: $T_c$ in units of $r_0$ as a function of pseudo-scalar mass for lattices with $N_t=4,6$.
The vertical line shows the physical value $m_{PS}r_0=0.321(5)$. From \cite{bbcr}. 
Right: Continuum extrapolations of $T_c$, determined from susceptibilities of 
the chiral condensate, the strange quark susceptibility and the Polyakov loop for physical quark masses. From \cite{fkt0}.} 
\label{fig:t0}  
\end{figure}

Having studied the flavour and quark mass dependence of $T_c$, 
the phenomenologically important quantity is of course $T_c$ for physical 
quark masses in the continuum limit. Two recent works
have pushed in this direction \cite{bbcr,fkt0}, and the comparison of the two is an interesting illustration of systematic effects in lattice calculations. 
The authors of \cite{bbcr} have worked with two lattice
spacings $N_t=4,6$, with the strange quark mass tuned to its physical value and a range 
of light quark masses, $am_{u,d}=0.1-0.4\, m_s$. Critical couplings are 
determined from peaks of chiral and Polyakov loop susceptibilities 
and found to be consistent with each other. They are then converted to 
temperature via a renormalization group improved two-loop beta function 
and the scale is set by $r_0$ (cf.~sec.~\ref{sec:syst}). Finally, 
a combined extrapolation to the continuum and chiral
limits is performed, as shown in fig.~\ref{fig:t0} (left), leading to the prediction $T_c=192(7)(4)$ MeV.

\begin{figure}[t]
\centerline{\includegraphics*[height=4.5cm,bb=90 620 492 816]{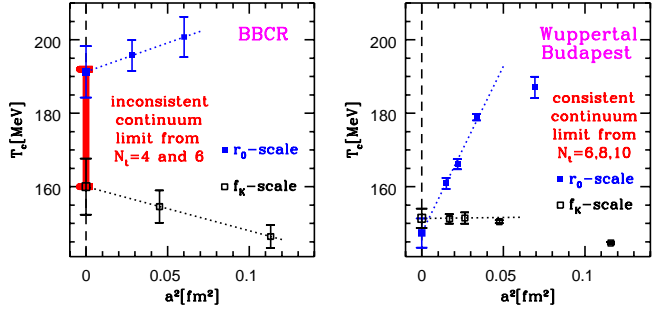}}
\caption[]{Left: Reanalysis of the data from \cite{bbcr} with different ways of setting the scale.
Right: Same for the data from \cite{fkt0}. Both plots from \cite{fkt0}.}
\label{fig:t0comp}
\end{figure}

A different number is quoted in \cite{fkt0}. In this work four lattice spacings are considered,
$N_t=4,6,8,10$, for which the quark masses are tuned along the lines of constant physics,
such that the meson masses correspond to nearly their physical values on all lattices.
The critical couplings are determined through the chiral and strange quark susceptibility, as well
as from the steepest change of the Polyakov loop. This is done for all lattice spacings, followed by a continuum extrapolation, as shown in fig.~\ref{fig:t0} (right). The scale in this work is
set by the kaon decay constant $f_K$, and the final value is based on the extrapolation of the
chiral condensate, $T_c=151(3)(3)$.
There are two striking observations to be made. Firstly, the $N_t=4$ data do not fall onto the straight $a^2$-extrapolation line to the continuum limit, {\it i.e.}~are not in the scaling region yet. 
Second, when finer lattices are considered, different observables lead to 
different critical temperatures. This is to be expected for a crossover, 
where $T_c$ is only pseudo-critical and not uniquely defined.
As we shall see in sec.~\ref{sec:pd}, this is the situation for physical QCD. 
Nevertheless, the gap is
larger than expected, and also the ordering is puzzling: 
it suggests that there is a temperature
range still displaying confinement while chiral symmetry is already restored, which goes against
conventional wisdom.
As a potential source for the discrepancy with \cite{bbcr}, the authors of \cite{fkt0} identify
cut-off effects, as is illustrated in fig.~\ref{fig:t0comp}. If the continuum extrapolation is performed
with two different methods to set the scale, then the data of \cite{bbcr} appear to lead to inconsistent
results. Likewise, a continuum extrapolation using only $N_t=4,6$ from the data in \cite{fkt0} would 
give a larger $T_c$ than using the full data set. On the other hand, comparing 
the results obtained from the analysis of Polyakov loops between the two works
gives only a small discrepancy. 
Clearly, cross checks with higher precision, more lattice spacings and other fermion discretizations are needed in order to come to fully conclusive results for the continuum. 
This discussion elucidates the difficulty of extracting accurate continuum information from lattice data, when statistical errors only appear to be small and under
control.

\subsection{The equation of state}

Energy density $\epsilon (T)$ and pressure $p(T)$ as a function
of temperature are certainly among the most
fundamental thermodynamic quantities of QCD
governing, {\it e.g.}, the expansion
of the plasma in the early universe as well as in heavy ion collisions.
The partition function for gases of free fermions and bosons are known exactly, and we recall
the results in the relativistic and non-relativisic limits, the former corresponding to the famous
Stefan-Boltzmann result:
\be
\begin{array}{rcl}
\textsl{Relativistic Boson, $m\ll T$ } 
& \textsl{$\times$ (Fermion) }&  \textsl{Non-relativistic, $m\gg T$}\\
 p_r = g\frac{\pi^2}{90} T^4 & \left(\frac{7}{8}\right) &
  p_{nr} =  gT\left( \frac{mT}{2\pi}  \right)^{\frac{3}{2}}\exp(-m/T)  \\
  \rho_r = g\frac{\pi^2}{30} T^4 & \left(\frac{7}{8}\right)& 
  \rho_{nr} = \frac{m}{T} p_{nr}\gg p_{nr}
\end{array}
\label{sb}
\ee
For the fully interacting case, we need to compute the free energy density $f=-T/V\ln Z(V,T)$ of the
QCD partition function, and the pressure follows as $p=-f$.
A technical obstacle here is that, in a Monte Carlo simulation, 
one cannot compute the partition function directly, 
since that provides the probability weights and is normalized to one. 
What one can compute are expectation values 
$\langle O\rangle$. The most frequently used detour 
is called the integral method \cite{integ}, which computes a suitable 
derivative of the free energy
resulting in a non-trivial expectation value, and then integrates the result,
\be
\frac{f}{T^4} {\biggl |}_{\beta_o}^{\beta} \; = \;  -\frac{N_t^4}{VN_t} 
\int_{\beta_o}^{\beta}
{\rm d}\beta ' \left(\left\langle \frac{\partial \ln Z}{\partial \beta'} \right\rangle - 
\left\langle \frac{\partial \ln Z}{\partial \beta'}\right \rangle_{T=0} \right).
\label{freediv}
\ee
Note that this introduces a lower integration constant, which needs to be fixed for the result to be
meaningful. While we do not know $f(\beta_0)$ from first principles, we can choose
$\beta_0$ corresponding to a temperature below the phase transition, where the free energy
should be well modelled by a weakly interacting glueball gas. Glueballs are heavy ($\gsim 1.6$ GeV), and according to eq.~(\ref{sb}) they produce exponentially small pressure which can be approximated by zero. 

\begin{figure}[t]
\vspace*{-0.8cm}
\hspace*{-0.5cm}
\includegraphics[height=4cm]{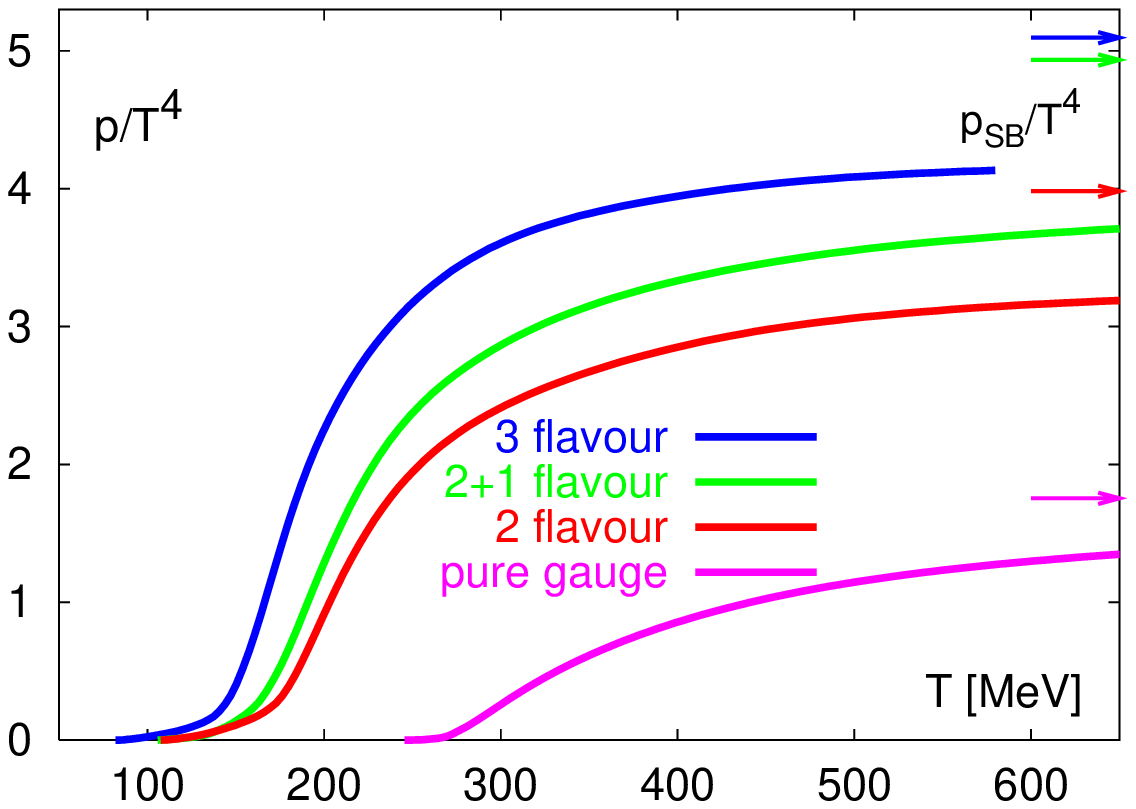}
\includegraphics[width=4.5cm]{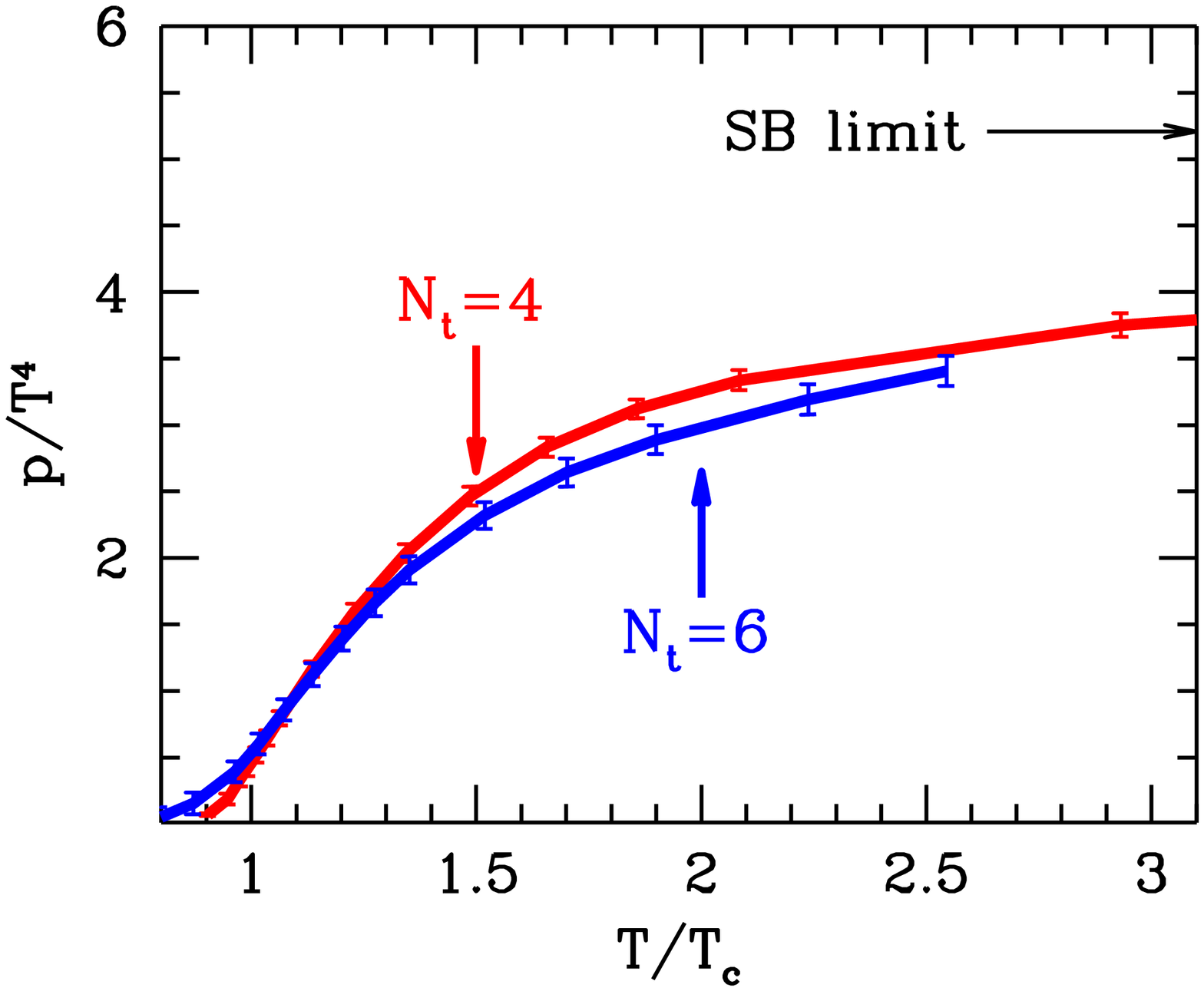}
\includegraphics[height=4cm]{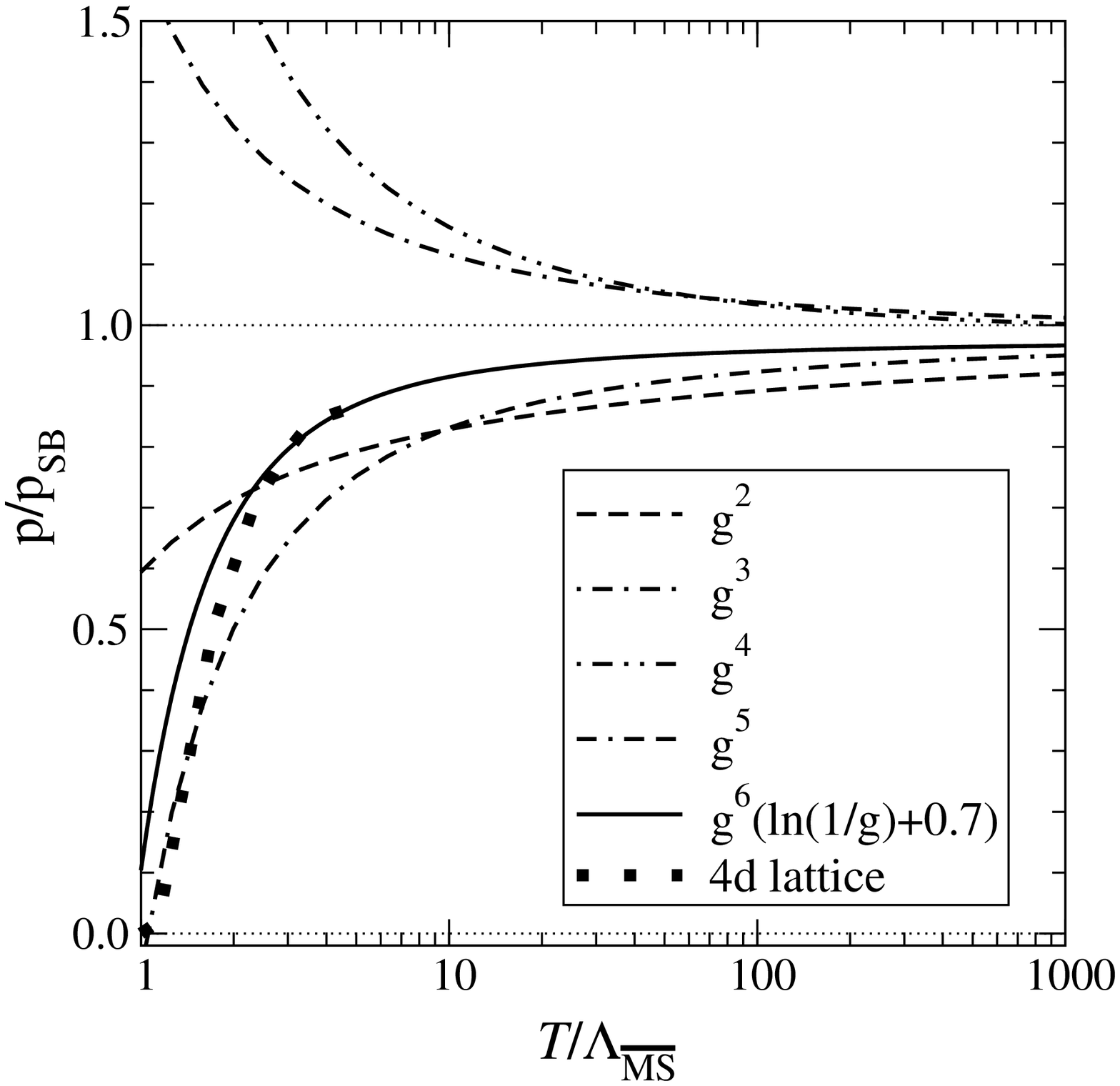}
\caption[]{Left:  Flavour dependence of the pressure
for $N_t=4$ lattices compared to a continuum extrapolated pure gauge result. From \cite{Peikert_eos}.
Middle: The pressure for $N_f=2+1$ with physical quark masses on $N_t=4,6$. From \cite{fkeos}.
Right: Various orders of perturbation theory and computation within the 3d effective field theory to order $g^6\ln(1/g)$. A missing contribution to $O(g^6)$ has been
adjusted by matching to the 4d lattice results, represented by the diamonds. 
From \cite{4loop1}.}
\label{fig:pressure}       
\end{figure}
Another difficulty is that strong discretization effects are to be expected.
At high temperature the relevant partonic degrees of freedom
have momenta of order $\pi T\sim \pi/(aN_t)$, {\it i.e.}~on the scale
of the lattice spacing, which thus strongly affects them.
For the equation of state it is therefore particularly important to gain control 
over these effects
and carry out the continuum limit $a \rightarrow 0$. This motivates the use of improved actions,
designed to minimize cut-off effects in the approach to the continuum.

The results of a computation of the pressure with an improved action \cite{Peikert_eos}
are shown in fig.~\ref{fig:pressure}. The data have been obtained for
$N_f=2,3$ with (bare) mass $m_q/T = 0.4$ as well as
for $N_f=2+1$ with a heavier mass $m^s_q/T = 1$.
For comparison the figure includes the
continuum extrapolated pure gauge result.
The figure shows a rapid rise of the pressure in the transition region.
The critical temperature as well as the
magnitude of $p/T^4$ reflect the number of degrees of
freedom liberated at the transition. 
This last conclusion is firm, since the pressure also rises for fixed temperature
when light quarks are added to the theory, consistent with the behaviour 
in the Stefan-Boltzmann
limit. 

An important question then is whether these features survive in the continuum limit.
Again, in pure gauge theory this can be firmly established by numerical 
extrapolation.
First steps towards a continuum extrapolation of dynamical simulations 
have been taken
in  \cite{fkeos,milc}, with 
consistent results. An example is shown in fig.~\ref{fig:pressure} (middle), which was computed with 
bare quark masses tuned to their physical values. 
The trend to fall short of the Stefan-Boltzmann limit turns out to be confirmed on 
a finer lattice,
and hence appears to be a robust result about the quark gluon plasma at moderate
temperatures.  

Investigations of the equation of state thus provide us with an intriguing picture of the complexity
of the quark gluon plasma. The sudden rise in pressure across the critical temperature has to
be interpreted as a release of degrees of freedom, 
and in this sense it is justified to speak of a deconfining transition.
On the other hand, the deconfined phase clearly shows remnant interactions and the released
degrees of freedom are
not appropriately characterized as a nearly free parton gas.

At asymptotically high temperatures, the pressure eventually must meet the free gas 
limit because of asymptotic freedom. This can be studied in the framework of the dimensionally reduced effective high $T$ theory. The pressure can be computed perturbatively to the
order $g^5$ before the prohibitive Linde problem sets in at $O(g^6)$ \cite{linde}. 
Fig.~\ref{fig:pressure} (right) shows the poor convergence behaviour of the 
series up to that order. The non-perturbative $g^6$ contribution is dominated by the magnetic modes $\sim g^2T$. However, it can be computed on the lattice by simulations of the 3d effective high temperature theory \cite{hie}. This requires matching of the 
coefficients between the effective and the full theory and converting from lattice to $\overline{\mbox{MS}}$
regularization schemes at four loop level, a task that has been recently completed \cite{4loop2}.
However, there is one last missing contribution to $O(g^6)$ 
coming from the perturbative scale $\sim T$. 
In the figure this coefficient has been chosen such that the 
results for the pressure computed in this framework match on to the lattice results at lower
temperatures. 

\subsection{Screening masses}
Essentially all static equilibrium properties of a thermal
quantum field theory are encoded in its equal time correlation functions. These are quantities
that are well defined and calculable to good precision by lattice methods.
Unfortunately, these quantities are not 
directly accessible in heavy ion collision experiments. Nevertheless, their theoretical knowledge
provides us with the relevant length scales in the plasma, from which
conclusions about the active degrees of freedom and their dynamics may be drawn.
The connected spatial correlation functions of gauge-invariant,
local operators $A(x)$,
\be
C(|{\bf x}|)=\langle A({\bf x})A(0)\rangle_c\sim \ex^{-M|\bfx|},
\label{eq:corrfct}
\ee
fall off exponentially with distance. The
``screening masses'' $M$
are the eigenvalues
of the spacewise Hamiltonian of the corresponding
lattice field theory, and classified by its symmetries.
Because of the shortening of the euclidean time direction at $T>0$, the continuum 
rotation symmetry of the hypertorus orthogonal to the correlation direction is broken down
from $O(3)$ to $O(2)\times Z(2)$, and its appropriate subgroup for the
lattice theory is $D_h^4$. The irreducible representations and the classification
of operators have been worked out for pure gauge theory \cite{gros,dg} as well as for
staggered quarks \cite{symg}.
Physically, the screening masses correspond to the inverse length scale over
which the equilibrated medium is sensitive
to the insertion of a static source carrying the quantum numbers of $A$. Beyond
$1/M$, the source is screened and the plasma appears undisturbed.
Technically, the computation of these quantities is equivalent to
spectrum calculations at zero temperature.

\begin{figure}[t]
\centerline{
\includegraphics[width=0.5\textwidth]{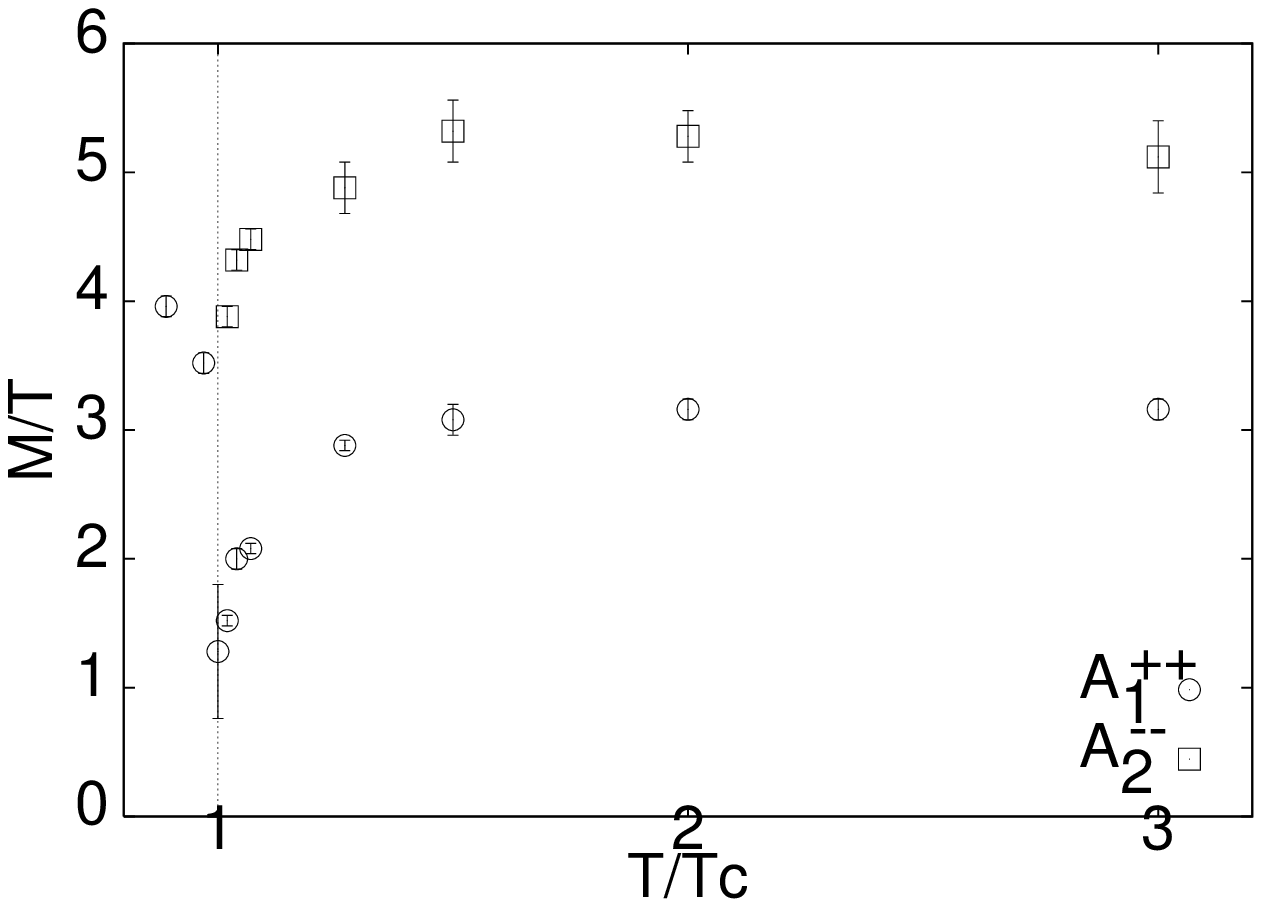}
\includegraphics[width=0.5\textwidth]{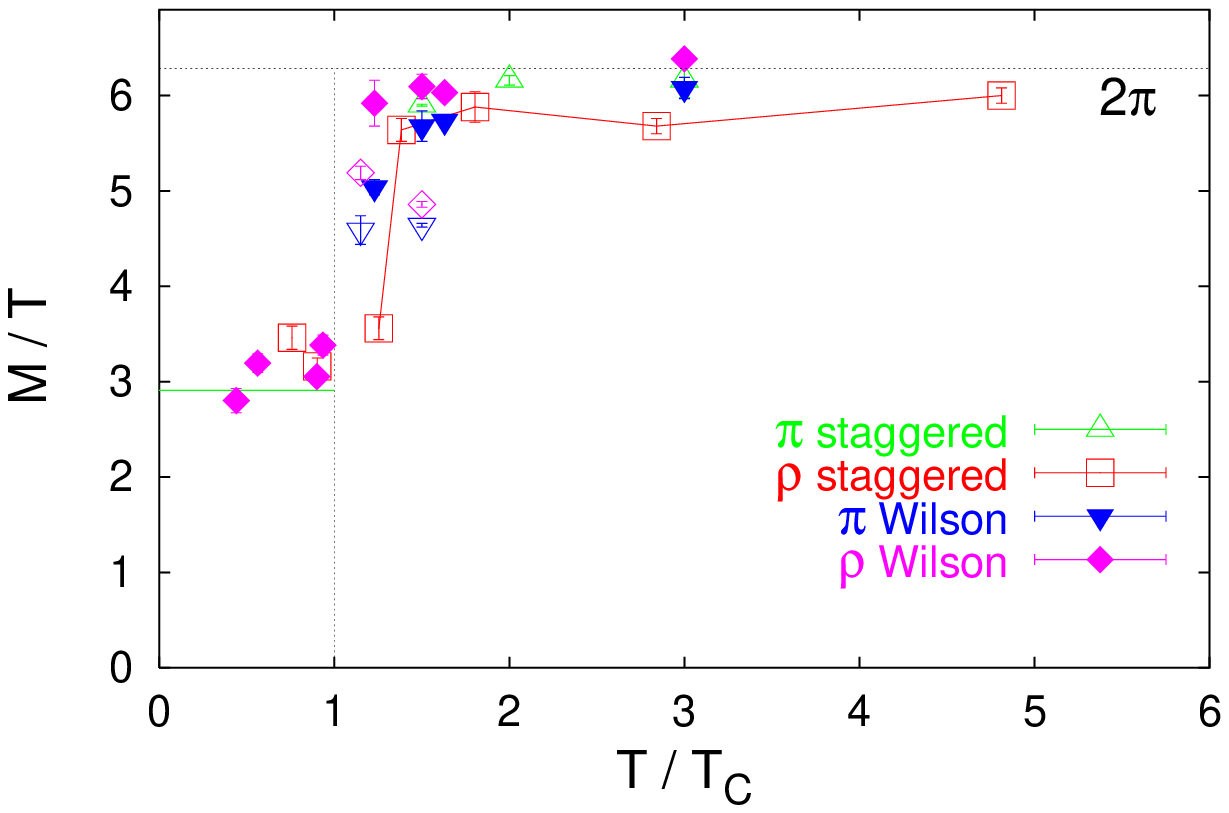}
}
\caption{Left: Screening masses for the pure gauge theory from $N_t=4$ lattices,
corresponding to the continuum $0^{++}_+$ (circles) and
$0^{+-}_-$ (squares) channels. From \cite{dg2}.
Right: Mesonic screening masses in the quenched chiral limit.
Below $T_c$, $M_{\rho}/T_c$ \protect\cite{PSchmidt} is plotted,
the line denotes the $T=0$ value. Open Wilson symbols denote anisotropic
lattices \protect\cite{Nucu_1}. Staggered pion data are extrapolated
to $a=0$ \protect\cite{Sourendu2}, Wilson and staggered rho are from
$N_t=8$ \protect\cite{PSchmidt} and 16 lattices.
The free quark limit has not been corrected for finite
         volume effects. From \cite{anrv}.
}
\label{fig:screen}       
\end{figure}

Fig.~\ref{fig:screen} (left) shows results for the lowest lying screening masses
corresponding to glueball operators around $T_c$.
In the range $0.8\, T_c<T<T_c$, the lowest scalar screening mass is observed to be roughly 20\%
lower than the lightest scalar glueball at zero temperature, $M(T)/M_G(T=0)\sim 0.8$.
At $T_c$ a sharp dip is observed, after which the screening masses appear
to be proportional to $T$.
Screening states with larger masses show the same qualitative behavior above $T_c$.

Apart from pure glue operators $A$ in eq.~(\ref{eq:corrfct}),
also correlations of meson operators
have been investigated, both in the quenched approximation
as with various numbers of dynamical fermions.
The picture which has emerged so far
is illustrated by some of the available data
shown in fig.~\ref{fig:screen} (right).
Below $T_c$, the screening masses do not show a marked
temperature dependence and roughly agree with the corresponding 
zero temperature masses.
At temperatures (slightly) above $T_c$ 
spatial (as well as temporal) correlation functions
reflect the restoration of the chiral $SU_L(N_f) \times SU_R(N_f)$ symmetry.
In particular, the vector and axial vector channel become
degenerate independent of the discretization and of the number of dynamical
flavors being simulated.
Moreover, the pion ceases to be a Goldstone boson
and acquires a screening mass.

At sufficiently high temperature, the screening masses are expected to approach
the limit of freely propagating quarks, $M \rightarrow 2 \pi T$.
In fact, already at temperatures of about 1.5 $T_c$,
the results for the vector channel are not far from this value.
If the finite volume effects for free quark propagation are
taken into account, the
deviations amount to about 15 \% and decrease slowly
with temperature. Thus, the hard fermionic modes $\sim \pi T$ behave 
quasi-perturbatively.
Quenching effects are found to be below 5\% for $T>T_c$ \cite{ggm}, so the computer time
saved on dynamical fermions can be invested in finer lattices and results are close to continuum
physics.

In the plasma phase, screening masses can also be investigated within the dimensionally
reduced theory, cf.~section \ref{sec:dr}.
In this framework they correspond to the spectrum of the transfer matrix for the 3d theory.
The associated Hamiltonian respects
SO(2) planar rotations, two-dimensional parity $P$,
$A_0$-reflections $R$, and the symmetry is again
$SO(2)\times Z(2)\times Z(2)=O(2)\times Z(2)$. Remember, however, that in this
setup one is interested in soft modes, while the Matsubara frequency
$\sim 2\pi T$ represents the UV cut-off for the effective theory.

\begin{figure}[t]
\vspace*{-1.5cm}

\includegraphics[height=40mm]{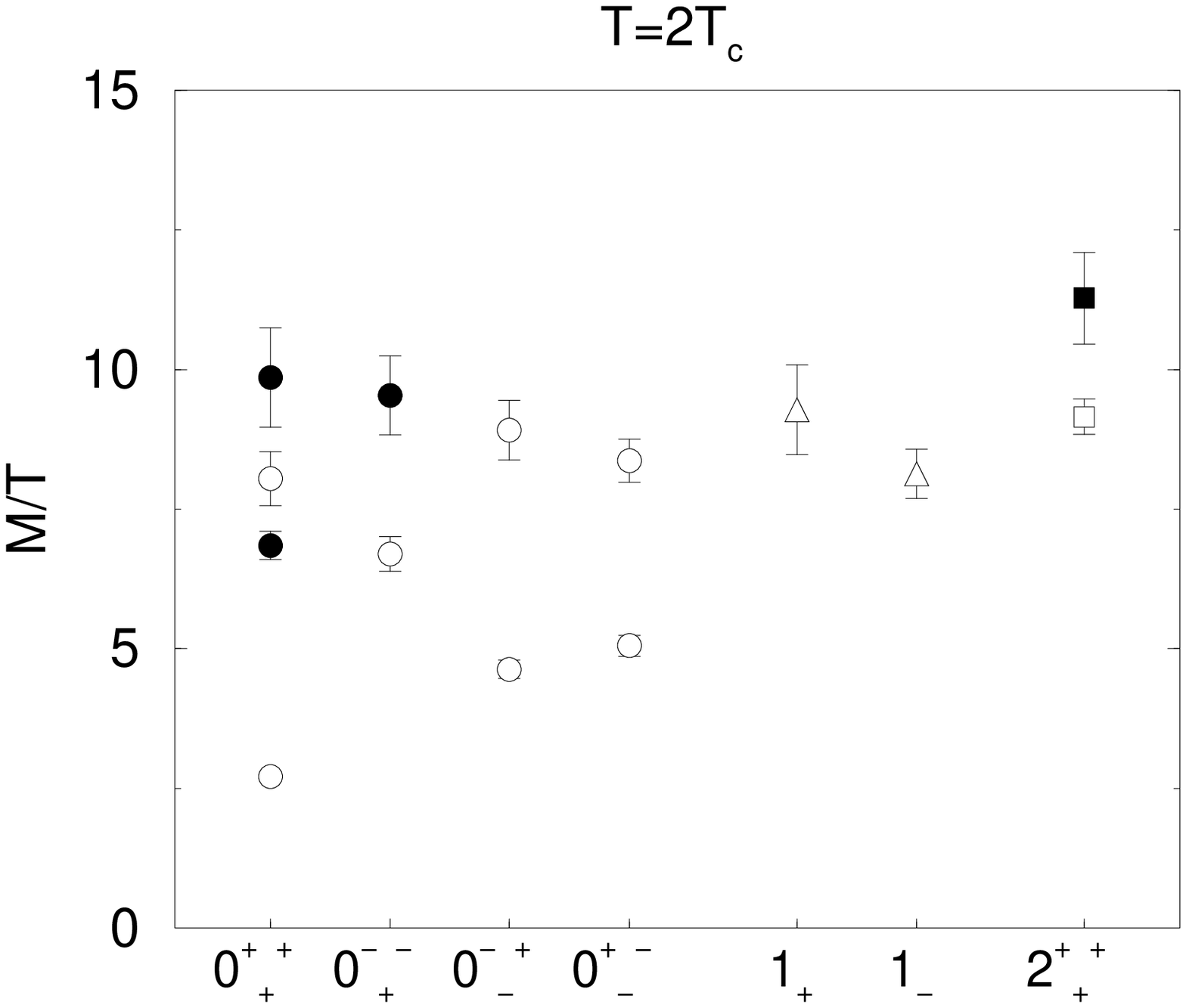}
\includegraphics[height=40mm]{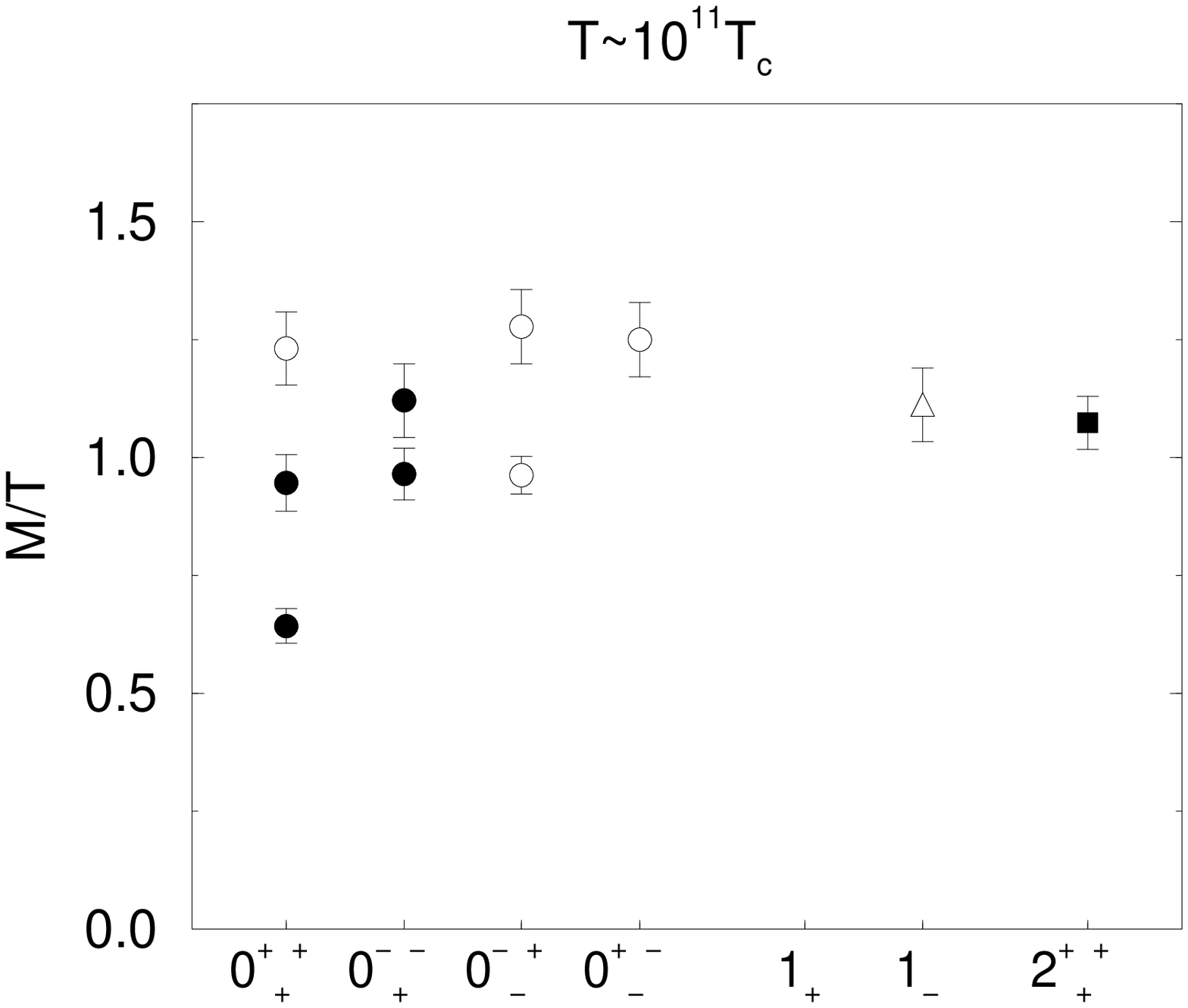}
\includegraphics[height=55mm]{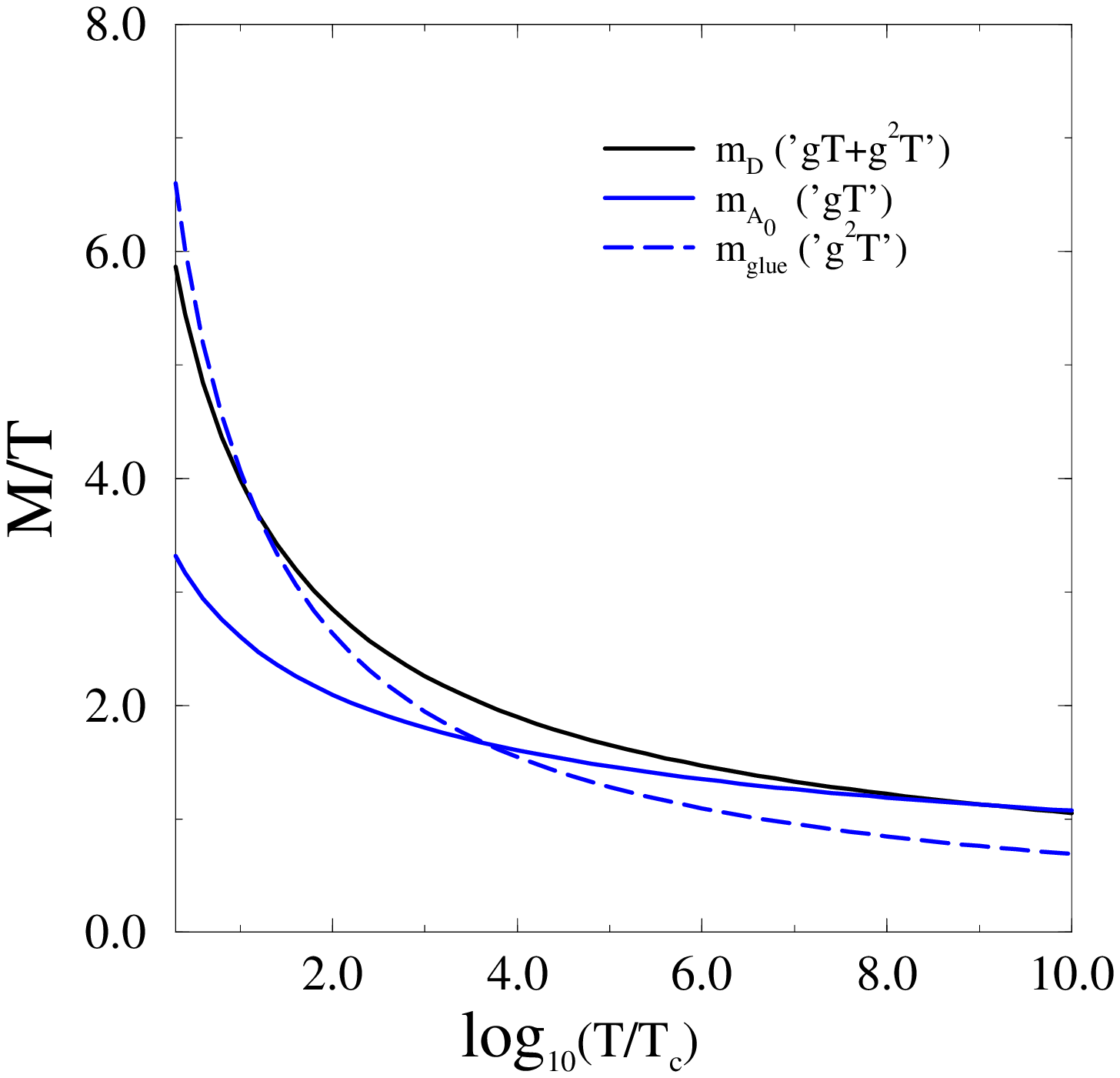}

\caption{Pure gauge screening masses from simulations of the effective 3d theory in various quantum number channels for $T=2T_c$ (left) and $T=10^{11}T_c$ (middle). Filled symbols refer to operators made of only magnetic gauge fields $A_i\sim g^2T$, while open symbols also contain electric gauge
fields $A_0\sim gT$. From \cite{hlp1}. Right: Interpolated temperature dependence of the two lowest $0^{++}_+$ and the lowest $0^{-+}_-$ states \cite{mik}.
}
\label{fig:drcomp}       
\end{figure}

The approach of dimensional reduction is particularly valuable in disentangling
contributions from different degrees of freedom by accurate mixing analyses,
as well as for treating larger temperatures $T\gg T_c$ which cannot be reached by 
4d lattices. This can be used to inspect to what extent the plasma behaves perturbatively.
fig.~\ref{fig:drcomp} then tells us that for any experimentally accessible temperature
the largest correlation length of gauge-invariant operators belongs to
the $A_0\sim gT$ degrees of freedom and not to the $A_i\sim g^2T$, in contrast to the naive parametric
ordering eq.~(\ref{eq:scale}). Only asymptotically is the perturbative ordering restored.
Thus, on the soft scales $\sim gT$ and $\sim g^2T$ the plasma is strongly coupled.

\subsection{The free energy of static quarks}

The free energy of a static quark-antiquark pair
is of interest for the physics of heavy quarkonia in the medium, 
and in particular to the question
of $J/\psi$-suppression due to screening of the confining
potential \cite{mats}. Potential models are then used to study bound 
state solutions, suitably generalized to finite 
temperature. More refined treatments try to establish a connection between
the static potential and quarkonium spectral functions, to be discussed in the 
next section. Non-perturbative free energies serve as input for such applications,
for recent work and references, see \cite{pots}. 

The $q\bar{q}$ free energy is defined \cite{mcls}
by the partition function
of the thermal heat bath containing a static quark and
antiquark source at separation $\vec r$,
\begin{equation}
\langle \; {\rm Tr} L(\vec r) \; {\rm Tr} L^\dagger(0) \; \rangle
= \exp\{- (F_{q\bar q} \,(|\vec r |,T) - F_0(T)\;)/T\},
\label{eq:poly_corrfct}
\end{equation}
where $F_0$ denotes the free energy of the heat bath.
At zero temperature, {\it i.e.}~$N_t\rightarrow \infty$, $F_{q\bar q}$ reduces to the
static quark potential.

\begin{figure}[t]
\centerline{
\includegraphics*[width=0.5\textwidth]{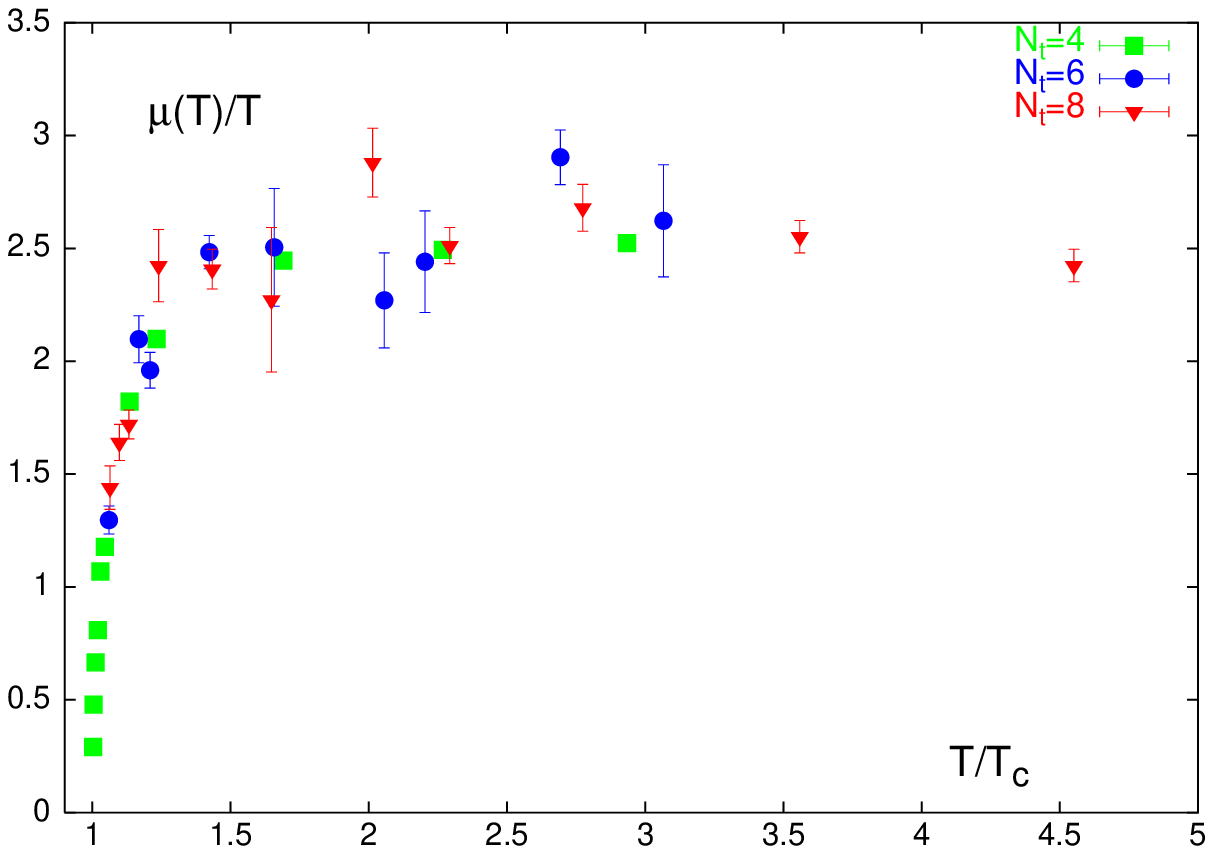}
\includegraphics*[width=0.5\textwidth]{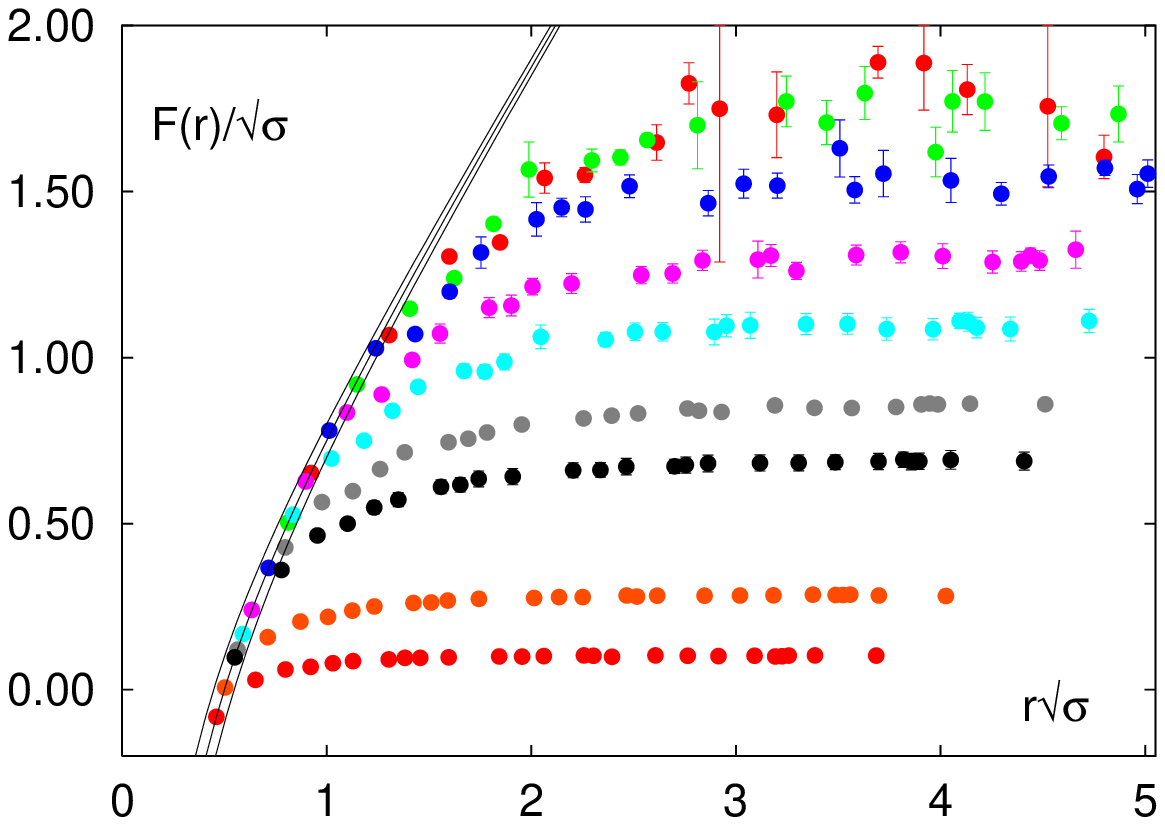}
}
\caption{
Left: Screening mass (quenched) for $T>T_c$ from fits
to eq.~(\ref{free_d}), $d=3/2$ \cite{potstat}.
Right: Static quark free energy for $N_f=3$
at temperatures $0.58<T/T_c<1.15$ \protect\cite{potdyn}.
$F(r)$ is
normalized at ($r=1/T$) to the $T=0$
Cornell potential, $V(r)/\sqrt{\sigma} = -\alpha /
r\sqrt{\sigma} + r\sqrt{\sigma}$ with $\alpha = 0.25\pm 0.05$ (solid band).
}
\label{fig:potdyn}       
\end{figure}

Let us begin our discussion in the pure gauge limit, where the expectation value of the
Polyakov loop is a true order parameter. 
Numerical results 
below and above the pure gauge phase transition are given in \cite{potstat}.
When the temperature is switched on and increased towards $T_c$, a linearly rising
free energy is maintained, implying infinite energy cost in separating the quark 
infinitely from the anti-quark, and thus
confinement. However, the slope decreases with temperature, 
and the corresponding effective string tension is well fitted by the form
\be
\frac{\sigma(T)}{\sigma(0)}=a\sqrt{1-b\frac{T^2}{T_c^2}},
\ee
(predicted by an effective Nambu-Goto string \cite{string}), 
with $a=1.21(5)$ and $b=0.990(5)$ \cite{potstat}.
Above $T_c$ an exponentially screened free energy is obtained, fitting an 
ansatz of the form
\be
 \frac{F_{q\bar{q}}(r,T)}{T} = - \frac{e(T)}{(rT)^d} e^{-\mu(T)r}.
\label{free_d}
\ee
In perturbation theory, the leading term originates
from two-gluon exchange and predicts
$d=2$ and exponential decay with
twice the Debye mass \cite{Nadkarni}.
Lattice investigations \cite{potstat,hkr}
have shown that this simple behavior does not apply in the temperature 
and distance range explored.
Rather, fits \cite{potstat} to eq.~(\ref{free_d}) favoured $d \simeq 3/2$ and
found screening masses $\mu(T)$, shown in fig.~\ref{fig:potdyn} (left),
to be compatible
with the lowest color singlet $0^{++}_+$ screening mass
shown in figs.~\ref{fig:screen}, \ref{fig:drcomp}. This is not surprising:
the Polyakov loop
is a gauge invariant operator, and since it is an exponential of gauge fields
it couples to all $J^{PC}$ sectors. Consequently, its correlator decays 
with the lightest screening mass of the spectrum.

When dynamical light quarks are present, the color charges of the heavy quarks 
are screened also below $T_c$ and one
observes \cite{potdyn,DeTar_pot,Bornyakov} the expected string breaking,
fig.~\ref{fig:potdyn} (right).
The distances where the free energies become flat in $r$
range from 1.5 to 1 fm, decreasing with temperature,
even at (bare) quark masses as large as $m/T = 0.4$.
Note that the deviations from the zero temperature quenched
potential set in already at distances of
${\mathcal O}(0.5 {\rm fm})$ for temperatures $\gsim \, 0.75 T_c$.
When the free energy is normalized to the short distance
zero temperature potential, its large $r$ asymptotic value
rapidly decreases with $T$.

The gauge invariant Polyakov loop 
averages over color singlet and octet configurations of the static source. 
Correspondingly, its correlator
is often written as a superposition \cite{mcls,Nadkarni,Nadkarni1}
\begin{equation}
e^{-F_{q \bar q}(r,T)/T } =
\frac{1}{9} \;  e^{- F_{1}(r,T)/T } +
\frac{8}{9}   \; e^{- F_{8}(r,T)/T },
\label{eq:average}
\end{equation}
with singlet and octet channels defined as
\ba
\ex^{-F_1(r,T)/T}&=&
\frac{1}{3}\langle \Tr L^\dag(\bfx)L(\bfy)\rangle,\nn\\
\ex^{-F_8(r,T)/T}
&=&\frac{1}{8}\langle \Tr L^\dag(\bfx) \Tr L(\bfy)\rangle
-\frac{1}{24}\langle \Tr L^\dag(\bfx) L(\bfy)\rangle.
\label{potdef}
\ea
While the singlet correlator is obviously gauge-dependent, 
the energy eigenvalues governing its
decay in a spectral analysis are not \cite{potgauge} and one might hope to gain 
useful insight
into the colour dynamics to be used in potential models. 
However, this is dangerous.
A careful spectral expansion making use of projection operators on the 
representations reveals that both, the singlet and the octet channel, 
decay as $\sim \sum_{n} |c_n[U]|^2\exp(-E_n/T)$ \cite{jahn},
where the energies $E_n$ represent the same static potential and its excitations 
which also
contribute to the ``average'' free energy, while the matrix elements $c_n[U]$
depend on the gauge or the operator used to describe the spatial string.
Thus, the gauge-invariant spectral information is the same for all channels, 
while $F_1, F_8$
and their difference from $F_{q\bar q}$ are properties of the operators used,
leaving their physical meaning in doubt.

\subsection{Dynamical quantities: quarkonium spectral functions and transport coefficients}

Quarkonium physics in the plasma can also be formulated more rigorously in 
a quantum field theoretical way. 
The physical excitations of the plasma with a given set of quantum numbers
are encoded in the retarded Green's functions, 
or real time correlators, of operators carrying those 
quantum numbers \cite{tbooks}.  
In general the poles of such (momentum space) Green's functions will be complex, 
thus they are {\it not} eigenvalues of the Hamiltonian. 
If the imaginary part is small compared to the real part, 
one speaks of a quasi-particle excitation, 
with the real part interpreted as its mass and the imaginary 
part as its decay width, due to Landau damping in the plasma.

\begin{figure}[t]
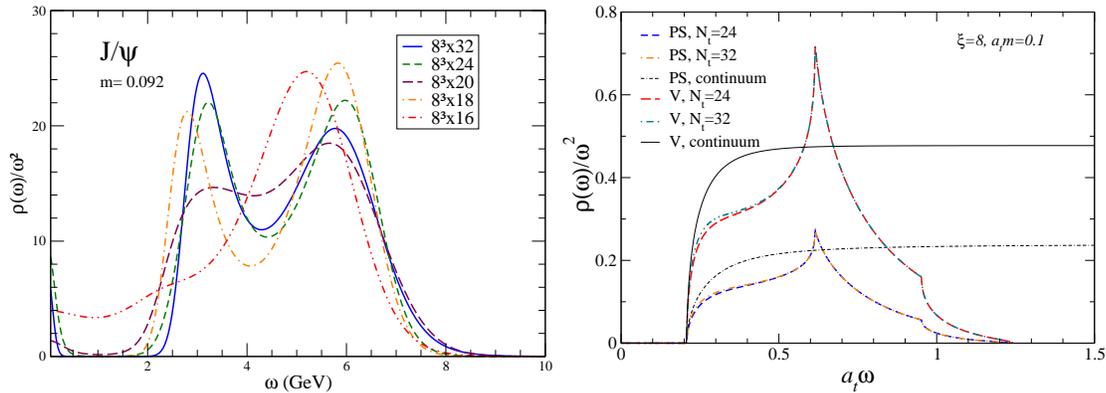

\centerline{
\includegraphics*[width=0.5\textwidth]{Jpsi_R6m092.eps}
\includegraphics*[width=0.5\textwidth]{plot_TCD_m0.1.eps}
}
\caption{Left: Spectral function in the vector channel for different temperatures on $8^3\times N_t$ lattices, \cite{aa2}. Right: Free fermion spectral functions in the pseudo-scalar and vector channels,
\cite{aa1}.}
\label{fig:spec}       
\end{figure}
Unfortunately, Monte Carlo methods only work for euclidean actions, and direct numerical simulations of real time correlators are impossible. However, one may try to statistically exploit the information 
encoded in euclidean Green's function as follows. Let $G_\Gamma(\tau) =
\sum_{\vec{x},\vec{y},t}\langle\bar{\psi}(\vec{x},t)\Gamma\psi(\vec{x},t)
\bar{\psi}(\vec{y},t+\tau)\Gamma\psi(\vec{y},t+\tau)\rangle$
be a correlator of some meson
operator in euclidean time. Then its Fourier transform is given by
\begin{equation}
G_\Gamma(\tau,\vec{p}) = 
\int_0^\infty\frac{d\omega}{2\pi}\rho_\Gamma(\omega,\vec{p})K(\tau,\omega)\,,\qquad
K(\tau,\omega) 
 = e^{\omega\tau}n_B(\omega) + e^{-\omega\tau}[1+n_B(\omega)]\,,
\label{spectral}
\end{equation}
where all the time dependence resides in the kernel $K(\tau,\omega)$, which 
contains only thermal distribution functions.  All the dynamical, 
spectral information of the theory is contained in $\rho(\omega,\vec{p})$, 
which will not change under analytic continuation to Minkowski time. 
Still the difficulties are formidable. Computing $G_\Gamma$  
by lattice simulations,
the left hand side is given only by a finite number of points and 
the inversion of the integral is
an ill-defined problem. The way out is to ``guess'' with statistical 
methods the most likely 
$\rho(\omega,\vec{p})$ which will fit the formula. 
This is known as maximum entropy method (MEM),
whose details are explained in \cite{mem}. Clearly, one needs as many points 
in the time direction as possible, hence anisotropic lattices with 
$a_t\ll a_s$ are useful. Another problem is that the MEM 
needs a prior, {\it i.e.}~a model function has to be provided to constrain 
the fitted function. Usually one
models spectral functions perturbatively at large $\omega$. For a discussion of 
this and other systematic difficulties, see \cite{aar}.

Early results based on this approach, obtained with quenched simulations, 
indicated that $J/\psi$ 
does not melt up to temperatures of $1.5-2T_c$ \cite{memquen}, thus providing another reason to 
term the QGP `strongly coupled'. 
There are now also dynamical simulations confirming this picture,
as shown in fig.~\ref{fig:spec} (left) for $N_f=2$ \cite{aa2}.
The temperatures parametrized by the shown $N_t$'s correspond to $T_c-2T_c$.
At low temperatures, the position of the first peak corresponds to the zero temperature mass of the 
$J/\psi$. As temperature increases, 
the peak shrinks but keeps existing until about $1.7T_c$, when
it finally disappears. 
Hence the conclusion that ``$J/\psi$ doesn't melt'' until then 
(for another interpretation see \cite{pots}). No physical meaning can be 
attributed to the second peak, which is interpreted as a lattice-distorted 
part of the free particle continuum. 
This is illustrated in fig.~\ref{fig:spec} (right), where the spectral 
funtion for free particles
in the scalar and vector channel are plotted in the continuum and on the lattice. The finite cut-off
is faking a peak, which should disappear in the continuum limit.

Other quantities of tremendous phenomenological interest are transport 
coefficients.  
In particular, the shear viscosity is defined as the slope of another 
spectral function,
\be
\eta=\left.\frac{1}{20}\frac{d}{d\omega}\rho_{\pi\pi}(\omega)\right|_{\omega=0},\quad
G_{\pi\pi}(\tau)=\int d^3x\langle \pi_{kl}(\tau,\bfx)\pi_{kl}(0,{\bf 0})\rangle=\int\frac{d\omega}{2\pi}\,
K(\tau,\omega)\rho_{\pi\pi}(\tau),
\ee
where $\pi_{kl}$ is the traceless part of the spatial energy-momentum tensor. 
The numerical procedure to compute this on a lattice is the same as above. 
Here accurate information, and hence high resolution, 
on the low frequency tail is required. An early numerical attempt \cite{naka} 
as well as a recent more refined analysis \cite{harv} for the pure gauge theory 
give small viscosity to entropy ratios, $\eta/s\sim 0.1-0.2$, 
consistent with observations in the quark gluon plasma. 
However, it has been pointed out that extracting the low frequency part 
is firstly sensitive to modelling the spectral function \cite{aar}, and second intrinsically 
unstable, though this problem could be removed by a simple rescaling trick \cite{aa3}. 
Building on this, ref.~\cite{harv} 
suggested a new method to systematically 
improve on the spectral function by expanding it in terms of an 
orthogonal function system and fitting its coefficients. Hence, 
MEM calculations might soon offer some control over sytematics.

Finally, a different approach to compute spectral functions is currently 
emerging, generalizing heavy quark 
effective field theory methods 
to real time correlation 
functions at finite temperature \cite{lpt}. 
In the heavy quark limit a certain meson correlator reduces to a Wilson loop in 
Minkowski time. 
Thus, the static potential reappears in a real time framework, 
but it develops an imaginary part 
which is responsible for the damping and melting of the bound state \cite{lpt}. 
It has been 
shown in HTL perturbation theory that indeed a broadening spectral function is 
obtained in this approach \cite{lv}.
The imaginary part of the potential has recently been confirmed non-perturbatively
by classical lattice simulations \cite{lpt2}.
It will be interesting to see if this approach can be generalized to the full
quantum theory in the future, thus providing a bridge between field theory
and potential models.

\section{The nature of the QCD phase transition \label{sec:pd}}

\begin{figure}[t]
\vspace*{-5cm}
\includegraphics*[width=0.5\textwidth]{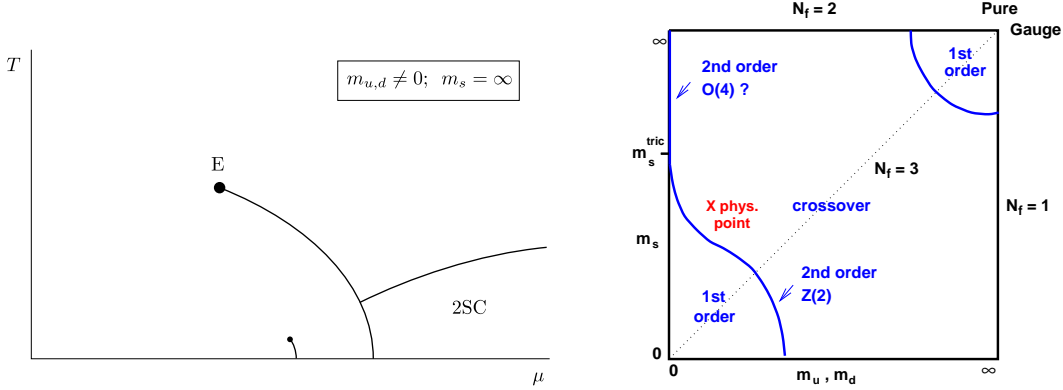}\hspace*{1cm}
\includegraphics*[width=0.5\textwidth]{phase_diagram_trunc.ps}
\caption{
Left: Conjectured phase diagram for $N_f=2$ QCD with finite light quark massees. The physical
case $N_f=2+1$ is believed to qualitatively look the same,
according to universality and continuity arguments as well as results from QCD-like 
models \cite{wilc}.
Right: Schematic phase transition behaviour of $N_f=2+1$ flavor QCD at $\mu=0$ for
different choices of quark masses $(m_{u,d},m_s)$, at $\mu=0$.}
\label{fig:1schem}      
\end{figure}

In the previous sections we discussed the properties of QCD matter on either side 
of the quark hadron
transition, but haven't yet addressed
what the nature of that transition is. To obtain an unambiguous answer to 
this question is in fact a 
very difficult task. In statistical mechanics, phase transitions are defined 
as singularities, or non-analyticities, in the free energy as a function of its thermodynamic parameters.
However, on finite volumes, free energies are always analytic functions and it can be proved that singularities only develop in the thermodynamic limit of infinitely many particles, or $V\rightarrow \infty$ \cite{ly}. 
This is particularly obvious in the case of lattice QCD, whose partition 
function is a functional integral
over a compact group with a bounded exponential as an integrand. 
It is thus a perfectly analytic function of
$T,\mu,V$ for any finite $V$. 
Hence,  a theoretical establishment of a true phase transition requires 
finite size scaling (FSS) studies on a series of increasing and 
sufficiently large volumes in order to extrapolate to the thermodynamic limit. 
Three different situations can emerge: a first order phase transition 
is characterized by coexistence of two phases, and hence a discontinuous jump 
of the order parameter (and other quantities)
when going through such a transition. A second order transition  
shows a continuous transition of the order parameter accompanied by a divergence of the correlation length 
and some other quantities, like the heat capacity. 
Finally, a marked change in the physical properties of a system may also occur 
without any non-analyticity of the free energy, in which case it is called 
an analytic crossover. 
A familiar system featuring all these possibilities is water, 
with a weakening first oder liquid-gas phase transition terminating in a 
critical endpoint with Z(2) universality, as well as a triple point 
where the first order liquid-gas and 
solid-liquid transitions meet. Similar structures are also conjectured to
be present in the QCD phase diagram \cite{wilc,tric}, fig.~\ref{fig:1schem} (left).
In this section we focus on the order of the transition for $\mu=0$
as a function of $N_f$ and quark masses, before tackling  $\mu\neq 0$
in sec.~\ref{sec:mu}. 
 
\subsection{Universality and finite size scaling}

A most fascinating phenomenon in physics is the ``universality'' exhibited by physical systems
near a critical point of second order phase transition. It is due to the divergence of the
correlation length, which implies that the entire system acts as a coherent collective.
Hence, microscopic physics becomes unimportant, the collective behaviour is determined by global symmetries.
With the divergence of the correlation length the characteristic length scale 
disappears from the problem, and thermodynamic observables in the critical 
region obey scale invariant power laws. For example, at a second order 
ferromagnetic phase transition, the order parameter magnetization vanishes as 
$M\sim |T-T_c|^{\beta}$, while the specific heat and correlation length diverge,
$C\sim |T-T_c|^{-\alpha}$ and $\xi \sim |T-T_c|^{-\nu}$, respectively.
The critical exponents $\alpha,\beta,\nu$ and similar ones for other quantities 
are determined by the effective 
global symmetries at the critical point, and thus are the same 
for all systems within a universality class. 
The latter are labeled by spin models, since those can be 
readily solved numerically. 
The power law behaviour for thermodynamic functions 
follows from the scaling form
of the singular part of the free energy, 
\be
f_s(t, h) = L^{-3} f_s(L^{y_t}\tau, L^{y_h}h),
\ee
where $L$ denotes the box size, and $\tau$ and $h$
are the relevant scaling fields. In the case of the Ising model these are the reduced
temperature and magnetic field.
Unfortunately, for most systems, and particularly QCD, it is not obvious which 
global symmetries the system will exhibit at a critical point, 
those are a dynamically
determined subset of the total symmetry of the theory.
Moreover, since there is no true order parameter in the case
of finite quark masses, fields and parameters of QCD map 
into those of the effective model in a non-trivial way.

\subsection{The nature of the QCD phase transition for $N_f=2+1$ at $\mu=0$: qualitative picture}

Before discussing numerical results, let us briefly outline the qualitative picture for
the order of the phase transition,
fig.~\ref{fig:1schem} (right).
As mentioned in sec.~\ref{sec:limits}, for $N_f=3$
in the limits of zero and infinite quark masses
(lower left and upper right corners),
order parameters corresponding to the breaking of a symmetry can be defined,
and one finds numerically at small and large quark masses
that a first-order transition takes place at a finite
temperature $T_c$. On the other hand, one observes an analytic crossover at
intermediate quark masses. Hence, each corner must be surrounded by a region of
first-order transition, bounded by a second-order line. The situation is less clear for the chiral
limit of the two flavour theory in the upper left corner.
If the transition is second order, then chiral symmetry
$SU_L(2)\times SU(2)_R \sim O(4)$ puts it in the universality
class of 3d $O(4)$ spin models.
In this case there must be a tricritical strange quark mass $m_s^{tric}$,
where the second order chiral
transition ends and the first order region begins. The exponents at such a
tricritical point would
correspond to 3d mean field \cite{tric}. On the other hand,
a first order scenario for the chiral limit
of $N_f=2$ is a logical possibility as well.

\subsection{The chiral transition for $N_f=2$:  is it O(4) or first order?}

\begin{figure}[t]
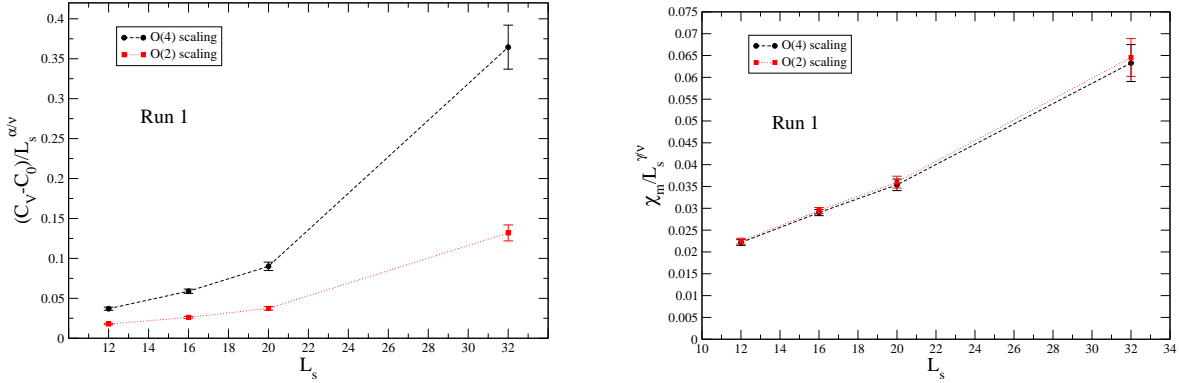

\centerline{
\includegraphics*[width=0.5\textwidth]{Cv_max_Run1.eps}\hspace*{1cm}
\includegraphics*[width=0.5\textwidth]{Chi_max_Run1.eps}
}
\caption{Finite volume scaling behaviour of specific heat and chiral susceptibility
on $N_t=4$. For O(4) viz.~O(2) behaviour, the data should fall on a horizontal line \cite{dig}.}
\label{fig:dig}
\end{figure}

On the lattice, O(4) will effectively look like O(2) as long as there are 
discretization effects \cite{o2}.
Wilson fermions appear to see O(4) scaling \cite{wil}, 
while staggered actions are inconsistent with both O(4) and O(2) \cite{s2}. 
(The staggered strong coupling limit, however, does display O(2) scaling 
\cite{c2}). 
An attempt to tackle this question by means of a finite size scaling analysis 
with unprecedented lattice sizes was made in \cite{dig}. The work simulates $L^3\times 4$ lattices with 
$L=16-32$,  using standard staggered fermions with several quark masses, 
the smallest being $m/T\lsim 0.055$.
In a critical region, the specific heat or the chiral susceptibility  
scale as
\ba
C_V - C_0 &\simeq & L^{\alpha/\nu} f_c \left(\tau L^{1/\nu}, am\, L^{y_h} \right),
\quad \tau=1-T/T_c\nonumber\\
\chi& \simeq &L^{\gamma/\nu} f_\chi \left(\tau L^{1/\nu}, am\,L^{y_h} \right).
\label{scale}
\ea
Here the non-singular part of the specific heat $C_0$ has been subtracted. 
The authors of \cite{dig} thus fix $y_h$ to its critical value, and then choose $L$ and
$m$ for a series of simulations such as to keep $(am \,L^{y_h})$ constant, reducing the scaling problem on one variable only.
The infinite volume limit in this procedure thus corresponds to the chiral limit and allows to check whether the data are consistent with the predicted scaling behaviour.

Fig.~\ref{fig:dig} shows simulation results from \cite{dig}. 
Scaling as in eq.~(\ref{scale}) would imply the data points to fall on a 
horizontal line, which is clearly not the case. After many variations on 
the analysis, the authors conclude that their data are closer to first order 
behaviour than to O(4)/O(2).

A different conclusion was reached in \cite{ksnf2}, in which $\chi$QCD was investigated numerically.
This is a theory modified by an irrelevant term ({\it i.e.}~one going to zero in the continuum limit) such as
to allow simulations in the chiral limit. The authors compare their data with those obtained from an 
$O(2)$ spin model on moderate to small volumes and find them to be compatible.
This might point at finite volume effects of current $N_f=2$ QCD simulations. 
The scaling region of the chiral
limit might be excessively small, which would require extraordinarily large volumes in order to see the correct scaling.
Another possibility for the failure of simulations
to unambiguously pin down the order of this transition is of course the
failure of the fermion discretizations to reproduce chiral symmetry.  
Future studies on either finer lattices or with chiral fermion actions should help to 
clarify this issue. 

Another possibility is to study the strength of the $U_A(1)$ anomaly
discussed in sec.~\ref{sec:limits}, which also enters the effective 
sigma model \cite{chiral}
for the critical region. 
It has recently been non-perturbatively demonstrated that a strong anomaly is 
required for the chiral phase transition to be second order \cite{chi}.

\subsection{$N_f=2+1$: 3d Ising universality and the critical line $m_s^c(m_{u,d})$}
\label{sec:m1m2c}

\begin{figure}[t]
\centerline{
\includegraphics*[width=0.5\textwidth]{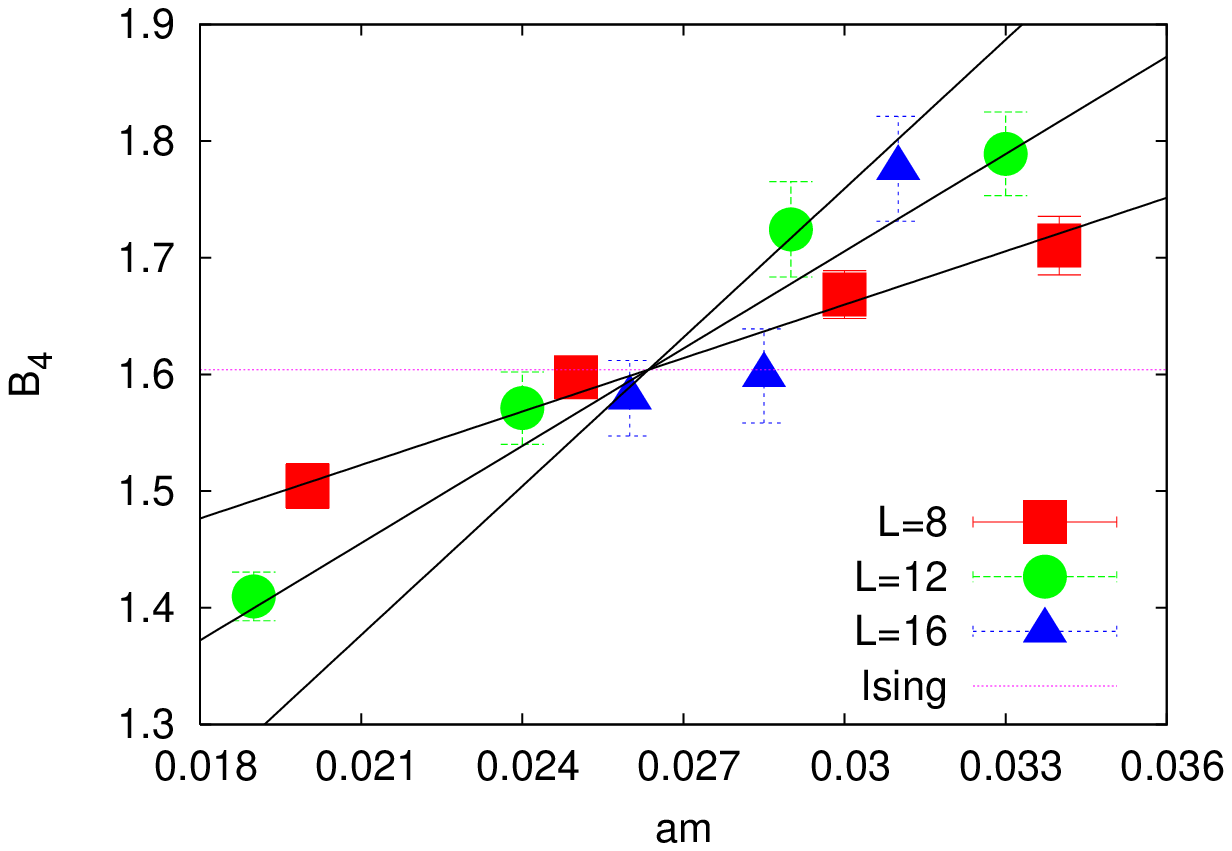}
\includegraphics*[width=0.5\textwidth]{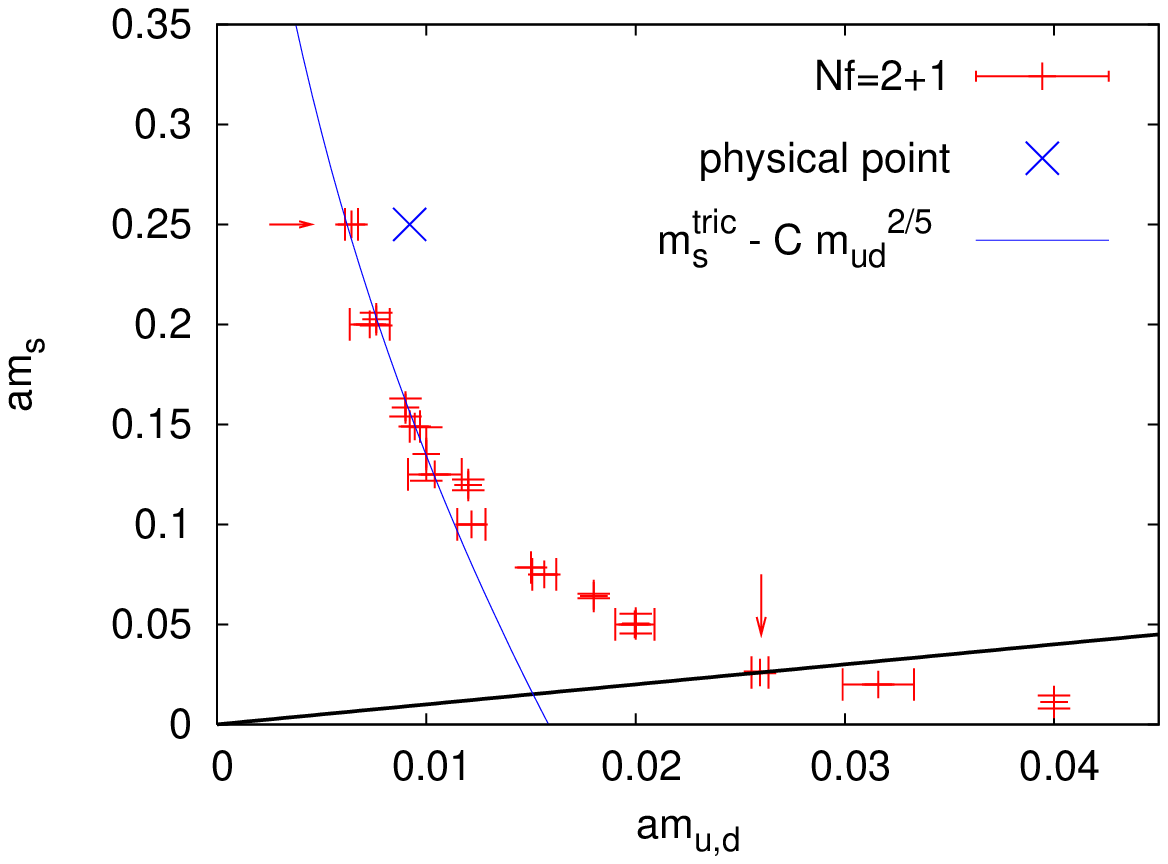}
}
\caption{Left: The Binder cumulant as a function or quark mass for $N_f=3, N_t=4$ and different
lattice sizes \cite{fp3}.
Right: The chiral critical line in the bare quark mass plane
at $\mu=0$, $N_t=4$.
$N_f=3$ is indicated by the solid line.
Also shown are the physical point according to
\cite{fk2}, and an extrapolation using mean field exponents to the chiral limit,
giving $m_s^{tric}\sim 2.8 T$. 
The point marked by the arrow has $m_\pi/m_\rho=0.148(2)$, compared to 
the physical value $0.18$. From \cite{fp3}.
}
\label{fig:B4}
\end{figure}

Next we move our attention to the second order boundary 
line separating the first order region from the crossover in 
fig.~\ref{fig:1schem} (right), starting with degenerate quark masses, $N_f=3$. 
In this case on $N_t=4$ lattices 
the critical quark mass marking the boundary 
between the crossover and the first order region is large enough for simulations to be carried out, and 
it was possible to determine it quite accurately \cite{kls,clm,fp2,fp3}, 
together with the Z(2) universality class of the 3d Ising model
to which it belongs \cite{kls}. The most practical observable to determine the 
latter is the Binder cumulant, constructed from moments of fluctuations of the order parameter,
\be
B_4(m,\mu)=\frac{\langle(\delta\bar{\psi}\psi)^4\rangle}
{\langle(\delta\bar{\psi}\psi)^2\rangle^2},\qquad 
\delta \bar{\psi}\psi=\bar{\psi}\psi - \langle \bar{\psi}\psi\rangle,
\ee
evaluated on the phase boundary $\beta=\beta_c(m,\mu)$.
In the infinite volume limit, $B_4\rightarrow 1$ or $3$ for a first order transition or crossover, respectively, whereas
it approaches a value characteristic of the universality class at a critical point. For 3d Ising
universality one has $B_4\rightarrow 1.604$ \cite{ising}.
Hence $B_4$ is a non-analytic step function, which gets smoothed out to an 
analytic curve on finite volumes,
with a slope increasing with volume to gradually approach the step function. 

A numerical example is shown in fig.~\ref{fig:B4} (left). 
The slope
increases with lattice size, and the data fit the 
finite volume scaling
formula predicted by universality,
\be
B_4(m,L)=b_0+b L^{1/\nu}(am-am^c).
\ee
One observes that all curves intersect at the critical $B_4$-value, and moreover $\nu$ is consistent with its 3d Ising value $\nu=0.63$. Hence we can read off the critical quark mass in lattice units, 
$am^c=0.0263(3)$ or $m^c/T_c=1.052(1)$ \cite{fp3}.

While the nature of the transition and its universality class are determined 
by long range fluctuations and thus should be insensitive to the cut-off effects 
on a coarse lattice, the critical quark mass is a
quantity requiring renormalization and is tremendously sensitive to such effects. 
Indeed, the critical
quark mass and the corresponding pion mass have been computed on $N_t=4$ lattices
with a standard staggered and a p4-improved staggered action. The critical pion 
mass in the improved 
case is almost a factor of four smaller than the non-improved 
result \cite{kls,milc}. Hence, the location
of the boundary line in the phase diagram is still far from the continuum result. 

Starting from the critical quark mass for $N_f=3$, the boundary line
has also been mapped for the non-degenerate $N_f=2+1$ theory, 
again with $N_t=4$ \cite{fp3}.
The question now arises which flavours
to couple to the quark chemical potential. Within QCD alone there are no 
flavour changing interactions,
while the initial state in heavy ion collisions has net $u,d$-quark 
numbers only. Thus it is sensible
to assign a chemical potential to the two degenerate light quarks only. 
In nature, where the weak interactions are switched on as well, the situation 
is somewhat more complicated than this \cite{exp}.  
The methodology then is the same as in the three flavour case: 
fix a strange quark mass $am_s$ and
scan the Binder cumulant in the light quark mass $am_{u,d}$ for the 
corresponding critical point.
Repeating for several strange quark masses produces the critical 
line $am_s^c(am_{u,d})$ shown
in fig.~\ref{fig:B4} (right). The results are in qualitative agreement with 
expectations from fig.~\ref{fig:1schem} (right).
They are also consistent with the possible existence of a 
tricritical point
$(m_{u,d}=0,m_s=m_s^{tric})$. Using its known, mean field exponents, the data 
favour a heavy $m_s^{tric} \sim 2.8 T_c$. Again large corrections,
presumably towards a smaller value, are expected in the continuum limit.
Also, $T_c$ is found to vary 
little along the critical line, in accordance with model calculations 
employing effective chiral lagrangians~\cite{Szep}.

\subsection{The nature of the transition at the physical point}
\label{sec:phys}

\begin{figure}[t]
\centerline{
\includegraphics*[bb=20 430 580 710,width=9cm]{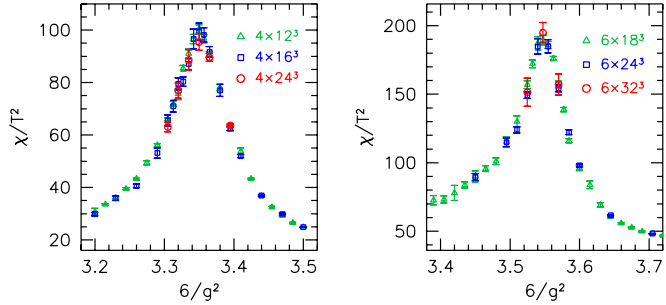}
}
\caption[]{Finite size scaling of the chiral susceptibility at the physical point for two lattice spacings$N_t=4$ (left) and $N_t=6$ (right). The peaks saturate at a finite height, consistent with an analytic crossover.}
\label{fig:fknat}
\end{figure}

A more immediate issue is whether the QCD physical point indeed lies on 
the crossover side of the critical line as expected. For that purpose spectrum 
calculations at $T\sim 0$ have been performed \cite{fp3} 
at the parameters indicated by the arrows
in fig.~ \ref{fig:B4} (right). They show that 
$m_s$ at the upper arrow is approximately tuned to its physical value 
($\frac{m_K}{m_\rho} \sim \frac{m_K}{m_\rho}|_{\rm phys}$=0.65),
while the pion is lighter than in physical QCD 
($\frac{m_\pi}{m_\rho}=0.148(2) < \frac{m_\pi}{m_\rho}|_{\rm phys}=0.18$).
This confirms that the physical point lies to the right of the critical line,
{\it i.e.} in the crossover region.  

An important question is whether this finding, calculated on coarse lattices $N_t=4$,
is stable under continuum extrapolations. This has been recently addressed and affirmed
in \cite{fknat}, using a Symanzik improved gauge and a stout-link improved staggered fermion action
for four different lattice spacings,  $N_t=4,6,8,10$. The simulations were carried out along the 
lines of constant physics, {\it i.e.}~the bare parameters were tuned such that the
hadron spectrum in physical units is constant for different lattice spacings.
The result of a finite size scaling analysis
of the susceptibility of the chiral condensate is shown in fig.~\ref{fig:fknat}. The peak height
appears to clearly saturate at some finite value, and the transition thus corresponds to a crossover. The finite peak height was calculated for all four 
lattice spacings and extrapolates to a finite value also in the continuum limit. 
Hence, one concludes that
the transition between hadronic
and quark gluon plasma physics at zero density indeed is an analytic crossover for 
physical QCD. In view of the concerns about the validity of the staggered fermion
actions \cite{sharpe}, it would of course be desirable to reprodce this result with 
alternative fermion discretizations.

\section{Lattice QCD at finite density}
\label{sec:mu}

Since 2001, significant progress was made towards simulating QCD with realistic parameter values
at small baryon densities. This is achieved by a number of methods that circumvent the sign problem,
rather than solving it: i) Multi-parameter reweighting, 
ii) Taylor expansion around $\mu=0$ and iii) simulations at imaginary 
chemical potential, either followed by analytic continuation or Fourier transformed to the canonical ensemble. 
It is important to realize that all of these approaches introduce some degree of 
approximation. However, their systematic errors are rather different, thus allowing for powerful crosschecks.  
All methods are found to be reliable as long as $\mu/T\lsim 1$, or $\mu_B\lsim 550$ MeV, which includes the region of interest for heavy ion collisions. 
Reviews specialized on this subject can be found in \cite{oprev,csrev}. 

\subsection{Two parameter reweighting}

Significant progress enabling finite density simulations was made a few years ago, by a generalization of the Glasgow method \cite{gla} to reweighting in two parameters \cite{fk0}.  
The partition function is rewritten identically as
\be 
Z=\left \langle \frac{\ex^{-S_g(\beta)}\det(M(\mu))}
{\ex^{-S_g(\beta_0)}\det(M(\mu=0))}\right\rangle_{\mu=0,\beta_0},
\ee
where the ensemble average is now generated at $\mu=0$ and a lattice gauge coupling 
$\beta_0$, while a reweighting factor takes us to the values
$\mu,\beta$ of interest. 
The original Glasgow method \cite{gla} reweighted in $\mu$ only and was suffering from the
overlap problem: while the reweighting formula is exact, its Monte Carlo evaluation is not. The integral
gets approximated by a finite number of the most dominant configurations, which are different for the reweighted and the original ensemble, and this difference grows with $\mu$.
When searching for a phase transition at some $\mu$, one-parameter reweighting uses a non-critical ensemble at $\mu=0$, thus missing important dynamics.
By contrast, two-parameter reweighting uses an ensemble generated at the pseudo-critical coupling 
$\beta_c(\mu=0)$, which is then reweighted along the pseudo-critical line of the phase change.  Thus one is working with an ensemble that probes both phases, improving the overlap with
the physical ensemble.
 
A difficulty in this approach is that the determinant
needs to be evaluated exactly, which is costly.  Moreover, because of the sign problem the reweighting factor is exponentially suppressed with volume and chemical potential, thus limiting the applicability to presently moderate values of those parameters. 
Moreover, since the statistical fluctuations are those of the simulated ensemble instead of the physical one,  all reweighted measurements are correlated and it is difficult to obtain reliable error estimates. For a proposed procedure see \cite{fk3}.  A further problem arises with staggered quarks, where the
root of the determinant has to be taken. For $\mu\neq 0$ this may enhance cut-off effects
to $O(a)$ instead of  $O(a^2)$ for $\mu=0$ \cite{gss}.

\subsection{Finite density by Taylor expansion}

Another way to gain information about non-zero $\mu$ is to compute the coefficients of a Taylor series expansion of observables in powers of $\mu/T$. 
Early attempts have looked at susceptibilities and the response of screening 
masses to chemical potential \cite{milc0,taro,hlp,gg}. 
More recently it has also been used to gain information on the phase transition 
and its nature itself \cite{bisw1,bisw11,bisw2,bisw3,ggpd}.
This idea exploits the fact that on finite volumes the partition function 
$Z(m>0,\mu,T)$ is an analytic function of the parameters of the theory.
For small enough $\mu/T$ one may then hope to get away with only a few terms, whose coefficients are calculated at $\mu=0$. 
Moreover, because of CP symmetry of the QCD action the partition function is even in $\mu$, 
$Z(\mu)=Z(-\mu)$, such that physical observables have series expansions in $(\mu/T)^2$. Thus, 
{\it e.g.}~the pressure density can be expressed as an even power series, 
\be
 p(T,\mu)=-\,\frac{F}{V} 
= \left(\frac TV\right)\log Z(T,\mu),\quad
{p\over T^4}=
\sum_{n=0}^\infty c_{2n}(T) \left({\mu\over T}\right)^{2n}.
\label{press}
\ee
The coefficients are equivalent to generalized quark number susceptibilities at $\mu=0$ and measurable with standard simulation techniques.
Since all the $\mu$-dependence of the partition function is in the fermion determinant, its derivatives  need to be computed,
\be
\frac{\partial \ln \det M}{\partial \mu}  ={\rm tr} \left( M^{-1} \frac{\partial M}{\partial \mu} \right),\quad
\frac{\partial {\rm tr} M^{-1}}{\partial \mu} =
- {\rm tr} \left( M^{-1} \frac{\partial M}{\partial \mu}
 M^{-1} \right), \quad \mbox{etc.},
 \ee
and iterations for higher orders. These expressions become increasingly complex and methods
to automatize their generation have been devised \cite{ggpd}.
Note that one now is dealing with traces of composite local operators, which greatly facilitates their numerical evaluation by statistical estimators compared to a computation of the full determinant as 
required for reweighting.

For high enough temperatures $T\gsim T_c$, the scale of the finite 
temperature problem is set by the 
Matsubara mode $\sim \pi T$, and one would expect coefficients of order one for an expansion in 
the `natural' parameter $\mu/(\pi T)$ \cite{fp2}. This is indeed borne out by numerical simulations
and explains the good convergence properties observed for $\mu/T\lsim 1$.

\subsection{QCD at imaginary $\mu$}

The hermiticity relation eq.~(\ref{gamma5}) tells us that the QCD fermion 
determinant with imaginary $\mu=i\mu_i$ is real positive, hence 
it can be simulated just as for $\mu=0$. 
It is then natural to ask whether such simulations can be exploited to learn 
something about physics at real $\mu$. 
The strategy to get back to real $\mu$ is to fit the full Monte Carlo results 
to a truncated Taylor series in $\mu_i/T$. In case of apparent convergence it is 
then easy to analytically continue the power series to real $\mu$. 
This idea was first used for observables like the chiral condensate and screening 
masses in the deconfined phase \cite{lom,hlp}. It was then shown to be 
applicable to the phase transition itself \cite{fp1}, which has recently been 
exploited in a growing number of works \cite{fp2,fp3,el1,el2,az1,cl,luo1}.

For complex $\mu=\mu_r+i\mu_i$, the QCD partition function eq.~(\ref{part})
is periodic in the imaginary direction, with period $2\pi/N_c$ for $N_c$ 
colours \cite{rw}. Hence, in addition to being even in $\mu$, 
the partition function has the additional symmetry $Z(\mu_r/T,\mu_i/T)=Z(\mu_r/T,\mu_i/T+2\pi/3)$. 
Because of the anti-periodic boundary conditions on fermions, this symmetry implies that a shift in $\mu_i$ by certain critical values is equivalent to a transformation by the $Z(3)$ centre of the gauge group. Thus, there are $Z(3)$ transitions between neighbouring centre sectors for all 
$(\mu_i/T)_c=\frac{2\pi}{3} \left(n+\frac{1}{2}\right), n=0,\pm1,\pm2,...$. It has been numerically verified that these transitions
are first order for high temperatures and a smooth crossover for low temperatures \cite{fp1,el1}. This limits the radius of convergence for analytic continuation, which is given by the first such transition, or $\mu/T=\pi/3$. Hence the approach is limited to $|\mu/T|\lsim 1$. 

Within this circle, the strategy then is
to measure expectation values of observables at imaginary $\mu$ and fit them by a Taylor series,
\be
 \langle O \rangle = \sum_n^N c_n \left(\frac{\mu_i}{\pi T}\right)^{2n}.
 \ee
 Working at imaginary $\mu$ has a couple of technical advantages. It is computationally simple and much cheaper than reweighting or computing coefficients of the Taylor expansion. Moreover, both parameters $\beta,\mu$ are varied and thus one obtains information from statistically independent ensembles. It also offers some control on the systematics by allowing a judgement on the convergence of the fits.
Furthermore, it is a good testing ground for effective QCD models: analytic results can always be continued to imaginary $\mu$ and be compared with the numerics there, as demonstrated for several examples in \cite{el2}.

A different approach making use of imaginary chemical potential is to employ 
the canonical ensemble at fixed quark number, which is related to the grand
canonical ensemble via the integral transform
\be
Z_C(T,B=3Q)=\frac{1}{2\pi}\int_{-\pi}^{\pi}d\left(\frac{\mu_i}{T}\right)\;\exp{(-i\mu_iQ/T)}\;
Z(\mu=i\mu_i,T,V).
\ee
One may then compute the grand canonical partition function at imaginary $\mu$
and perform the Fourier transform numerically \cite{kap}. This will only 
work for moderate $Q$ and small volumes, as the sign problem is reintroduced by the 
oscillations of the exponential. For small lattices $6^3\times 4,4^4$, 
first interesting results on the phase diagram have been obtained, 
both for staggered \cite{slavo} and Wilson
fermions \cite{andre}. However, this approach has no overlap problem, and one
might hope to push to larger chemical potentials once computational resources 
are available.

\subsection{Plasma properties at finite density}

Having developed computational tools for finite density, 
one can repeat the studies discussed in the
previous sections and see how finite baryon densities affect the screening 
masses \cite{milc,taro,hlp,gg}, the equation of state \cite{bisw2,bisw3,fk4} 
or the static potential \cite{fk5}. In all those cases the influence of the 
chemical potential is found to be rather weak, and for lack of space we will not further discuss 
those calculations here.
We likewise have to pass over the work done on certain modifications of QCD, 
for which the sign problem is manageable or absent, 
such as QCD in the static limit, two-colour QCD or QCD at finite isospin.
Such studies give important qualitative insights and some are covered in 
previous reviews \cite{latrevs,oprev}.   
Instead we concentrate here on calculations of the QCD phase diagram at finite 
density, where the order of the phase transition is expected to change  
as $\mu$ is increased, fig.~\ref{fig:1schem} (left).

\subsection{The critical temperature at finite density}

\begin{figure}[t!]
\vspace*{-1cm}
\begin{minipage}{0.5\textwidth} 
\includegraphics[angle=-90,width=7cm]{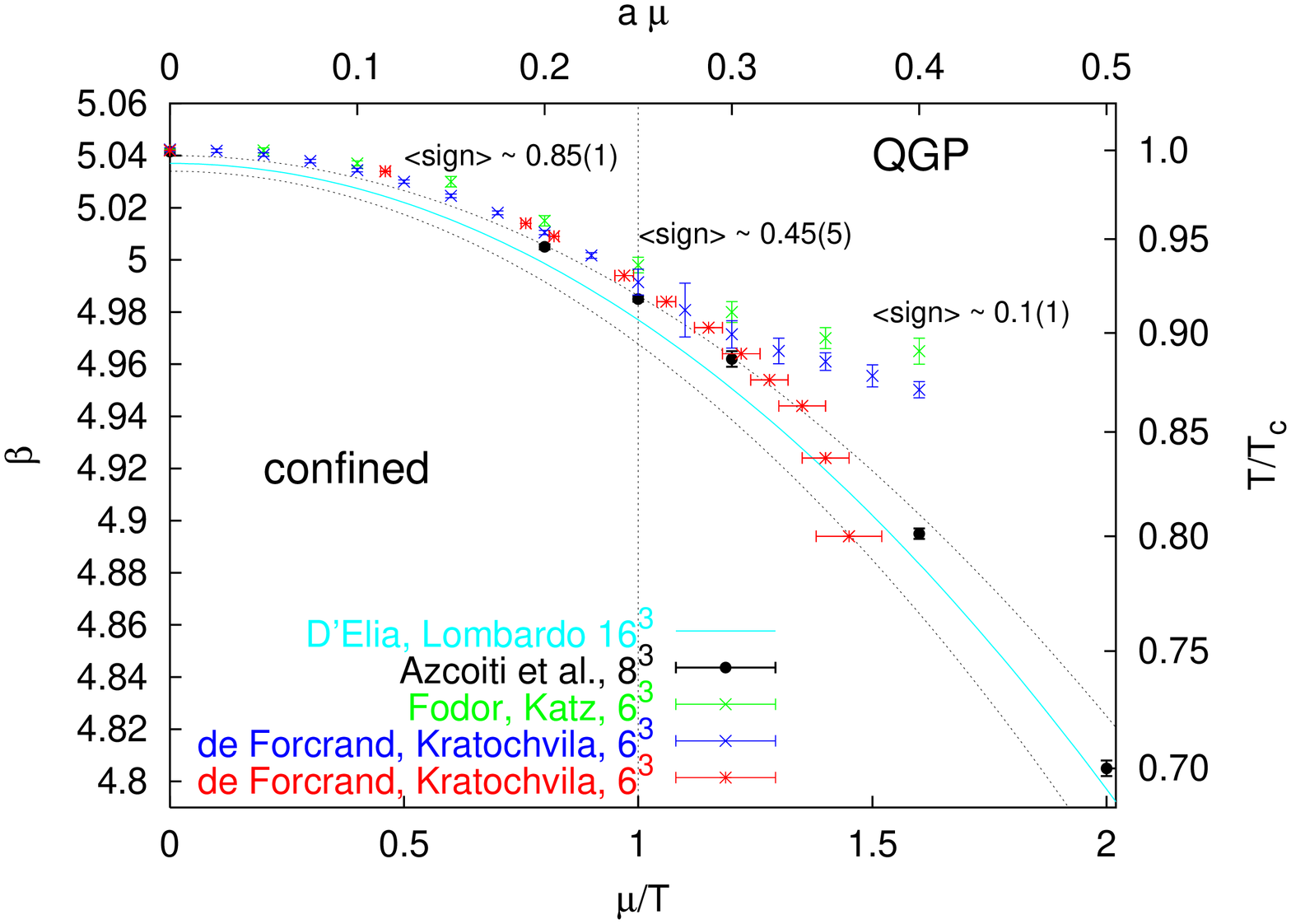}
\end{minipage}
\begin{minipage}{0.5\textwidth} 
\includegraphics[width=7cm]{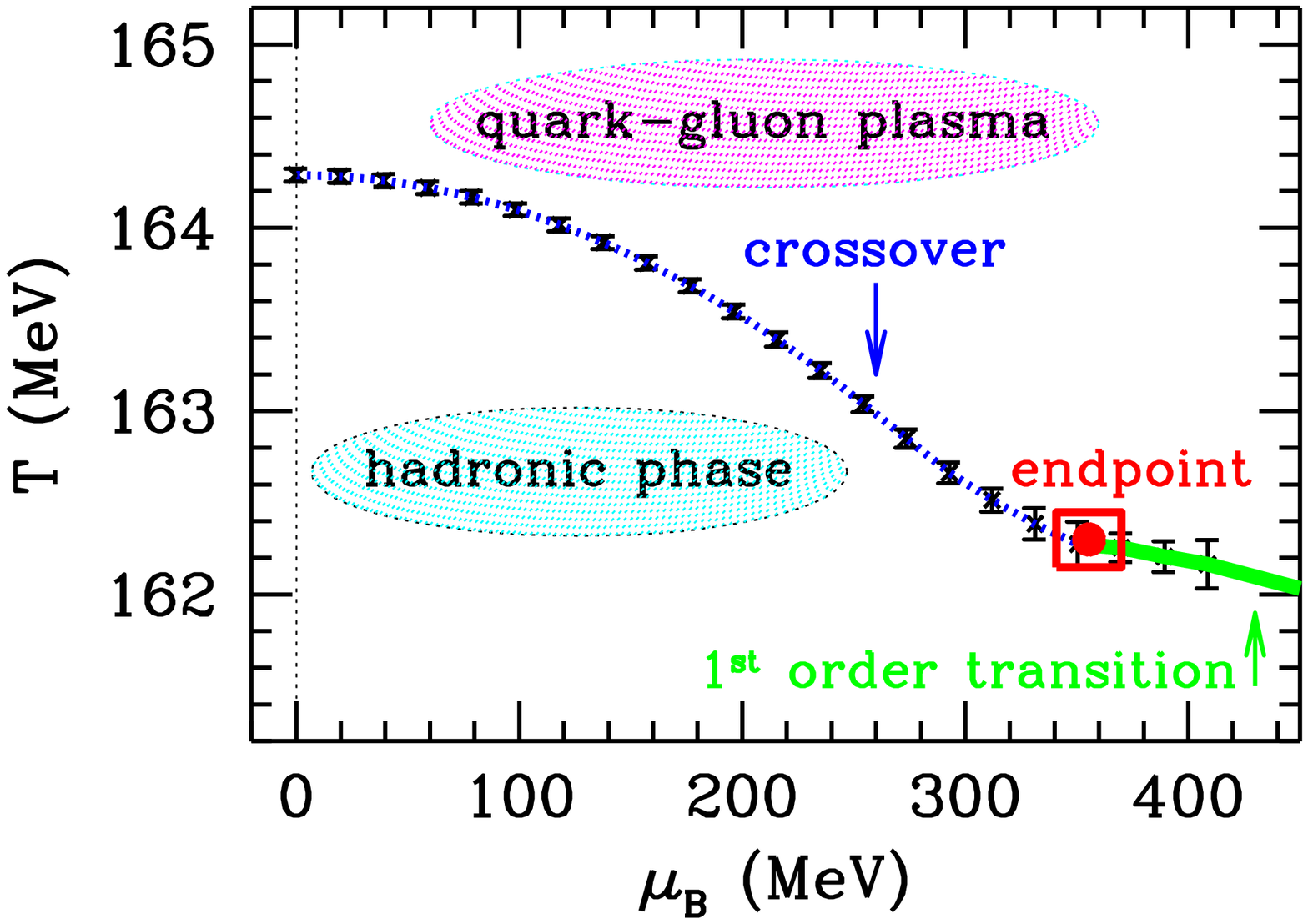}
\end{minipage}
\vspace*{-1cm}
\caption[]{Left: Comparison of different methods to compute the critical couplings
\cite{slavo}.
Right: The phase diagram for physical quark masses as predicted by the two parameter reweighting method \cite{fk2}.}
\label{fig:tccomp}
\end{figure}

As in the case of zero density, let us first discuss the phase boundary, 
{\it i.e.}~the (pseudo-)critical 
temperature $T_c(\mu)$, before dealing with the order of the phase transition. 
The critical line has been calculated for a variety of flavours and 
quark masses using different methods. For a quantitative comparison one 
needs data at one fixed parameter set and also eliminate the uncertainties of 
setting the scale. Such a comparison is shown
for the critical coupling in fig.~\ref{fig:tccomp} (left), 
for $N_f=4$ staggered quarks with the same action and quark mass $m/T\approx 0.2$.
(For that quark mass the transition is first order
along the entire curve).
One observes quantitative 
agreement up to $\mu/T\approx 1.3$, after which the different results start 
to scatter. This vindicates our earlier statement that all methods appear 
to be reliable for $\mu/T\lsim 1$. 
Note that strong coupling results at $\beta=0$ 
predict $a\mu_c(\beta=0)=0.35-0.43$ \cite{kawa}, requiring the line to bend down
rapidly, and thus favour the data from the canonical approach.

The case of physical quark masses, after conversion to continuum units, 
is shown in fig.~\ref{fig:tccomp} (right) \cite{fk2}.
One observes that the critical temperature is decreasing only very 
slowly with $\mu$. This is consistent with a description by a 
Taylor expansion in powers of $(\mu/\pi T)^2$ with
coefficients of order one,
\begin{equation}
\frac{T_c(\mu)}{T_c(0)}=1-t_2(N_f,m_f)\left(\frac{\mu}{\pi T}\right)^2
+{\mathcal O}\left( \left( \frac{\mu}{\pi T}\right)^4 \right) \quad .
\label{texp}
\end{equation}   
The leading coefficients for various cases have been collected from the literature in \cite{csrev} and are
reproduced in table \ref{tab:tccomp}. The curvature gets stronger with increasing $N_f$, which is 
consistent with the observation at zero density that $T_c$ is lowered by light flavours, cf.~sec.~\ref{sec:tc}. Subleading coefficients are also beginning to emerge but at present not statistically significant yet. 
It has been noted that the convergence should speed up when constructing Pad{\'e}
approximants for the series, which also tends to increase the curvature 
towards larger $\mu$ \cite{el2,pade}.
Finally one should note that the continuum conversions relying on the two-loop beta function 
are certainly not reliable for these coarse lattices, while fits to 
non-perturbative beta functions tend to also increase the curvature.
\begin{table}
\begin{center}
\begin{tabular}{cccccccc}\hline \hline
$N_f$&$am$&$N_s$&$t_2$&Action&$\beta$-Function&Method&Reference    \\ \hline
2  &0.1   & 16      &0.69(35) &p4   &non-pert.   &Taylor+Rew.& \cite{bisw1} \\
   &0.032 & 6,8     &0.500(54)&stag.&2-loop pert.&Imag.      & \cite{fp1}\\
3  &0.1   & 16      &0.247(59)&p4   &non-pert.   &Taylor+Rew.& \cite{bisw11}\\
   &0.026 & 8,12,16 &0.667(6) &stag.&2-loop pert.&Imag.      & \cite{fp3}\\
   &0.005 & 16      &1.13(45) &p4   &non-pert.   &Taylor+Rew.& \cite{bisw11}\\
4  &0.05  & 16      &1.86(2)  &stag.&2-loop pert.&Imag.      & \cite{el1}\\ \hline
2+1&0.0092,0.25&6-12&0.284(9) &stag.&non-pert.   &Rew.       & \cite{fk2}\\
\hline \hline
\end{tabular}
\caption{Coefficient $t_2$ in the Taylor
expansion of the transition line, eq.~(\ref{texp}). 
All results have been obtained with
$N_t=4$. \label{tab:tccomp}}
\end{center}
\end{table}

\subsection{The QCD phase diagram for $\mu\neq 0$ and the critical point}

\begin{figure}[t!]
\vspace*{-1cm}
\centerline{
\scalebox{0.65}{\includegraphics{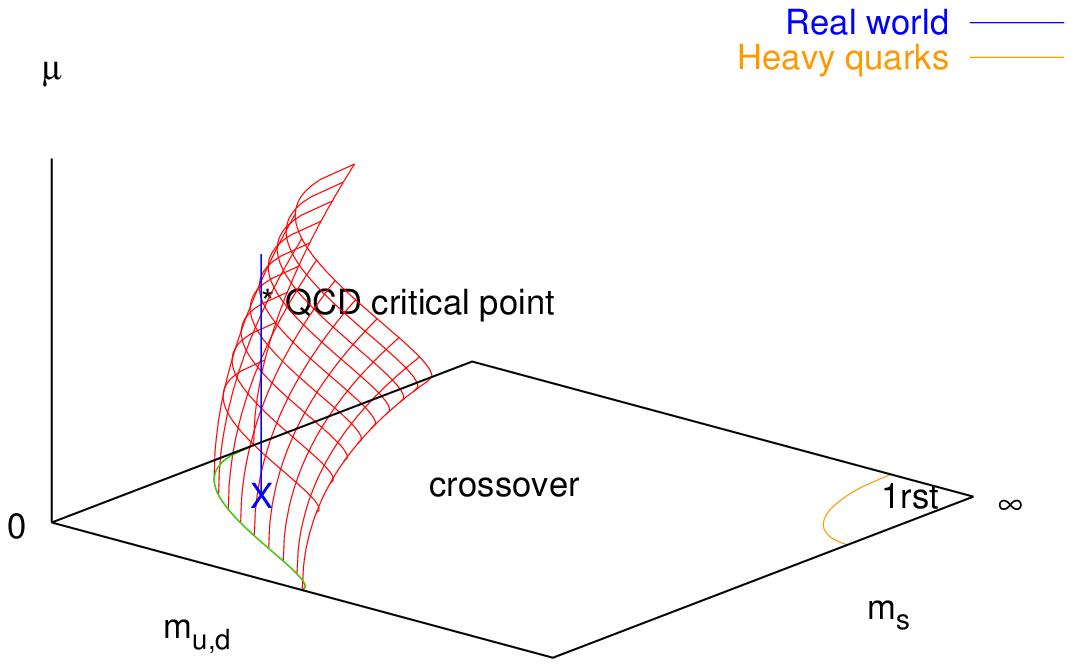}}
\scalebox{0.65}{\includegraphics{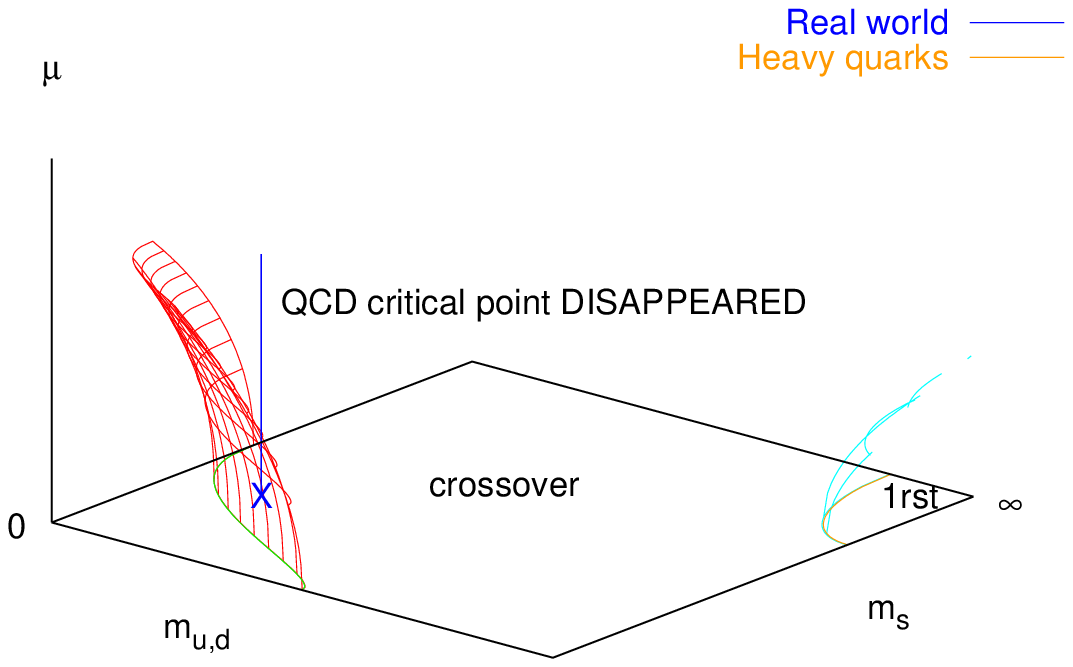}}
}
\centerline{
\scalebox{0.58}{\includegraphics{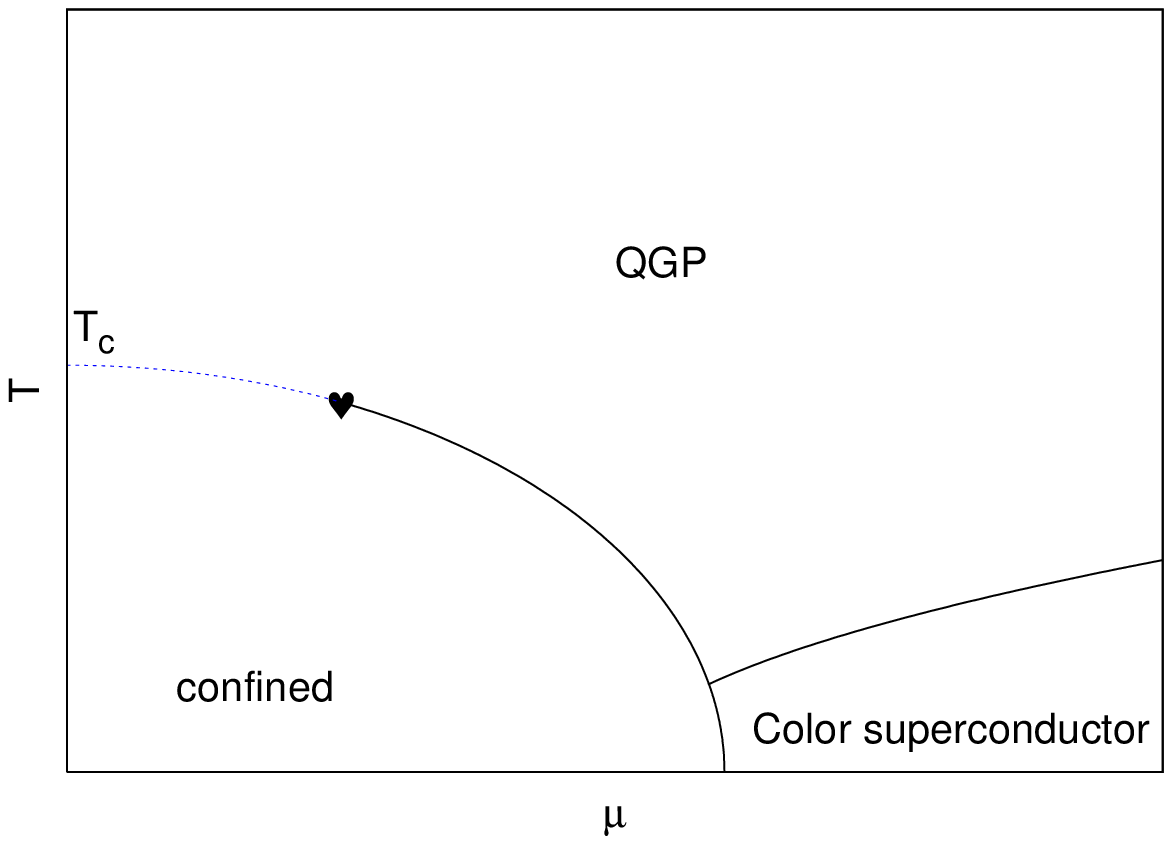}}
\scalebox{0.58}{\includegraphics{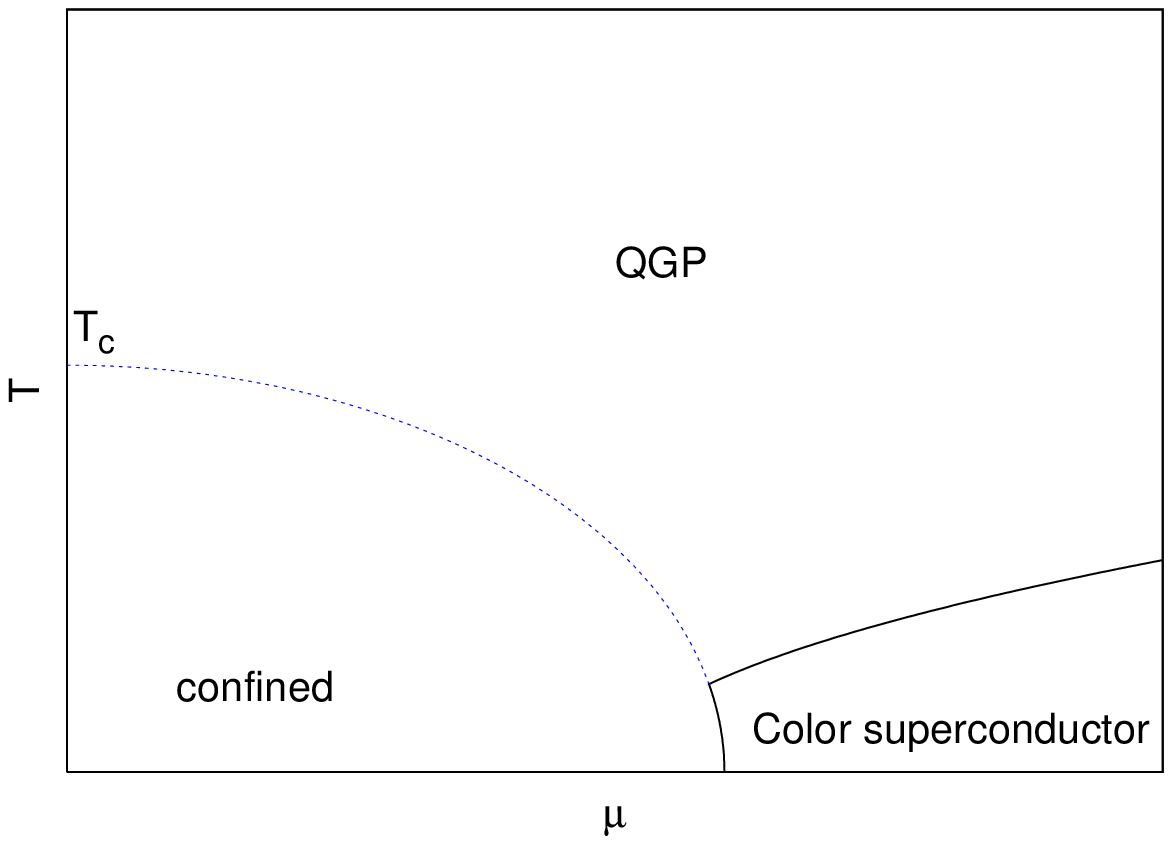}}
}
\caption{Upper panel: 
The chiral critical surface in the case of positive (left) and negative (right) curvature.
If the physical point is in the crossover region for $\mu=0$, a finite $\mu$
phase transition will only arise in the scenario (left) with positive curvature,
where the first-order region expands with $\mu$.
Note that for heavy quarks, the first-order region shrinks with $\mu$,
cf.~sec.~\ref{sec:potts}.
Lower panel: phase diagrams for fixed quark mass (here $N_f=3$) corresponding to 
the two scenarios depicted above.}
\label{fig:2schem}
\end{figure}

As in the case of $\mu=0$, a determination of the order of the transition, an hence the search for the critical endpoint, is much harder, and we begin by discussing the qualitative picture.
If a chemical potential is switched on for the light quarks, there is an additional parameter requiring
an additional axis for our phase diagram characterizing the order 
of the transition, fig.~\ref{fig:1schem} (right).
This is shown in fig.~\ref{fig:2schem}, where the horizontal plane is spanned by the $\mu=0$ phase diagram in $m_s, m_{u,d}$ and the vertical axis represents $\mu$. 
The critical line separating the first order section from the crossover will now extend to finite $\mu$ and
span a surface. A priori it is, of course, not known whether this surface might be leaning to the right or the left, or even have a more complicated behaviour as a function of the quark masses. 
%
However, the expected QCD
phase diagram is only obtained if the left situation of fig.~\ref{fig:2schem} 
is realized (unless there are additional critical surfaces yet unknown). 
The first order region expands as $\mu$ is turned on, so that the
physical point, initially in the crossover region, eventually belongs to the
critical surface. At that chemical potential $\mu_E$, the transition is second order:
that is the QCD critical point. Increasing $\mu$ further makes the transition
first order. 
A completely different scenario arises if instead the first-order
region shrinks as $\mu$ is turned on. In that case the physical
point remains in the crossover region for any $\mu$, fig.~\ref{fig:2schem} (right). 

There are then different strategies to determine the QCD phase diagram. 
One can fix a particular set of quark masses and for that theory switch on and increase the chemical potential to see whether a critical surface is crossed or not. Such calculations are covered in 
sec.~\ref{sec:cp}.
Alternatively, sec.~\ref{sec:crit} discusses how to start from the known critical line at $\mu=0$
and study its evolution with a finite $\mu$. That will give information on the whole phase diagram
in fig.~\ref{fig:2schem}, including, eventually, physical QCD. 

\subsection{Critical point for fixed masses from reweighting and Taylor expansion}
\label{sec:cp}

\begin{figure}[t]    
\vspace*{-1cm}
\includegraphics[width=0.44\textwidth]{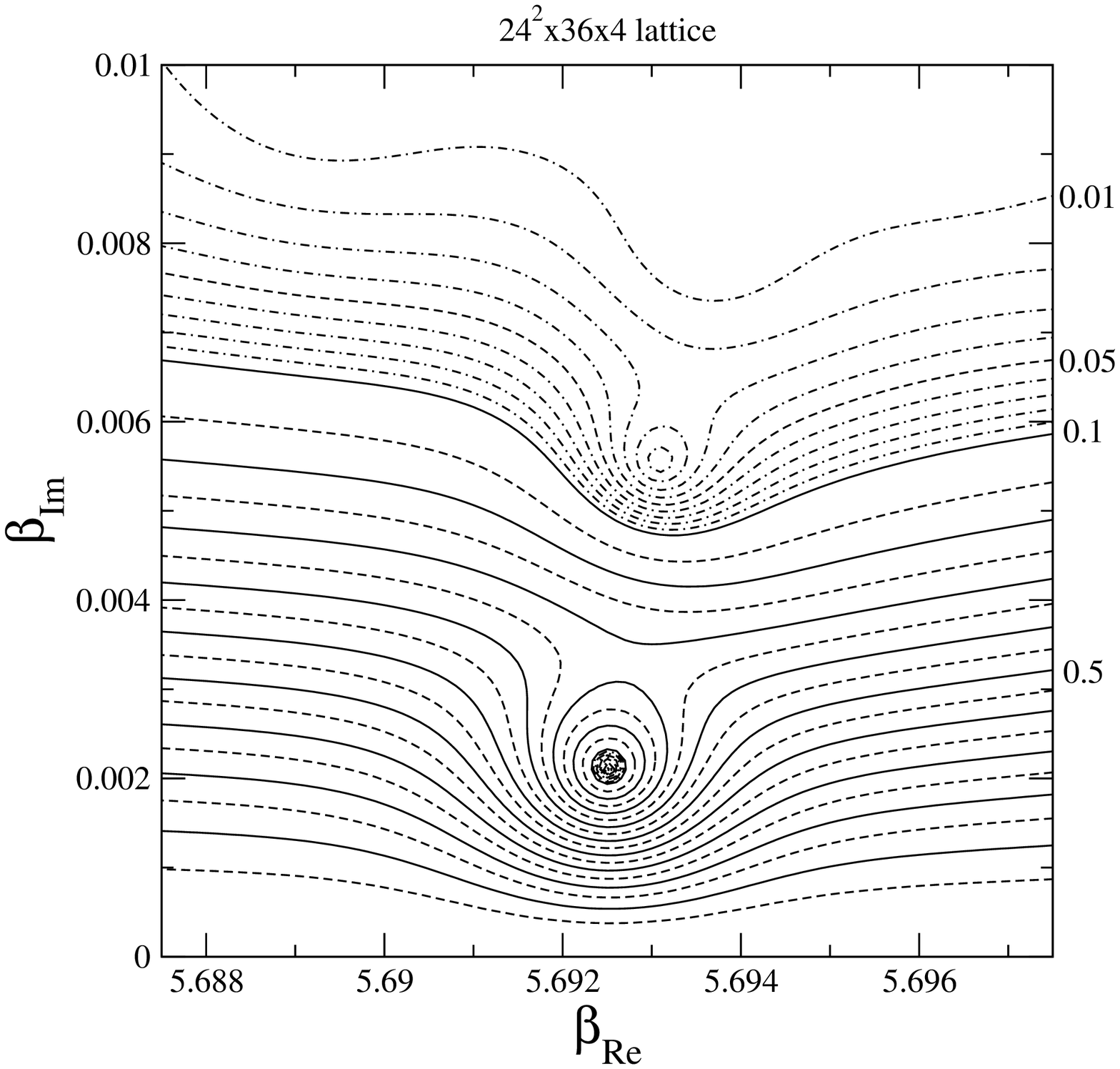}\hspace*{0.5cm}
\includegraphics[width=0.5\textwidth]{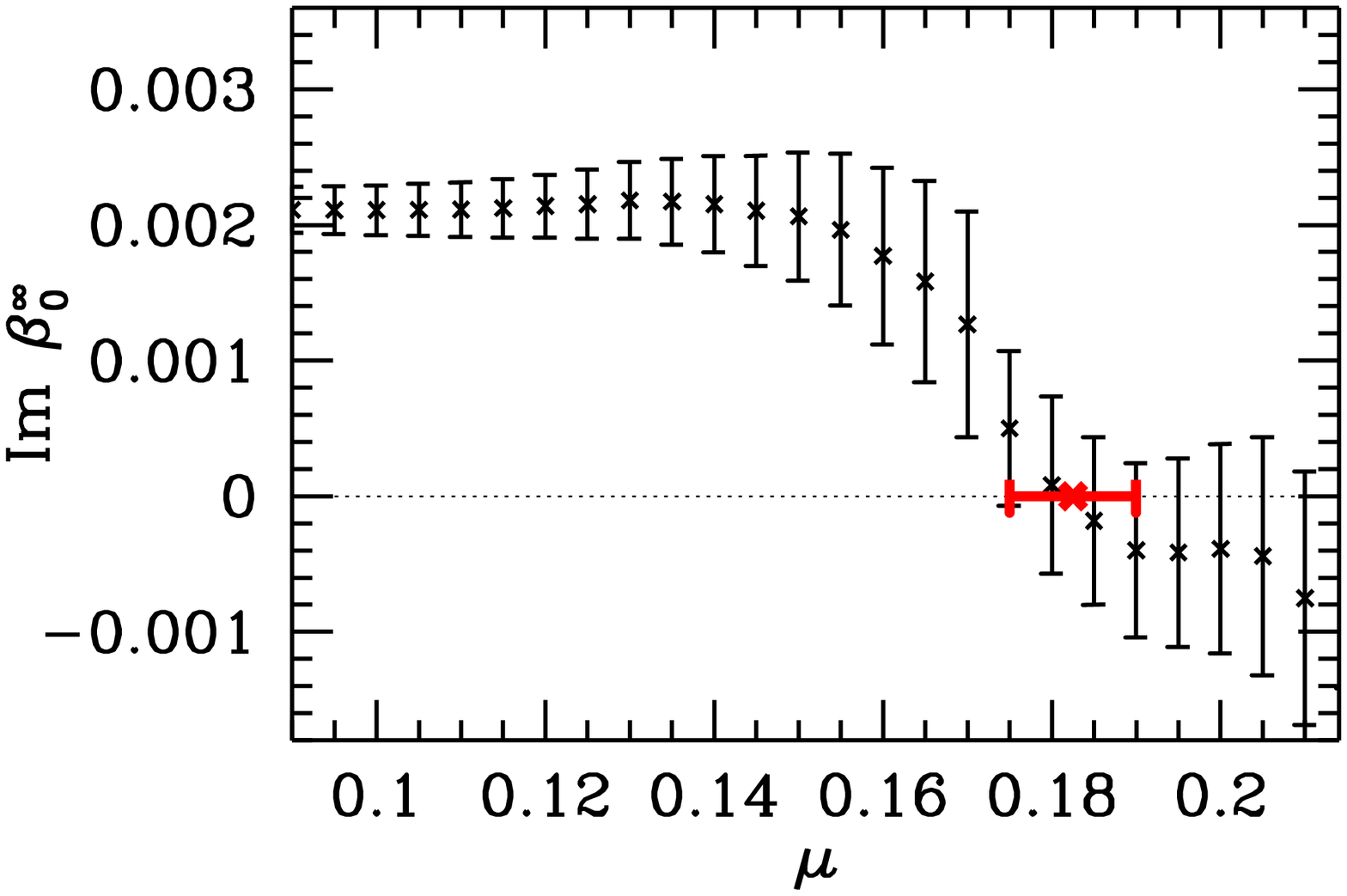}

\caption[]{ Left: Lee-Yang zeroes in the complex $\beta$-plane for SU(3) 
pure gauge theory \cite{shinji}.
Right: Imaginary part of the Lee-Yang zero closest to the real axis as a 
function of chemical potential \cite{fk2}. }
\label{fk}
\end{figure}

Reweighting methods produced the first finite density phase diagram 
from the lattice, 
obtained for light quarks corresponding to $m_\pi\sim 300$ MeV \cite{fk1}. 
A later simulation at physical quark masses puts the critical point 
at $\mu_B^E\sim 360$ MeV \cite{fk2}, fig.~\ref{fk} (left). 
In this work $L^3\times 4$ lattices with $L=6-12$
were used, working with the standard staggered fermion action. 
Quark masses were tuned to $m_{u,d}/T_c\approx 0.037, m_s/T_c\approx 1$, 
corresponding to the mass ratios 
$m_{\pi}/m_{\rho}\approx 0.19, m_{\pi}/m_K\approx 0.27$, 
which are close to their physical values.
A Lee-Yang zero analysis \cite{ly} was employed in order to find the 
change from crossover behaviour at $\mu=0$ to a first order 
transition for $\mu>\mu_E$.
This is shown in fig.~\ref{fk}. For a crossover the partition 
function has zeroes only off the real axis,
whereas for a phase transition the zero moves to the real axis 
when extrapolated to infinite volume.  
For a critical discussion of the use of Lee-Yang zeros in 
combination with reweighting, see \cite{shinji}.
Recently, reweighting has been combined with the density of states 
method \cite{dos}, in order to extend the applicability of reweighting 
to larger values of $\mu/T$. 
First interesting results have been obtained for $N_f=4$,  
indicating a new high density phase and a possible triple point, 
where the high density transition line meets the deconfinement line \cite{dos}. 
Unfortunately, the method is computationally very expensive and so far 
limited to coarse and small lattices, so that it is difficult to 
unambiguously establish those findings at present.

In principle the determination of a critical point is also possible 
via the Taylor expansion.
In this case true phase transitions will be signalled by an 
emerging non-analyticity, or a finite radius of convergence for the pressure series about $\mu=0$ as the volume is increased. 
The radius of convergence of a power series gives the distance between the expansion point and the nearest singularity, and may be extracted from the high order behaviour of the series. A possible 
definition is
\be
\rho = \lim_{n\rightarrow\infty}\rho_n \qquad \mbox{with}\quad
    \rho_n=\left|\frac{c_0}{c_{2n}}\right|^{1/2n}.
\label{rad}
\ee
General theorems ensure that if the limit exists and asymptotically all coefficients of the series are positive, then there is a singularity on the real axis. 
More details as well as previous applications to strong coupling expansions 
in various spin models can be found in \cite{series}.
In the series for the pressure such a singularity would correspond to the critical point in the $(\mu,T)$-plane.
 
A critical endpoint for the $N_f=2$ theory, based on this approach, was reported  in \cite{ggpd}. The
authors perfomed simulations on $L^3\times 4$ lattices with $L=8-24$, using the standard staggered action. The quark mass was fixed in physical units to $m/T_c=0.1$. The aim of the simulations was to bracket the critical point by computing the Taylor coefficients of the quark number susceptibility up to sixth order ({\it i.e.}~8th order for the pressure) for various temperatures in the range $T/T_c=0.75-2.15$, and extrapolate to finite $\mu$.
This was done for different lattice volumes in order to get an estimate of finite voulme effects.

\begin{figure}[t]
\includegraphics*[height=45mm]{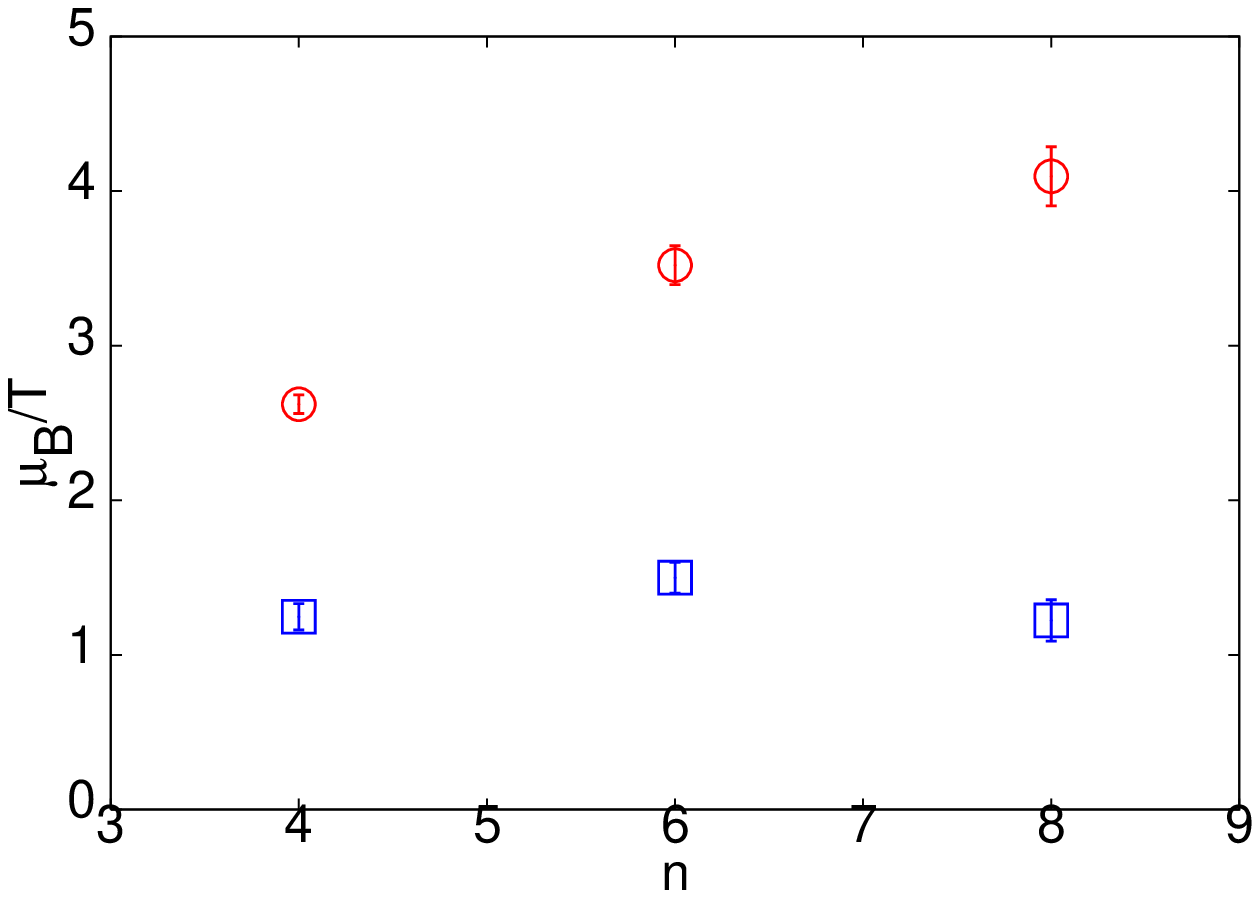}\hspace*{1cm}
\includegraphics*[height=45mm]{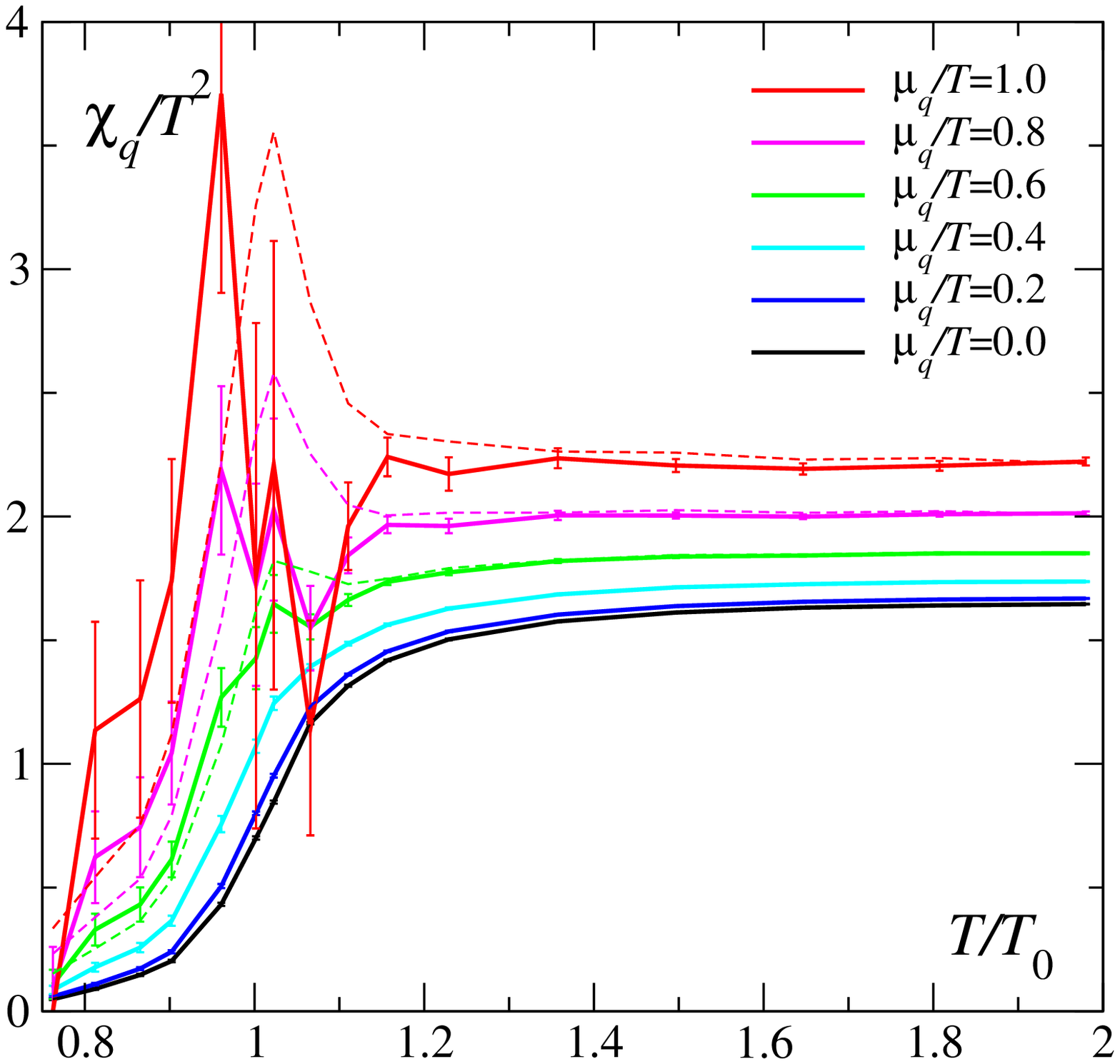}
\caption[]{Left: Estimators of the radius of convergence $\rho_n$, 
eq.~(\ref{rad}), at $T/T_c=0.95$ on $N_t=4$ lattices. 
Circles represent $L=8$, squares $L=24$ \cite{ggpd}.
Right: Quark number susceptibility computed through $O(\mu^4)$ (dashed lines) and $O(\mu^6)$ (solid lines) \cite{bisw3}.}
\label{ggfig}
\end{figure}

The results for the convergence radius eq.~(\ref{rad}) are shown in 
fig.~\ref{ggfig} (left). A rather strong volume dependence is apparent. 
While for the smaller $8^3$ lattice the estimator $\rho_n$ does not 
seem to converge to a finite radius of convergence, the results on the larger $24^3$ lattice are consistent with settling at a limiting value. 
Taking the large volume result at face value and extrapolating to all 
orders the estimate 
for the location of the critical point is $\mu^E_B/T_E=1.1\pm 0.2$ at 
$T_E/T_c(\mu=0)=0.95$ \cite{ggpd}.

Another investigation of the two-flavour theory,  also using the Taylor expansion of the pressure,
is reported in \cite{bisw2,bisw3}. 
This group works with a $16^3\times4$ lattice with p4-improved staggered fermions and a Symanzik-improved Wilson action,  
the quark mass is set to $m/T\approx0.4$.  The calculation to order 
$\mu^4$ was performed in \cite{bisw2} while results on $\mu^6$ are presented in 
\cite{bisw3}. The last work also contains detailed discussions of analytic 
calculations to compare with,
namely the pressure in high temperature perturbation theory \cite{vuo}, 
which is going to hold at very high temperatures, as well as the 
hadron resonance gas model, which gives a rather good description of 
the pressure in the confined phase \cite{krt}.

In agreement with \cite{ggpd} and qualitative expectations, 
their detailed results for the coefficients in the pressure series 
satisfy $c_6\ll c_4 \ll c_2$ for $T>T_c$, {\it i.e.}~one would have 
coefficients of order one for an expansion in $(\mu/\pi T)$. 
An impression of the convergence of the series can be obtained by 
looking at the quark number susceptibility calculated to consecutive orders, 
as shown in fig.~\ref{ggfig} (right).
For $T\lsim 1.2 T_c$, the series seems to converge rapidly and the $\mu^6$-result 
is compatible with the one through order $\mu^4$. Around the 
transition temperature $T_c$, the $\mu^4$-results show a peak emerging 
with growing $\mu/T_c$, which in \cite{bisw2}  was interpreted as evidence for
a critical point. However, the $\mu^6$ contribution suggests that in this region results do not yet converge,  and the structure is hence not a significant feature of the full pressure.
Furthermore, estimates for the radius of convergence through that order agreed with predictions
from the hadron gas model, which however has infinite radius of convergence. Hence the conclusion
in \cite{bisw3} that there is no signal for a critical point at that quark mass.

\subsection{The change of the critical line with $\mu$ \label{sec:crit}}

\begin{figure}[t!]
\centerline{
\scalebox{0.62}{\includegraphics{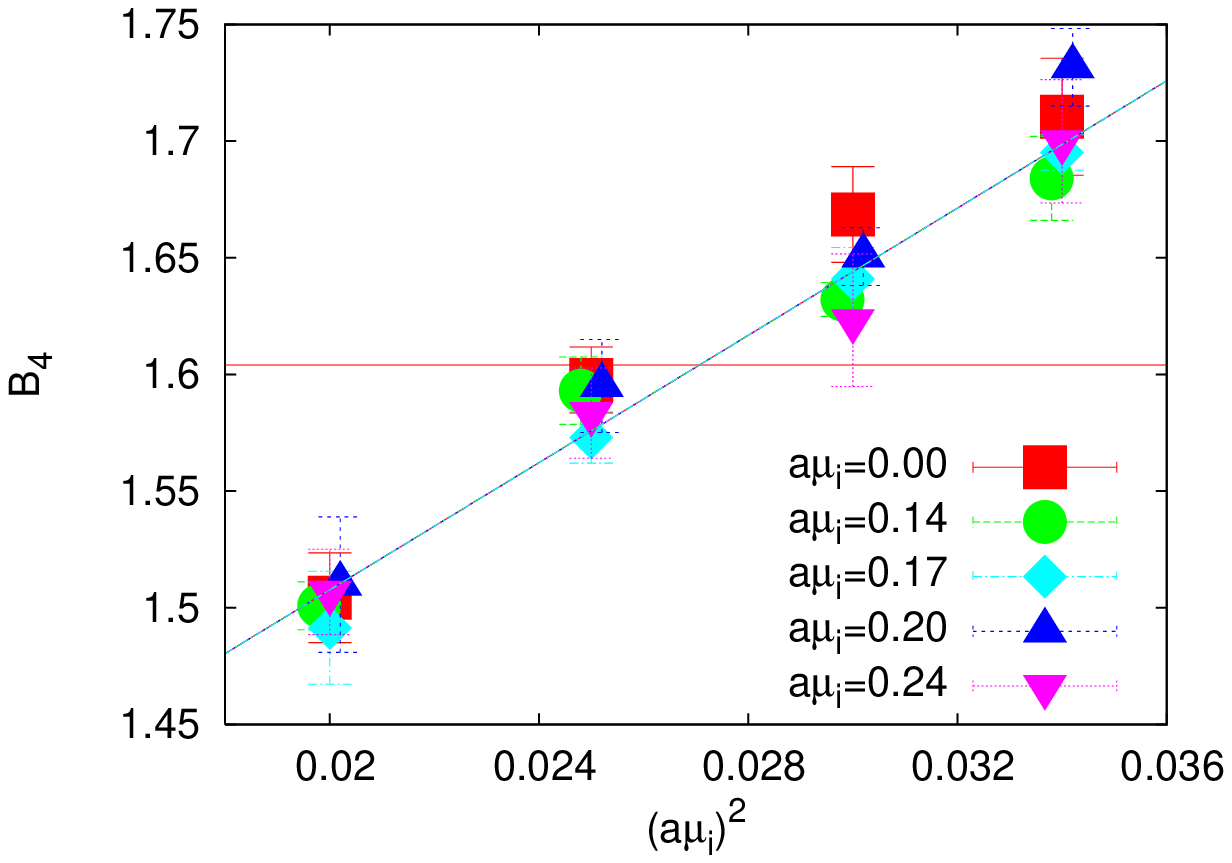}}
\scalebox{0.62}{\includegraphics{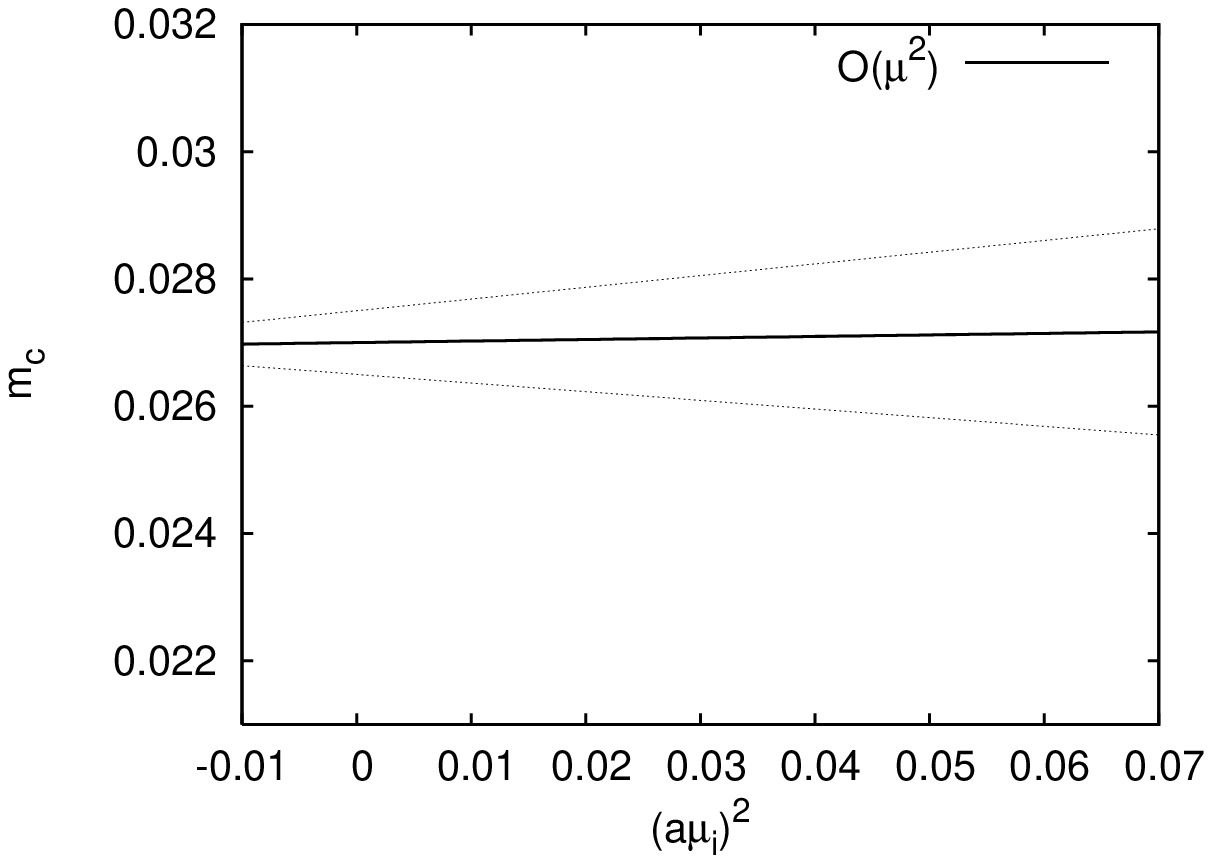}}
}
\caption{Left: $B_4(am,a\mu_i)$ for $N_f=3$, $N_t=4$ and
imaginary chemical potentials.
Right: One-sigma error band for the critical mass $a m^c(a \mu_i)$ resulting
from a linear fit to the data on the left. From \cite{fp3}.}
\label{fig:immu}
\end{figure}

Rather than fixing one set of masses and considering the effects of $\mu$, 
one may map out the critical surface in fig.~\ref{fig:2schem} by measuring
how the $\mu=0$ critical boundary line changes under the influence of $\mu$.
This question was adressed using an imaginary 
chemical potential in 
\cite{fp2,fp3}. The methodology employed is the same 
as in sec.~\ref{sec:m1m2c}, {\it i.e.}~measurement
of the Binder cumulant of the chiral condensate. 
The results for $N_f=3$ are summarized in fig.~\ref{fig:immu} (left). 
The chemical potential is found to have almost no influence on $B_4$ as a function
of quark mass. 
A lowest-order fit, linear in $a m$ and $(a \mu)^2$, 
gives the error band fig.~\ref{fig:immu} (right), 
corresponding to
\be
a m^c(a \mu) = 0.0270(5) - 0.0024(160) (a \mu)^2 \; .
\label{c_prime1}
\ee

Care must be taken for the conversion to physical units. The crucial
point is that the gauge coupling $\beta$ is tuned with changing $\mu$ to 
stay on the critical line, so that $a(\beta)$ decreases.
Expressing the change of the critical quark mass with chemical potential in lattice
and continuum units as
\be
\label{c1prime}
\frac{a m^c(\mu)}{a m^c(0)} = 1 + \frac{c'_1}{a m^c(0)} (a \mu)^2 + ..., \qquad
\frac{m^c(\mu)}{m^c(0)} = 1 + c_1 \left( \frac{\mu}{\pi T} \right)^2 + ...
\label{c1}
\ee
then $c_1$ and $c'_1$ are related by
\be
c_1 = \frac{\pi^2}{N_t^2} \frac{c'_1}{a m^c(0)} + \left( \frac{1}{T_c(m,\mu)} \frac{d T_c(m^c(\mu),\mu)}{d(\mu/\pi T)^2} \right)_{\mu=0}.
\ee
Since $c_1'$ is observed to be nearly zero, it is the second term that dominates, leading
to an overall negative coefficient $c_1=-0.7(4)$ \cite{fp3}.

This is evidence that, in the $N_f=3$ theory on an $N_t=4$ lattice,
the region of first-order transitions {\em shrinks} as a baryon chemical potential
is turned on, and the ``exotic scenario'' of fig.~\ref{fig:2schem} (right) is realized.
Interestingly, similar qualitative conclusions are obtained from simulations of the three flavour theory
with an isospin chemical potential~\cite{DKS_mc}, as well as simulations at imaginary $\mu$
employing Wilson fermions \cite{luo1}.

This investigation has also been extended to the case of non-degenerate quarks \cite{fp3}.
Fig.~\ref{potts} (left) shows a comparison of the critical line for $\mu=0$  with some critical masses
calculated for a sizeable baryon chemical potential. The data show the same trend as for $N_f=3$: the critical mass is nearly unchanged 
or slightly increasing with $\mu_i$, in lattice units. 
The conversion to physical units
proceeds as in the three flavour case and gives
a dominant negative contribution to $c_1$.
Together with a very small value for $c'_1$, 
it implies again that the first-order region shrinks as the baryon chemical
potential is turned on, and the ``exotic scenario'' of fig.~\ref{fig:2schem} (right) is the correct one.

\subsection{The heavy quark limit: Potts model}
\label{sec:potts}

In light of these surprising results, it is also interesting to consider
the heavy quark corner of the schematic phase diagram of fig.~\ref{fig:2schem}.
Simulations of quenched QCD at finite baryon number have been done in \cite{quenb}. 
As the quark mass goes to infinity, quarks can be integrated out and QCD reduces to a gauge theory of Polyakov lines. First simulations of this theory with Wilson valence quarks can be found in \cite{nucu}.
At a second order phase transition, universality allows us to neglect the 
details of gauge degrees of freedom, so the theory reduces to the 
3d three-state Potts model, which is in the 3d Ising universality class. 
Hence, studying the three-state Potts model should teach us about the behaviour of QCD in the neighbourhood of the critical line separating the quenched first order region from the crossover region.
For large $\mu$ the sign problem in this theory has actually been solved by means 
of cluster algorithms \cite{clust}. 

\begin{figure}
\centerline{
{\includegraphics*[width=0.5\textwidth]{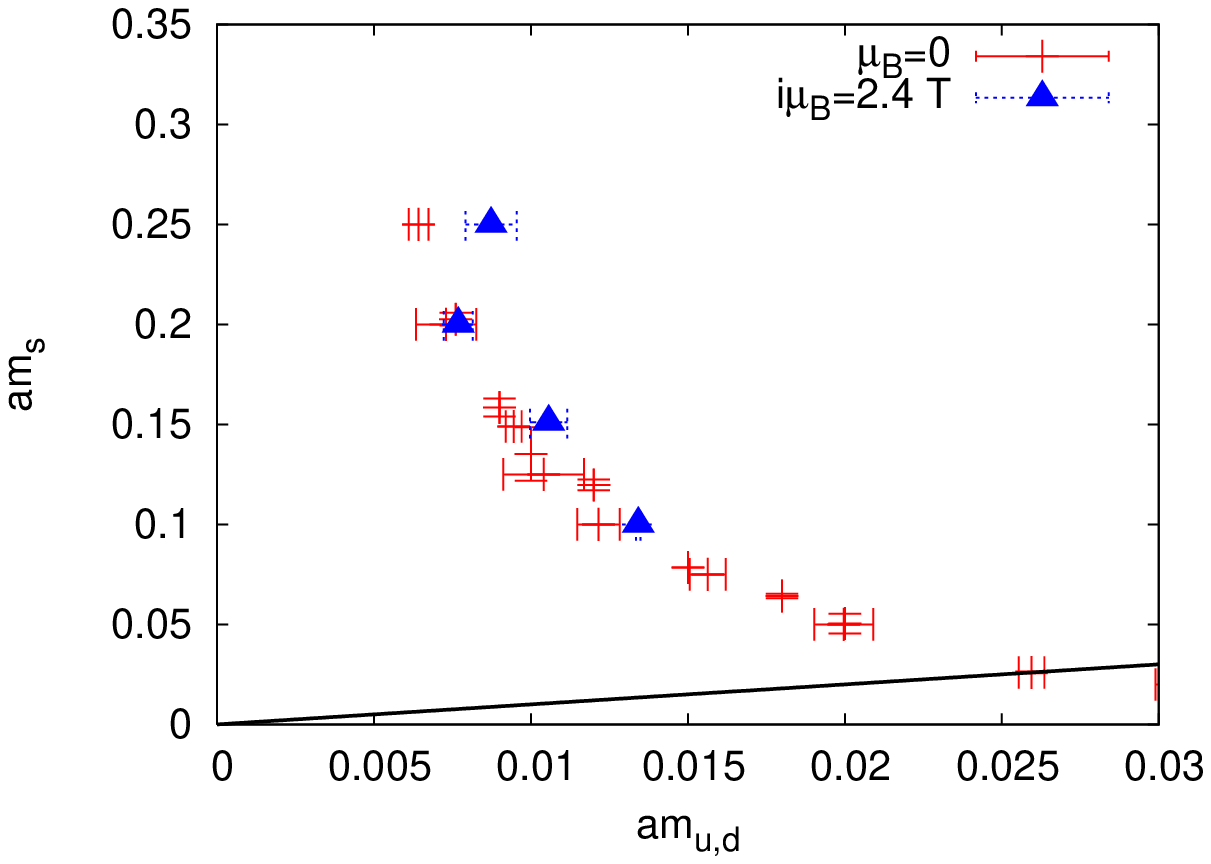}}
\includegraphics*[width=0.45\textwidth]{potts.eps}
}
\caption{Left: Comparison of the critical line at $\mu=0$ and $a \mu_I = 0.2$.
Right: The critical heavy quark mass separating first order from crossover as a function of $\mu^2$ from the Potts model \cite{Potts}.}
\label{potts}
\end{figure}

Here we are interested in simulations at small $\mu/T$ \cite{Potts}.  In this case, the sign problem is mild enough for brute force simulations at real $\mu$ to be feasible. 
In \cite{Potts}, the change of the critical heavy quark mass is 
determined as a function of real as well as imaginary $\mu$, 
as shown in fig.~\ref{potts} (right). 
Note that $M^c(\mu)$ rises with real chemical potential. {\it i.e.}~the first order region in fig.~\ref{fig:2schem} shrinks as finite baryon density is switched on. This system is thus an example of the non-standard scenario discussed in the previous section. 
The calculation also gives some insight in the problem of analytic continuation: while fig.~\ref{potts} clearly endorses the approach in principle, an $O(\mu^8)$-fit was required to reproduce the data
on both sides of $\mu^2=0$. Similar accuracy is much more difficult to achieve 
around the chiral line with present
resources.  

\subsection{Discussion of critical end point results}

The results about the critical surface from sec.~\ref{sec:crit} appear to be in qualitative contradiction 
with those of \cite{fk2}, \cite{ggpd}, which both conclude for the existence of a critical point
$(\mu_E,T_E)$ at small chemical potential $\mu_E/T_E \lsim 1$.  However,
in considering the reasons for such disagreement, one can see that the different data sets
are actually not inconsistent with each other, and the differing pictures can be explained
by standard systematic effects.

\begin{figure}[t]
\vspace*{-0.5cm}
\centerline{
\scalebox{0.62}{\includegraphics{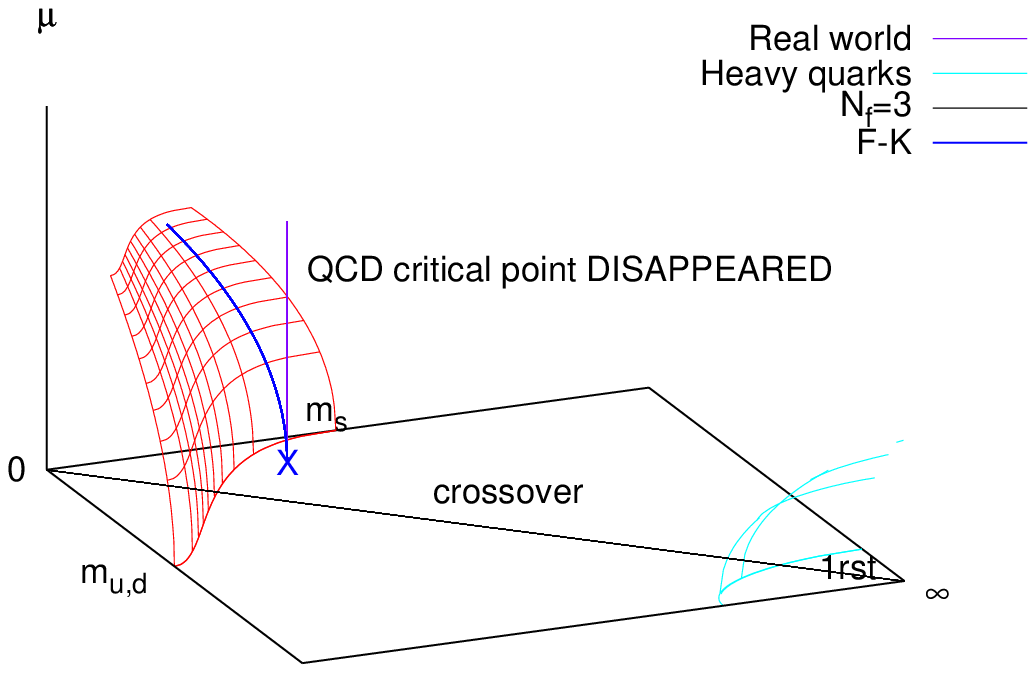}}
\scalebox{0.62}{\includegraphics{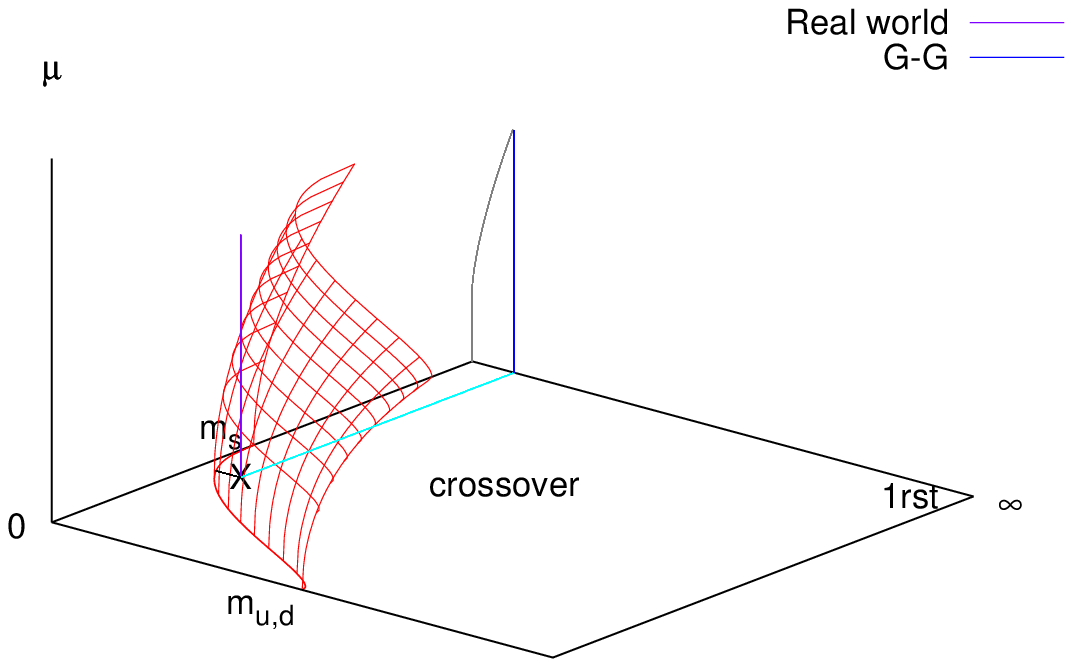}}
}
\caption{Left: Effect of keeping the quark mass fixed in lattice units while changing $\mu$, as in \cite{fk2}.
Right: Location of critical points for the $N_f=2+1$ and the $N_f=2$ theory, as considered in \cite{ggpd}.
Knowing the latter does not predict the former.}
\label{fig:syst}
\end{figure}

In \cite{ggpd} the critical point was inferred from an estimate of
the radius of convergence of the Taylor expansion of the free energy.
Regardless of the systematics when only 4 Taylor coefficients
are available, the estimate is made for $N_f=2$. 
The $(\mu,T)$ phase diagram of this theory might well be qualitatively different
from that of $N_f=2+1$ QCD, as illustrated in fig.~\ref{fig:syst} (right). 
Its critical endpoint point, obtained 
by intersecting a critical surface when going up vertically from the 
$N_f=2$ quark mass values,
is clearly a long way from the critical endpoint of physical QCD, and nothing follows quantitatively
from the value of one for the other.

In \cite{fk2} the double reweighting approach was followed.
By construction, this reweighting is performed at a quark mass fixed in lattice 
units:
$a m_{u,d}=\frac{m_{u,d}}{T_c} = const$. Since the critical temperature 
$T_c$ decreases 
with $\mu$, so does the quark mass. This decrease of the quark mass pushes
the transition towards first order, which might be the reason why a critical
point is found at small $\mu$. This effect is illustrated in the 
sketch fig.~\ref{fig:syst} (left), where the bent
trajectory intersects the critical surface, while the vertical line
of constant physics does not.
Put another way, \cite{fk2} measures the analogue of $c_1'$ instead of $c_1$
in eq.~(\ref{c1}).
From fig.~\ref{fk} (right) we see that the distance of the physical theory from
criticality stays initially constant, consistent with a coefficient $c'_1\approx 0$ 
as that in \cite{fp3}, sec.~\ref{sec:crit}. 
Taking the variation $a(\mu)$
into account could then make a dominant contribution, which might possibly change the results qualitatively.

Conversely, in \cite{fp3} only the leading order coefficient of $m^c(\mu)$ has been determined
from imaginary $\mu$. 
It cannot yet be excluded that there are cancelling terms
of higher order, alternating in sign. Such contributions would no 
longer cancel after continuation to real $\mu$, leading to a different picture.
Konwledge of the next term in the series will clarify this.

On the other hand, 
a robust finding is the high quark mass sensitivity
of the critical point: irrespective of the sign, if 
$c_1\sim O(1)$ in eq.~(\ref{c1}), 
$m^c(\mu)$ is a slowly varying function of $\mu$, just as the pressure, screening
masses or $T_c$. Hence $\mu_E(m)$ is rapidly
varying. A change of quark masses by a few percent will then imply
a change of $\mu_E$ by $O(100\%)$.
One should also remember that most investigations so far have used unimproved 
staggered quarks on coarse $N_t=4$ lattices only. This might be worrisome
given the exceedingly light quarks involved, cf.~sec.~\ref{latte} and \cite{sharpe,gss}.
Finally, a more complicated picture with additional
critical surfaces in fig.~\ref{fig:2schem} is yet another possibility.
In light of these circumstances even the qualitative features of the QCD phase 
diagram cannot be regarded as settled until they are probed with better
accuracy on finer lattices.

\section{Conclusions}

We have summarized our current understanding of finite temperature and density QCD
from numerical studies on the lattice. 
While many results in the pure gauge theory are
available in the continuum limit, simulations with
dynamical fermions still suffer from systematic errors.
These are mainly due to the finite lattice spacing, and quark masses
which don't meet their physical values yet.
For the critical temperature and the equation
of state these problems are now being tackled and the first continuum extrapolations
become available.

Existing results
provide us with a detailed picture
of how equilibrium plasma properties change through the deconfinement
transition up to a few $T_c$.
Combinations of perturbative calculations and numerical
simulations have produced insight into the regime of very high
temperatures as well as the dynamics and mixing of hard and soft momentum modes.
Altogether this provides a quantitative understanding of the relevant
length scales in the plasma, as well as tests of the applicability
of thermal perturbation theory.

The naive picture of the deconfined phase as a weakly interacting
parton gas is not supported. For temperatures relevant to
heavy ion collisions, the plasma displays strong residual interactions
through soft gluonic modes, which cannot be treated
perturbatively, and which influence different
quantities in different ways. 
This gives a consistent explanation to the
various observed features: the equation of state,
susceptibilities and fermionic correlators
are dominated by hard modes and not far
from their ideal gas values, but the corrections are significant.
Gluonic correlators, on the other hand,
are dominated by soft modes and entirely off their leading perturbative
predictions. An ideal gas is established
only at asymptotically high temperatures.

Significant progress was made over the last five years regarding 
dynamical quantities as well as finite density simulations. 
In both fields methods have been developed, that
are currently being scrutinized for their reliability. 
Calculations of spectral functions provide
a picture of remnant binding forces in the plasma phase as well as first semi-quantitative
results for the transport coefficients. Calculations at finite density are 
now possible for $\mu/T\lsim 1$,
with good agreement between all methods whenever equal parameter sets are compared.
However, the way to a quantitative understanding of the QCD phase 
diagram is still long, mainly due to
the high quark mass and cut-off sensitivity of the critical endpoint. 
Current developments, also regarding computing speed-up for 
different fermion discretizations and improvement programmes,
give reason to believe that these questions can be significantly 
improved upon in the near future.

\end{document}